\documentclass[12pt]{article}
\usepackage{fullpage,epsfig,graphics,amsbsy,amssymb,cancel,slashed,mathrsfs}
\usepackage{psfrag,hyperref}
\usepackage{graphicx,color}
\usepackage{wrapfig}
\usepackage{subfigure}

\newcommand{\be}{\begin{eqnarray}}
\newcommand{\ee}{\end{eqnarray}}
\usepackage{amsmath}
\numberwithin{equation}{section}
\usepackage{caption}
\usepackage[nosort]{cite}
 \usepackage[bulletsep]{collref}

\newcommand{\bea}{\begin{eqnarray}}
\newcommand{\eea}{\end{eqnarray}}  
\newcommand{\nn}{\nonumber}
\newcommand{\Tr}{\textrm{Tr}}

\newcommand{\NN}{\mathcal{N}}

\newcommand{\tphi}{\tilde\phi}
\newcommand{\tm}{\tilde{m}}

\begin{document}

\thispagestyle{empty}
\begin{flushright} \small
UUITP-06/15
 \end{flushright}
\smallskip
\begin{center} \LARGE
{\bf  Phase transitions in 5D super Yang-Mills theory}
 \\[12mm] \normalsize
{\bf  Anton Nedelin} \\[8mm]
 {\small\it
  Department of Physics and Astronomy,
     Uppsala university,\\
     Box 516,
     SE-75120 Uppsala,
     Sweden\\
   }
\end{center}
\vspace{7mm}
\begin{abstract}
 \noindent In this paper we study a phase structure of $5D$ ${\cal N}=1$ super Yang-Mills 
 theory with massive matter multiplets and $SU(N)$ gauge group. In particular, we are interested  
 in two cases: theory with $N_f$ massive hypermultiplets in 
 the fundamental representation and theory with one adjoint massive hypermultiplet. 
 If these theories are considered on $S^5$ their partition 
 functions can be localized to matrix integrals, which can be 
 approximated by their values at saddle points in the large-$N$ limit.  We solve 
 saddle point equations corresponding to the decompactification limit of both theories. 
 We find that in the case of the fundamental hypermultiplets theory experiences 
 third-order phase transition when coupling is varied. We also show that in the case of one adjoint 
 hypermultiplet theory experiences 
 infinite chain of third-order phase transitions, while interpolating between weak and strong coupling regimes.
 
 \end{abstract}

\eject
\normalsize

\tableofcontents

\section{Introduction and Main Results}

Recently, $5D$ super Yang-Mills (SYM) theory attracted much attention due to its relation with $6D$ $(2,0)$
superconformal theory discussed in \cite{Lambert:2010iw,Douglas:2010iu}. Using localization \cite{Pestun:2007rz}
it is possible to reduce full path integral of $5D$ $SU(N)$ SYM to a finite-dimensional 
matrix integral \cite{Kallen:2012cs,Kallen:2012va}. The last one appears to be solvable in certain 
limits. In particular the free energy of $5D$ ${\cal N}=1$ SYM with the adjoint hypermultiplet 
was derived using localization in \cite{Kim:2012av,Kallen:2012zn,Minahan:2013jwa}. It was shown that the free energy 
of this theory behaves as $N^3$ in the strong coupling limit. This result is consistent with the well-known $N^3$
behavior of the free energy of $6D$ theory obtained from the supergravity considerations 
\cite{Klebanov:1996un,Henningson:1998gx}. Localization results were later generalized to 
  theories living on the manifolds with more complicated 
 geometries in \cite{Qiu:2013pta,Qiu:2013aga,Qiu:2014oqa},
 revealing the same $N^3$ behavior of the free energy with only difference in the prefactor.

Another $5D$ theory of interest
is the super Yang-Mills with the $USp(N)$ gauge group and infinite coupling constant, which is 
conformal fixed point in five dimensions.
This theory is especially interesting because it has known holographic dual in $AdS_6$ space \cite{Bergman:2012kr}.
This gauge theory was studied using localization as well in \cite{Jafferis:2012iv} and
\cite{Assel:2012nf}. It was found that the free energy of this SYM theory has $N^{5/2}$ 
behavior in  the planar limit. This result matches the result obtained from the $AdS/CFT$ correspondence. 
Finally this $5D$ CFT was also studied on the backgrounds with more general geometries, e.g.
on the squashed spheres,
in \cite{Alday:2014rxa,Alday:2014bta}. These results allow one to evaluate
supersymmetric R\'{e}nyi in five dimensions \cite{Alday:2014fsa,Hama:2014iea}.

One can also introduce a Chern-Simons term into $SU(N)$ SYM theory and consider 
the limit of the infinite Yang-Mills coupling. Then this theory is also  
 $5D$ superconformal fixed point, which 
was studied using localization in \cite{Minahan:2014hwa}.  
It was found that in this case the 
free energy of theory also behaves as $N^{5/2}$ when $N$ is large. This suggests the existence 
of the holographic dual of $5D$ supersymmetric Chern-Simons theory, though the dual is not known yet. 

Using localization, it was recently found that different supersymmetric theories have a very 
interesting phase structure. The matrix models
obtained from the localization of supersymmetric filed theories with the massive hypermultiplets were shown to
 experience a phase transition at some critical values of the couplings in the decompactification limit. First evidences 
of these phase transitions were obtained for $4D$ ${\cal N}=2^*$ SYM on $S^4$. It was found that 
in the limit of infinite radius this theory experiences infinite chain of 
phase transitions as coupling is varied from weak to strong. In particular,
 the authors of \cite{Russo:2013qaa} solved the matrix model obtained from the localization of  ${\cal N}=2^*$ SYM. 
They have found that the support of the eigenvalue density in this case consists of many intervals 
of the length equal to the mass $m$ of the adjoint hypermultiplet. On the boundaries of these intervals 
density distribution has cusps. When the 't Hooft coupling is increased the length of the support grows 
as well and at some points new cusps appears in the distribution. Emergence of new 
cusps signals about the phase transition at the corresponding point. 
Finally, at very strong coupling cusps of the distribution smooth out and the solution approaches the one 
obtained in \cite{Buchel:2013id}. This interesting phase structure of ${\cal N}=2^*$ SYM on $S^4$ was investigated 
in details in the series of papers 
\cite{Russo:2013qaa,Russo:2013kea,Russo:2013sba,Zarembo:2014ooa,Chen:2014vka,Chen-Lin:2015dfa}. Later these 
results were generalized to the decompactification limit of ${\cal N}=2^*$ SYM on the ellipsoids 
in \cite{Marmiroli:2014ssa}. 

Another $4D$ theory experiencing large-$N$ phase transition is ${\cal N}=2$ SYM with $2N$ massive 
hypermultiplets in the fundamental representation considered on $S^4$
\cite{Russo:2013kea}. However, phase structure of this theory appears to be much simpler. It
experiences only one phase transition, while interpolating between weak and strong coupling in the 
decompactification limit. From the matrix model point of view this phase transition takes place when 
the length of the support of the eigenvalue distribution becomes larger than two times the mass of the hypermultiplet.

Similar phase structures were also obtained in $3D$ supersymmetric theories.  
In \cite{Barranco:2014tla,Russo:2014bda} authors considered matrix model obtained 
from the localization of ${\cal N}=2$ supersymmetric Chern-Simons theory on $S^3$ coupled to 
$2N_f$ massive hypermultiplets in the fundamental representation. It was shown that 
in the decompactification limit this theory experiences third-order phase transition 
similar to the one in  ${\cal N}=2$ SYM with $2N$ massive fundamental hypermultiplets on $S^4$.
Finally these results were generalized to the case of massive 
deformations of the $ABJM$ theory on $S^3$ \cite{Anderson:2014hxa,Anderson:2015ioa}, for which the 
phase structure of the theory appears to be similar to the case of ${\cal N}=2^*$ SYM on $S^4$.
Namely, it was found that the theory experiences an infinite 
chain of third-order phase transitions, while interpolating between weak and strong coupling 
regimes. As in $4D$ case, these phase transitions are related to the emergence of new cusps 
in the distribution of the eigenvalue density.

In this paper we try to generalize the results of $3D$ and $4D$ to the case of $5D$ gauge theories\footnote{ 
Some phase transitions were already found before in $5D$ ${\cal N}=1$ Chern-Simons theory \cite{Minahan:2014hwa}
and in $5D$ ${\cal N}=1$ SYM theory with adjoint hypermultiplet \cite{Borla}. However these phase transitions
are different from the transitions described in this work.}. 
We are particularly interested in the $SU(N)$ ${\cal N}=1$ SYM coupled to either one adjoint or $N_f$ fundamental 
massive hypermultiplets. To work with these theories we use the matrix model obtained by localizing ${\cal N}=1$ SYM 
on $S^5$ with arbitrary hypermultiplet content \cite{Kallen:2012cs,Kallen:2012va}.
In general it is not possible to evaluate these  
matrix integrals. However in this paper we focus only on the large-$N$ limit where the matrix integrals are dominated 
by their saddle points. In the case of adjoint hypermultiplet this saddle point equations have been solved 
in the weak and strong coupling limits \cite{Kallen:2012zn,Minahan:2013jwa}. Here we send 
the radius of the five-sphere to infinity, which will simplify the saddle point equations and allow us 
to solve them exactly at any coupling. When the solution is observed we can evaluate 
supersymmetric observables in ${\cal N}=1$ SYM. In particular, we consider the free energy and the expectation 
value of the circular Wilson loop. Studying both these observables we can draw conclusions about details of 
the phase structure of the theory.

This analysis leads to the following picture of the phase structure of considered theories. 
In the case of the theory coupled to $N_f$ fundamental hypermultiplets there is only one 
third-order phase transition taking place at
\be
t_c=-\frac{8\pi^2}{m}\left(1-\frac{N_f}{2 N}\right)^{-1}\,,
\ee
where $t=g_{YM}^2 N$ is the definition of the 't Hooft coupling used throughout in this paper and $m$ is the 
mass of the hypermultiplets. Notice that in order to reach this critical point one should consider 
negative $g_{YM}^2$. As explained in section \ref{section:matrix:model} this does not 
contradict anything. Negative coupling arises in the theory due to the renormalization and doesn't spoil 
the convergence of the matrix integral. 

The phase structure of the $SU(N)$ ${\cal N}=1$ SYM with the adjoint hypermultiplet is even more interesting.
For this theory we observe an infinite chain of third-order phase transitions at the following  
critical points
\be
t_c^{(n)}=\frac{8\pi^2}{m}(n+1)\,.
\ee
In the strong coupling limit these phase transitions smooth out and approach the 
strong coupling solution found in \cite{Minahan:2013jwa}. A nature of these phase transitions 
is the same as in the mass-deformed $ABJM$ theory and in $4D$ ${\cal N}=2^*$ SYM. As we show, 
the eigenvalue density at the saddle point of the matrix model has cusps. The number of these 
cusps grows as we increase the 't Hooft coupling, and each time a new cusps appear the system experiences 
a phase transition. 

This paper is organized as follows. In section 2 we review some properties of the matrix model 
obtained by localizing $5D$ SYM on $S^5$. In particular we consider issue of renormalization 
of the coupling and find the decompactification limit of this model. In section 3 we solve the saddle point 
equations of the matrix model corresponding to the theory coupled to $N_f$ fundamental hypermultiplets 
in the decompactification limit. Then calculating the free energy of the matrix model we show that 
the theory experiences the third-order phase transition at the critical value of the coupling. In section 4 
we solve the matrix model with adjoint massive hypermultiplet in the decompactification limit and show 
how the chain of phase transitions described above arise in this case. Finally, in section 5 we 
discuss the effect of the squashing of  five-sphere on the phase transition. We find that in contrast to the 
$4D$ case discussed in \cite{Marmiroli:2014ssa}, phase transitions are not affected by the deformations 
of the sphere. Some technical details of calculations are included in the appendicies of the paper.

\section{Matrix Model}
\label{section:matrix:model}

To study the phase structure of $5d$ SYM we use the result of supersymmetric localization \cite{Kallen:2012cs,Kallen:2012va},
that reduces full field theory path integral to the finite-dimensional matrix integral given by

\be
Z&=&\int [d\tphi]~e^{-  \frac{8\pi^3 r}{g_{YM}^2}  \text{Tr}(\tphi^2)-\frac{\pi k}{3}\text{Tr}
(\tphi^3)}  Z_{\rm 1-loop}^{\rm vect} (\tphi)    Z_{\rm 1-loop}^{\rm hyper} (\tphi) + \mathcal{O} (e^{-\frac{16 \pi^3 r}
{g_{YM}^2}})~,\label{vh1loop-intro}
\ee
 Here $g_{YM}$ is Yang-Mills coupling, $k$ is Chern-Simons (CS) level, $r$ is the radius of $S^5$ and integration variable $\tphi$ is related to the expectation value of the scalar field $\sigma$ from vector multiplet 
 as the following $\tphi=-i r\sigma$.\footnote{Note that in order to have well defined 
 matrix integral we should integrate over real $\tphi$ or equivalently imaginary $\sigma$. In all our conventions we 
 follow \cite{Kallen:2012va}.}. One-loop contributions of the vector- and hypermultiplets are given by 
\bea
 Z_{\rm 1-loop}^{\rm vect} (\tphi) =\prod\limits_\beta\prod\limits_{t \neq 0}\left( t - \langle \beta,i\tphi\rangle  
 \right)^{(1+\frac{3}{2}t+\frac{1}{2}t^2)}~,\label{vect1-loop-beg}
\eea
 and
\begin{equation}\label{hyper-loop}
Z_{\rm 1-loop}^{\rm hyper} (\tphi) = \prod\limits_\mu \prod\limits_{t}\left( t - \langle \mu, i\tphi 
\rangle -i m+\frac{3}{2} \right)^{-(1+\frac{3}{2}t+\frac{1}{2}t^2)}~,
\end{equation}
where $\beta$ are the roots, $\mu$ are the weights of the representation $R$ and $m=-i M r$ with $M$ being the mass of the 
the hypermltiplet. The matrix model described above was well studied in some regimes. In particular, planar limit of the $SU(N)$ 
SYM was studied in  \cite{Kallen:2012zn,Minahan:2013jwa,Kim:2012qf}. In \cite{Jafferis:2012iv,Assel:2012nf} this model 
was studied for the case of $5d$ superconformal theory. And, finally, planar limit of the CS theory was 
studied in \cite{Minahan:2014hwa}.

In general (\ref{vh1loop-intro}) should also include instanton contributions. However we, in this paper will be interested
in the large-$N$ limit of the theory, which suppresses instanton contributions. Hence we omit 
all instanton contributions and consider only classical action together with one-loop determinants
in the partition function.

In this paper we put CS term to zero and concentrate on the decompactification limit of the $SU(N)$ SYM with different hypermultiplet content.
In all cases hypermultiplets are massive which leads to the phase transitions similar to the ones obtained 
 in \cite{Russo:2013qaa,Anderson:2014hxa,Russo:2013kea}. Before moving to the solutions of
the matrix model we should discuss general properties of this model. More detailed analysis 
of this matrix model can be found in \cite{Minahan:2013jwa}.

\subsection{Renormalization of the Coupling Constant}

As we can see the one-loop contributions for both vector- and hypermultiplets can be written in the following 
form
\be
Z_{\rm 1-loop}^{\rm vect} (\tphi)&=&\frac{1}{\langle \beta,i\tphi\rangle}{\cal P}(\langle \beta,i\tphi\rangle)\,,\\
Z_{\rm 1-loop}^{\rm hyper} (\tphi)&=&{\cal P}^{-1}\left(\langle \mu, i\tphi\rangle +i \tm-\frac{3}{2}\right)\,,
\ee
where ${\cal P}(x)$ is the infinite product
\be
{\cal P}(x)=  x \prod_{t=1}^\infty  \left ( t- x \right)^{(1+\frac{3}{2}t+\frac{1}{2}t^2)} \left ( t+ x \right)^{(1-\frac{3}{2}t+\frac{1}{2}t^2)}~.
\label{infin-prod}
\ee
This product is divergent with the divergent part given by
\be
\log {\cal P} = \sum_{t=1}^\infty \left ( -3x - \frac{x^2}{2} \right ) + {\rm convergent~ part}\,,
\ee
where we have also omitted $x$-independent divergent part. To regularize the product (\ref{infin-prod}) 
we introduce the UV cut-off at $t_0=\pi \Lambda_0 r$ leading to 
\be
\log {\cal P} = -\pi\Lambda_0 r\left ( \frac{x^2}{2} + 3 x \right )\,.
\ee
Using this expression we extract divergent terms in the one-loop determinants
\be
\log Z_{\rm 1-loop}^{\rm vect} (\tphi)&=& -\pi \Lambda_0 r
 \sum\limits_{\beta} \left[\frac{1}{2}(\langle \beta,i\tphi\rangle )^2+3 \langle \beta,i\tphi\rangle \right]+ {\rm convergent~ part}=\nn\\
& &\pi \Lambda_0 r~ C_2 ({\rm adj}) \text{Tr}(\phi^2 ) + {\rm convergent~ part}\label{loop-div-vect}\\
\log Z_{\rm 1-loop}^{\rm hyper} (\tphi)&=& \pi \Lambda_0 r
 \sum\limits_{\mu} \left[\frac{1}{2}\left(\langle \mu, i\tphi\rangle +i \tm-\frac{3}{2}\right)^2+3 \left(\langle \mu, i\tphi\rangle +i \tm-\frac{3}{2}\right) \right]+\nn\\
 & & {\rm convergent~ part}=-\pi \Lambda_0 r~ C_2 ({R}) \text{Tr}(\tphi^2 ) + {\rm convergent~ part}\label{loop-div-hyper}
\ee
Here $C_2(R)$ is the quadratic Casimir element for representation $R$, defined by
$\Tr (T_A T_B) = C_2(R) \delta_{AB}$. With these conventions  $\sum\limits_{\mu} ( \langle \tphi , \mu\rangle)^2 = 2 C_2(R) \Tr (\tphi^2)$.
Note that we have omitted terms linear in $\tphi$, because the gauge group is $SU(N)$. In this case the 
tracelessness condition implied for $\tphi$ forces all linear terms to be zero. 

The divergent terms in the one-loop contributions (\ref{loop-div-vect}) and (\ref{loop-div-hyper}) are 
proportional to $\Tr(\tphi^2)$. So in the full matrix integral (\ref{vh1loop-intro}) these divergent terms 
can be absorbed into the renormalization of the Yang-Mills coupling
\be\label{coup-renorm}
 \frac{1}{g_{eff}^2} = \frac{1}{g_{YM}^2} - \frac{\Lambda_0}{8\pi^2} \left (C_2 ({\rm adj}) - 
 \sum\limits_{I} C_2(R_I) \right)~,  
\ee
where the sum in the second term is over all hypermultiplets of the theory.
Equation (\ref{coup-renorm}) reproduces the renormalization obtained in flat space using
perturbation theory in \cite{Flacke:2003ac}.

Further we will consider theories with particular matter contents, leading to different effective coupling 
constants. In each case we will come back to (\ref{coup-renorm}) and consider particular examples of 
this renormalization.

After the regularization described above finite parts of the one-loop contributions 
(\ref{loop-div-vect}) and (\ref{loop-div-hyper}) can be written in the form 
\be
\log Z_{\rm 1-loop}^{\rm vect(reg)}(\tphi)\equiv -{\rm tr}_{Ad} ~F_{V}(\tphi)\,,\label{1-loop-vect-reg}\\
\log Z_{\rm 1-loop}^{\rm hyper(reg)}(\tphi)\equiv -\sum\limits_{I}{\rm tr}_{R_I}~ F_{H}(\tphi)\,,\label{1-loop-hyp-reg}
\ee
where $F_V(\phi)$ and $F_H(\phi )$ are functions defined as follows
\be
F_V(x)&=&-\frac{1}{2}\log\left(\sinh^{2}(\pi x)\right)-\frac{1}{2}f(i x)\,,\label{FV-funct}\\
F_H(x)&=&\frac{1}{4} l\left(\frac{1}{2}-i\tm -ix\right)+\frac{1}{4} l\left(\frac{1}{2}+i\tm +ix\right)\nn\\
&+&\frac{1}{4} f\left(\frac{1}{2}-i\tm-ix\right)+\frac{1}{4} f\left(\frac{1}{2}+i\tm+ix\right)\,.\label{FH-funct}
\ee
Here functions $f(x)$ and $l(x)$ are the ones introduced in  \cite{Kallen:2012cs} and \cite{Jafferis:2010un}
correspondingly. For the definitions and properties of these functions see appendix \ref{spec_functions}.
Using results of the regularization it is convenient to write down the matrix integral (\ref{vh1loop-intro})
in the following form
\be\begin{aligned}
Z&=\int\limits_{\rm Cartan} [d\tphi] e^{-F(\tphi)}\,,\\
F(\tphi)&=\frac{8\pi^3 r}{g_{eff}^2}{\rm tr}\left(\tphi^2\right)+{\rm tr}_{Ad} F_V(\tphi)+
\sum\limits_{I}{\rm tr}_{R_I} F_H(\tphi)\,.\label{partition-reg}
\end{aligned}\ee

Another observable that can be evaluated using localization technique is 
the expectation value of the circular Wilson loops 
\be
\langle W(C)\rangle\equiv\left\langle\frac{1}{N}\Tr ~\mathcal{P}~ \exp\left(\oint\limits_C d\tau\left(i A_{\mu}\dot{x}^{\mu}
+\sigma|\dot{x}|\right)\right)\right\rangle\,,
\label{wilson:def}
\ee
where $C$ is wrapping the equator of $S^{5}$. Such loops were considered in  
\cite{Young:2011aa} and \cite{Minahan:2013jwa} for $5D$ $SU(N)$ SYM theory and in \cite{Assel:2012nf} for 
$5D$ superconformal theories. After the localization expectation value (\ref{wilson:def}) in the 
case of fundamental representation 
turns into the expectation value of the matrix model (\ref{partition-reg}) \cite{Pestun:2007rz}
\be
\langle W\rangle=\frac{1}{N}\langle \Tr \,e^{2\pi\tilde\phi}\rangle\,.
\label{wilson:matrix}
\ee

\subsection{Decompactification limit}

The matrix integral (\ref{partition-reg}) is not solvable even in the planar limit. 
Some particular limits of this matrix model were studied in \cite{Kallen:2012zn,Minahan:2013jwa}
 for the case of one adjoint hypermultiplet and $N_f$ fundamental hypermultiplets. 
In this paper we study the behavior of the theory with $SU(N)$ gauge group and different 
hypermultiplets content in the decompactification (infinite radius) limit. 

In all equations above we have used the dimensionless variables $\tphi$ and $\tm$, or equivalently 
we can assume that the radius of the sphere was set to one. To reproduce the $r$-dependence 
of the matrix integral we should rescale our variables as follows\footnote{From now on we will denote all dimensionless 
parameters with tilde and corresponding dimensionfull ones - without tilde}:
\be
\tphi\to \phi r\,,\qquad \tm\to m r\,.
\label{r:depend}
\ee
The decompactification limit then corresponds to the limit $r\to\infty$. 
Using asymptotic expressions (\ref{asymptotic_F})  we can find that in this limit  
$F(\phi)$ in (\ref{partition-reg}) can be written as
\be\begin{aligned}
\frac{1}{2\pi r^3}F(\phi)&=\frac{4\pi^2}{g_{eff}^2}{\rm tr}\left(\phi^2\right)+{\rm tr}_{Ad}\left(\frac{1}{12}|\phi|^3-
\frac{1}{2 r^2}|\phi|\right)\\
&-\sum\limits_I {\rm tr}_{R_I}\left(\frac{1}{12}|\phi +m|^3+\frac{1}{16 r^2}|\phi +m|\right)\,.
\label{asymptotic_free}\end{aligned}
\ee
Here we also included $1/r^2$ terms linear in $\phi$.
These terms will be important when we discuss the effects of finite $r$ in the large radius limit. 

We see that in the decompactification limit the matrix model simplifies a lot, as (\ref{asymptotic_free}) 
does not contain any complicated functions and, as we show further, can be solved explicitly in the planar limit.
However, this is not the only reason to consider this limit. As mentioned in the introduction, studies
of the decompactification limit in different $3D$ and $4D$ theories revealed interesting phase 
structure of these theories. 
For instance, phase transitions  were found in 
$3D$ Chern-Simons coupled to the massive fundamental hypermultiplet \cite{Barranco:2014tla},
 in $4D$ $\mathcal{N}=2$ super-QCD with massive quarks in the Veneziano limit
\cite{Russo:2013kea}, in mass-deformed $ABJM$ theory \cite{Anderson:2014hxa,Anderson:2015ioa} and finally in 
 $4D$ $\mathcal{N}=2^*$ SYM theory \cite{Russo:2013qaa}. In each case theory was considered in 
 the decompactification limit and it was shown that in the finite volume phase transitions disappear.
We expect to observe similar picture in $5D$ SYM theory and hence we should also concentrate 
on the decompactification limit of the corresponding matrix integral (\ref{partition-reg}).

All considerations above are general. They can be applied to any gauge group and hypermultiplet content. 
From now on we will concentrate on the $SU(N)$ gauge group, meaning that $C_2(adj)=N$ and $C_2(fund)=\frac{1}{2}$.
Inspired by the previous results on the phase structure of the supersymmetric gauge theories we consider 
two type of theories: ${\cal N}=1$ SYM coupled to $N_f$ massive fundamental hypermultiplets and 
${\cal N}=1$ SYM coupled to one massive adjoint hypermultiplet.

\section{$\NN=1$ SYM with $N_f$ fundamental hypermultiplets}
\label{N=1:fund:sec}

In this section we consider $\mathcal{N}=1$ SYM containing vector multiplet and $N_f$ hypermultiplets with the mass
$m$. Half of the hypermultiplets is in fundamental, while another half is in antifundamental representation.\footnote{For shortness
we often address this case as the case of $N_f$ fundamental hypermultiplets in this paper.} 
Before working out matrix model equations of motion and its solutions we should discuss the renormalization 
issues, as they play very important role in this case. According to (\ref{coup-renorm}) effective coupling in the 
case of $N_f$ fundamental hypermultiplets have the form 
\be
\frac{1}{g_{eff}^2}=\frac{1}{g_0^2}-\frac{\Lambda_0 N}{8\pi^2}\left(1-\frac{1}{2}\zeta\right)\,,
\label{coup-renorm-fund}
\ee
where $\zeta\equiv N_f/N$ is the Veneziano parameter. Note that the theory has constraint 
on $\zeta$ that can be found from the decompactification limit expression 
(\ref{asymptotic_free}). We want $F(\phi)$ to be positive at large $\phi$ in 
oder for the matrix integral (\ref{partition-reg}) to be convergent. In the case 
of the $SU(N)$ gauge group with $N_f/2$ fundamental and $N_f/2$ antifundamental hypermultiplets  we obtain:
\be\begin{aligned}
\frac{1}{2\pi r^3}F(\phi)=\sum_{j\neq i}\sum_i \frac{1}{12}|\phi_i-\phi_j|^3
-N_f\sum_i \frac{1}{24}\left(|\phi_i+m|^3+|\phi_i-m|^3\right)+\dots=\\
\frac{1}{12}\left(2 N_c-N_f\right)\sum_{i}|\phi_i|^3+\dots
\label{fund-free-asymp}
\end{aligned}\ee
where by the ellipses  we mean all terms subleading in $1/|\phi|$. From this expression we see that 
 in order to have well defined matrix integral, parameters of the theory should satisfy
 $ N_f\leq 2 N$ or $\zeta\leq 2$ in terms of the Veneziano parameter. 
 The same restriction was obtained from the flat space prepotantial in
\cite{Intriligator:1997pq}.

According to (\ref{coup-renorm-fund}) YM coupling $g_{eff}^2$ can be negative, except in the case of $\zeta=2$.
Further in our considerations we will always assume $\zeta<2$.
Then we can redefine UV cut-off $\Lambda_0$ so that 
\be
\frac{1}{g_{eff}^2}=-\frac{\Lambda N}{8\pi^2}\,.
\ee
and use it in the matrix model further. 

Now we are ready to solve the  matrix model with described matter content and coupling 
renormalization. In the planar limit the matrix integral (\ref{partition-reg}) is dominated by the saddle point
satisfying 
\be
\frac{d F(\tphi)}{d\tphi_k}=\frac{d}{d\tphi_k}\left(\frac{8\pi^3 r}{g_{eff}^2}\sum_i \tphi_i^2+
\sum_{j\neq i}\sum_i F_V(\tphi_i-\tphi_j)+\frac{N_f}{2} \sum_i \left(F_H(\tphi_i)+F_H(-\tphi_i)\right)\right)&=&0\,.\nn\\
&&
\ee
Using expressions (\ref{derivtives-F}) for the derivatives of $F_V$ and $F_H$ we obtain the 
saddle point equations
\be\begin{aligned}
\sum_{j\neq i}\left(2-(\tphi_i-\tphi_j)^2\right)\coth(\pi(\tphi_i-\tphi_j))+\frac{N_f}{4}\left(\frac{1}{4}+
(\tphi_i+\tm)^2\right)\tanh(\pi(\tphi_i+\tm))+\\\frac{N_f}{4}\left(\frac{1}{4}+(\tphi_i-\tm)^2\right)\tanh(\pi(\tphi_i-\tm))=
\frac{16\pi^2 r}{g_{eff}^2}\tphi_i=-2\Lambda N r \tphi_i\,.
\label{eq-fund-full}
\end{aligned}\ee
After taking the decompactification limit we arrive to
\be\begin{aligned}
-2\Lambda N \phi_i=\frac{1}{4}N_f(\phi_i +m)^2{\rm sign}(\phi_i +m)+\frac{1}{4}N_f(\phi_i -m)^2{\rm sign}(\phi_i -m)
-\\
\sum_{j\neq i}(\phi_i-\phi_j)^2 {\rm sign}(\phi_i-\phi_j)\,.
\end{aligned}\label{eom-fund-disc}\ee
This equation can be also observed from (\ref{asymptotic_free}). Introducing usual eigenvalue density 
\be
\rho(\phi)\equiv \frac{1}{N}\sum_i \delta(\phi-\phi_i)\,,
\label{density:def}
\ee
we rewrite the saddle point equation (\ref{eom-fund-disc}) as the integral equation
\be\begin{aligned}
-2\Lambda\phi=\frac{1}{4}\zeta(\phi +m)^2{\rm sign}(\phi +m)+\frac{1}{4}\zeta(\phi -m)^2{\rm sign}(\phi -m)
-\\
\int d\psi \rho(\psi)(\phi-\psi)^2 {\rm sign}(\phi-\psi)\,.
\label{eom-fund-int}
\end{aligned}\ee

We assume that the eigenvalue density $\rho(\phi)$ has finite support $[-a,a]$. Then 
we should consider two separate cases. In the first case, which corresponds 
to the phase of the theory below the
transition, $a<m$ is satisfied.
In the second case we have $a>m$, which corresponds to the phase above the transition point.

To solve equation (\ref{eom-fund-int}) we differentiate it three times:
\be
-2\Lambda=\frac{1}{2}\zeta(\phi +m){\rm sign}(\phi +m)+\frac{1}{2}\zeta(\phi -m){\rm sign}(\phi -m)-\nn\\
2\int d\psi\rho(\psi)(\phi-\psi){\rm sign}(\phi-\psi)\label{1st:der:fund}\,,\\
0=\frac{1}{2}\zeta{\rm sign}(\phi +m)+\frac{1}{2}\zeta{\rm sign}(\phi -m)-2\int d\psi\rho(\psi){\rm sign}(\phi-\psi)
\label{2nd:der:fund}\,,\\
0=\zeta\left(\delta(\phi +m)+\delta(\phi-m)\right)-4\rho(\phi)\label{3rd:der:fund}\,,
\ee
where (\ref{1st:der:fund}),(\ref{2nd:der:fund}) and (\ref{3rd:der:fund}) correspond to 
the first, second and third derivatives of the equation (\ref{eom-fund-int}) respectively.
Now we consider these equations for the cases $a<m$ and $a>m$ separately.

\subsection{Solution below the transition point}

When the theory is below the transition point (Phase I) terms with masses in (\ref{2nd:der:fund}) 
cancel each other and we are left with
\be
0=\int\limits_{-a}^{a}d\psi\rho(\psi){\rm sign}(\phi-\psi)=
\int\limits_{-a}^{\phi}d\psi\rho(\psi)-\int\limits_{\phi}^{a}d\psi\rho(\psi)\,.
\label{fund:below:equation:2nd}
\ee
This equation should be satisfied for any $\phi\in[-a,a]$, hence we conclude that 
$\rho(\phi)$ is sharply peaked near the endpoints of the distribution at $\phi=\pm a$.
Then the solution to (\ref{fund:below:equation:2nd}) is given by
\be
\rho^{(I)}(\phi)=\frac{1}{2}\left(\delta(\phi-a)+\delta(\phi+a)\right)\,,
\label{density:fund:below}
\ee
where the overall coefficient $1/2$ comes from the standard normalization 
condition 
\be
\int d\phi \rho(\phi)=1\label{normalization}\,.
\ee
To be more precise the arguments of the $\delta$-functions in (\ref{density:fund:below}) should look like ($\phi\pm a\mp \epsilon$)
with $\epsilon\to0$. This small shift should be done in order to include the whole delta function 
into the density support $[-a,a]$. Everywhere further we imply these shifts of 
the $\delta$-function arguments at the distribution endpoints.

Note that (\ref{density:fund:below}) does not satisfy (\ref{3rd:der:fund}) at the 
endpoints of distribution. The best way to make the solution (\ref{density:fund:below}) more rigorous is to consider 
equations of motion (\ref{eq-fund-full}) in the limit of large but finite $r$, 
taking into account terms subleading in $1/r$. These terms correspond to the 
repulsion of the eigenvalues at small separations, which washes out $\delta$-functions and 
turn them into the peaks of $1/r$ width. The form of these peaks can be found analytically. 
Then the solution (\ref{density:fund:below}) can be obtained in the limit $r\to\infty$. Details 
of these calculations can be found in appendix \ref{finite:R:fund}.

To determine the position of the endpoint $a$, we substitute solution (\ref{density:fund:below})
back into the original equation (\ref{eom-fund-int}) which leads to
\be
a=\Lambda+\frac{1}{2}\zeta m
\label{endp:fund:below}\ee
By construction the  solution (\ref{density:fund:below}) works when $a<m$ or 
$\Lambda<m\left(1-\frac{1}{2}\zeta\right)$. Notice that the r.h.s of (\ref{endp:fund:below}) is 
always positive due to the condition $\zeta<2$, we have discussed previously.

\subsection{Solution above the transition point}

Now we increase $\Lambda$, pass the point $\Lambda_c=m\left(1-\frac{1}{2}\zeta\right)$ and 
arrive to the phase above the transition point (Phase II), where we have $a>m$.
To solve (\ref{eom-fund-int}) we again use its three derivatives (\ref{1st:der:fund})-(\ref{3rd:der:fund}).
First we address (\ref{3rd:der:fund}). In the case $a<m$ terms $\delta(\phi\pm m)$ do not contribute into 
this equation, because arguments of these $\delta$-functions are never zero. 
However, if $a>m$ $\delta$-functions start contributing into the eigenvalue density. Then it is natural to assume 
 that solution contains these $\delta$-functions on top of the solution (\ref{density:fund:below})
\be
\rho^{(II)}(\phi)=\frac{1}{4}\left[(2-\zeta)\delta(\phi +a)+\zeta\delta(\phi +m)
+\zeta\delta(\phi -m)+(2-\zeta)\delta(\phi-a)\right]\,.
\label{ansatz:fund}
\ee
Here the coefficients in front of $\delta(\phi\pm m)$ are chosen to satisfy  
(\ref{3rd:der:fund}) and the coefficients in front of $\delta(\phi\pm a)$ are found from the normalization 
condition (\ref{normalization}). 


\begin{figure}[!h]
\begin{center}
  \subfigure[$\Lambda=2.7 \left(\Lambda<\Lambda_c\right)$]{\label{fund:below}\includegraphics[width=52mm,angle=0,scale=1.5]{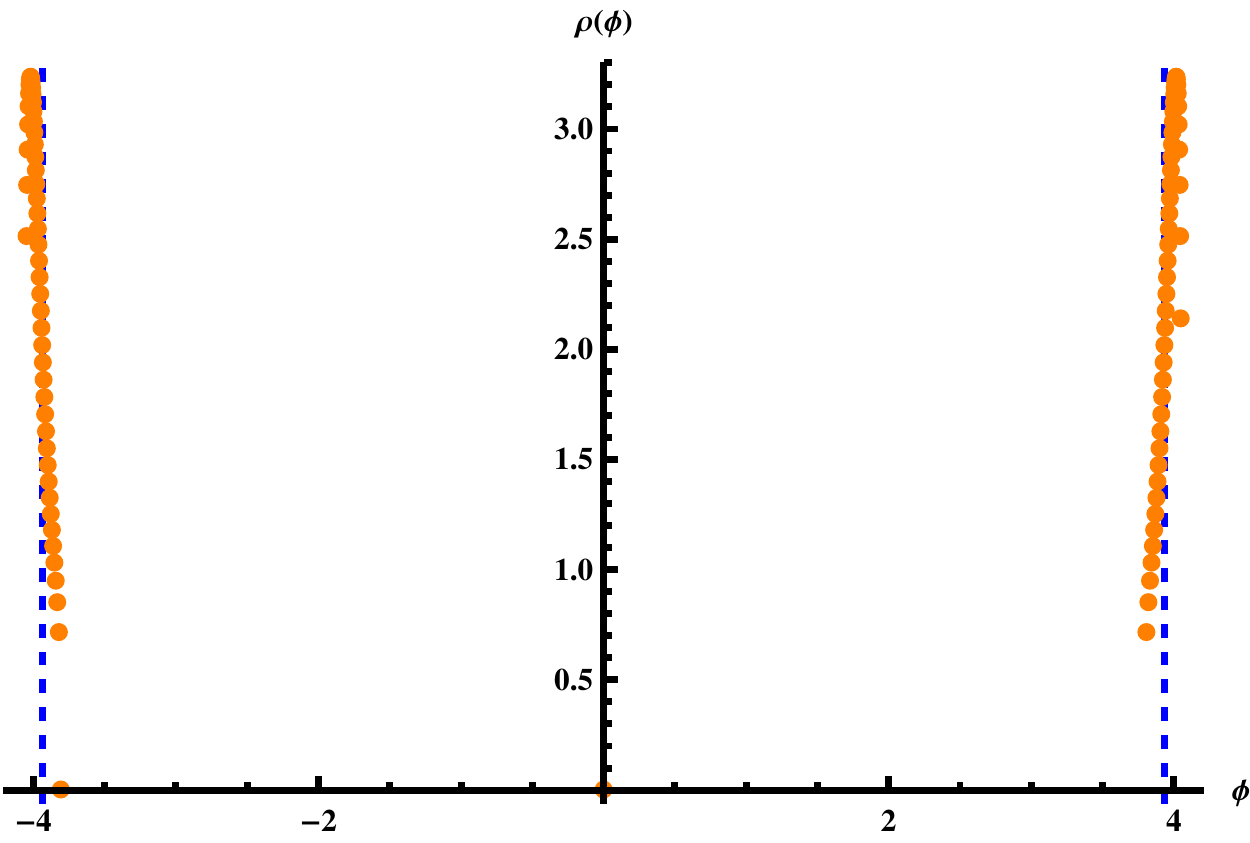}}
  \subfigure[$\Lambda=4.7 \left(\Lambda>\Lambda_c\right)$]{\label{fund:above}\includegraphics[width=52mm,angle=0,scale=1.5]{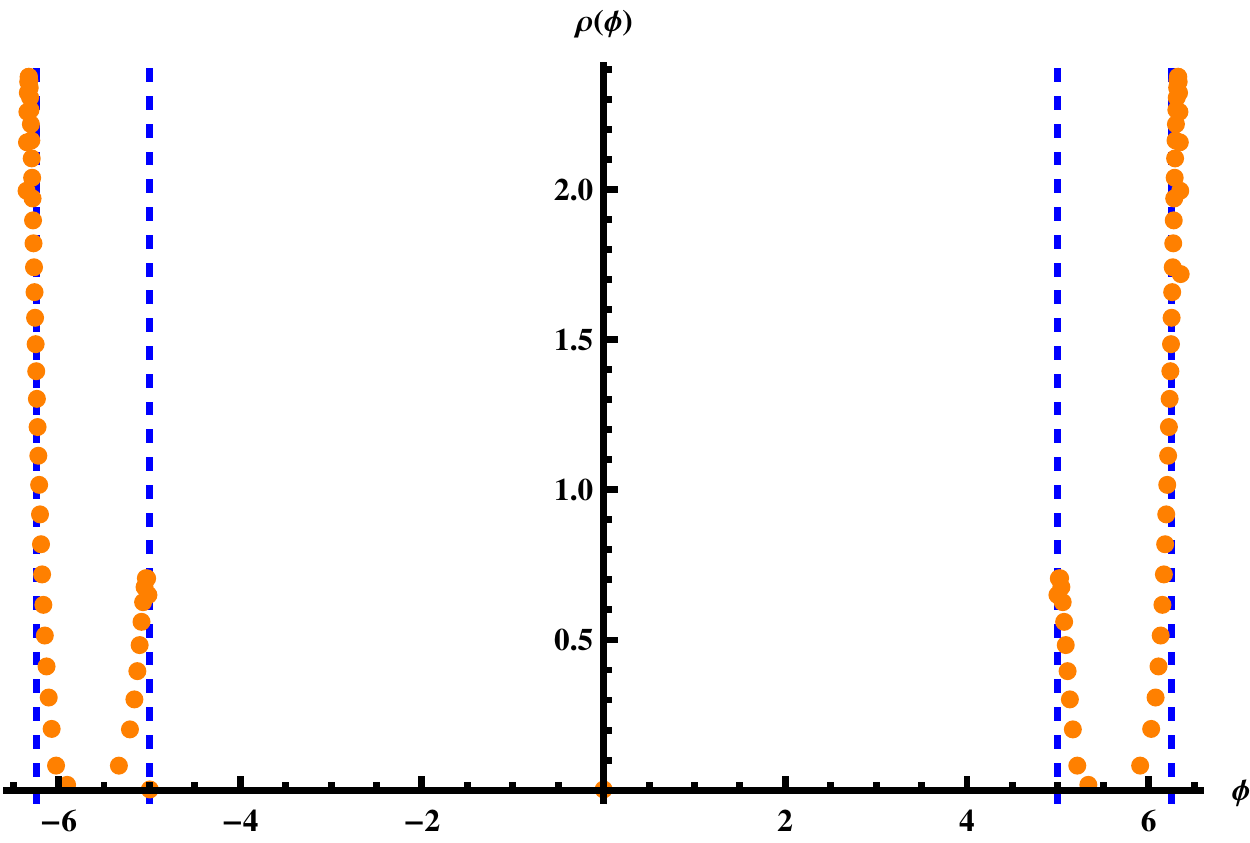}}
\end{center}
\caption{Eigenvalue density $\rho(\phi)$ calculated with the following value of the parameters: $r=10,\,m=5,\, \zeta=
\frac{1}{2},\,N=100$ corresponding to $\Lambda_c\approx 3.76$. The orange dots show the 
results for the numerical solution, while the dashed blue 
lines show the positions of the $\delta$-functions in the analytical 
solutions (\ref{density:fund:below}) and (\ref{ansatz:fund}).}
\label{fund:dens}
\end{figure}

Now we substitute the ansatz (\ref{ansatz:fund})
into original equation of motion (\ref{eom-fund-int})
in order to obtain the position of the distribution 
endpoint $a$
\be
a=\left(1-\frac{1}{2}\zeta\right)^{-1}\Lambda\,.
\label{endp:fund:above}
\ee
This solution works when $\Lambda>m\left(1-\frac{1}{2}\zeta\right)$. Thus we covered 
all the values of $\Lambda$ and, unlike in $3D$
Chern-Simons with $N_f$ fundamental hypermultiplets \cite{Barranco:2014tla},  we 
do not observe any intermediate phase. The only transition takes place at the 
critical point
\be
\Lambda_c=m\left(1-\frac{1}{2}\zeta\right)
\label{critical:pt:fund}
\ee
The solution found here suffers the same problems as (\ref{density:fund:below}). To justify 
it we again consider equations of motion with large but finite $r$ and take the decompactification limit 
in the very end. This calculation is done in appendix \ref{finite:R:fund}, where we 
reproduce the eigenvalue density (\ref{ansatz:fund}) in the limit $r\to\infty$.

We also check solutions (\ref{density:fund:below}) and (\ref{ansatz:fund})
numerically. In Fig.\ref{fund:dens} the orange dots show the results 
for the numerical solution  of the full equation of motions (\ref{eq-fund-full}). On the same plots the 
dashed blue lines denotes the positions of the $\delta$-functions in our analytical solutions for the densities 
below (\ref{density:fund:below}) and above (\ref{ansatz:fund}) the transition point. As we see from
these plots, our analytical solutions reproduce the result of the numerical evaluation very well. 
For some technical details of the numerical simulation see appendix \ref{numerix:appendix}.

\subsection{Free energy and the order of transition}

We now determine if the transition between  phases I and II is phase transition and 
, if so, what is the order of this transition. To answer these questions we 
find the behavior of the free energy near the critical point $\Lambda_c$. 
To calculate the free energy in the decompactification limit we directly use 
the asymptotic formula (\ref{asymptotic_free}) with an appropriate matter content. 
In the large-$N$ limit we express the free energy in the integral form
\be\begin{aligned}
\frac{1}{2\pi r^3 N^2}F=-\frac{1}{2}\Lambda\int d\phi \rho(\phi)\phi^2+\frac{1}{12}\int\int d\phi d\psi
\rho(\phi)\rho(\psi)|\phi-\psi|^3-\\
\frac{\zeta}{24}\int d\phi \rho(\phi)\left(|\phi +m|^3+|\phi-m|^3\right)\,.
\end{aligned}\label{free:fund:int}\ee
These integrals can be evaluated both below and above the phase transition point using 
corresponding expressions (\ref{density:fund:below}) and (\ref{ansatz:fund}) for the eigenvalue density.
After some calculations we obtain
\be\begin{aligned}
F^{(I)}&=-\frac{\pi r^3 N^2}{24}\left(8\Lambda^3 +12 m\zeta\Lambda^2+6m^2\zeta^2\Lambda+m^3\zeta (4+\zeta^2) \right)\,,\\
F^{(II)}&=-\frac{\pi r^3 N^2}{6}\left(\frac{4}{2-\zeta}\Lambda^3+3 m^2 \zeta\Lambda+m^3\zeta^2\right)\,.
\end{aligned}\label{free:en:fund}\ee
where $F^{(I)}$ and $F^{(II)}$ are the free energies of phase I ($\Lambda<\Lambda_c$) and phase II ($\Lambda>\Lambda_c$)
respectively. Using (\ref{free:en:fund}) we find that the free energy has a discontinuity 
 in its third derivative at the critical point
\be
\left.\partial_{\Lambda}\left(F^{(II)}-F^{(I)}\right)\right|_{\Lambda=\Lambda_c}&=&
\left.\partial^{2}_{\Lambda}\left(F^{(II)}-F^{(I)}\right)\right|_{\Lambda=\Lambda_c}=0\,,\\
\left.\partial^{3}_{\Lambda}\left(F^{(II)}-F^{(I)}\right)\right|_{\Lambda=\Lambda_c}&=&2\pi r^3 N^2\frac{\zeta}{\zeta-2}\,.
\ee
Thus at the critical point $\Lambda_c=\left(1-\frac{1}{2}\zeta\right) m$ the theory experience 
third-order phase transition.
Third-order phase transitions are typical for matrix models in the planar limit and has been
obtained for many different
systems \cite{Gross:1980he,Boulatov:1986sb,Douglas:1993iia,Anderson:2014hxa}. As we have mentioned before there are 
examples of the phase transitions similar to the one we have described in this section. These are phase transitions
in $3D$ Chern-Simons-Matter theory \cite{Barranco:2014tla,Russo:2014bda} and $4D$ super-QCD with $2N_f$ massive hypermultiplets
\cite{Russo:2013kea}. In both examples the phase transition of the theory in the decompactification 
limit at large-$N$ is of third order. 

\begin{figure}[!h]
\begin{center}
  \subfigure[Free Energy $F$]{\label{free:energ:fund:pic}\includegraphics[width=51mm,angle=0,scale=1.6]{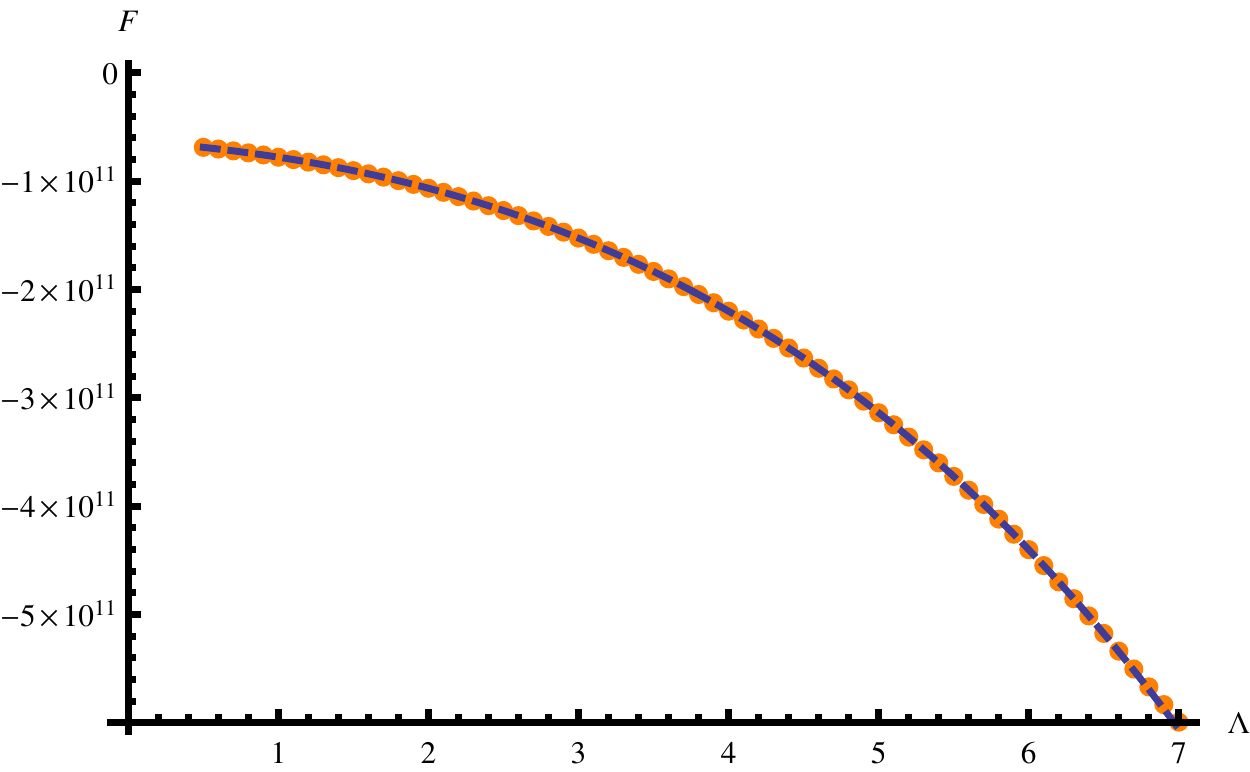}}
  \subfigure[Third derivative of the free energy $\partial_{\Lambda}^3 F$]{\label{free:energy:der:fund:pic}\includegraphics[width=51mm,angle=0,scale=1.6]{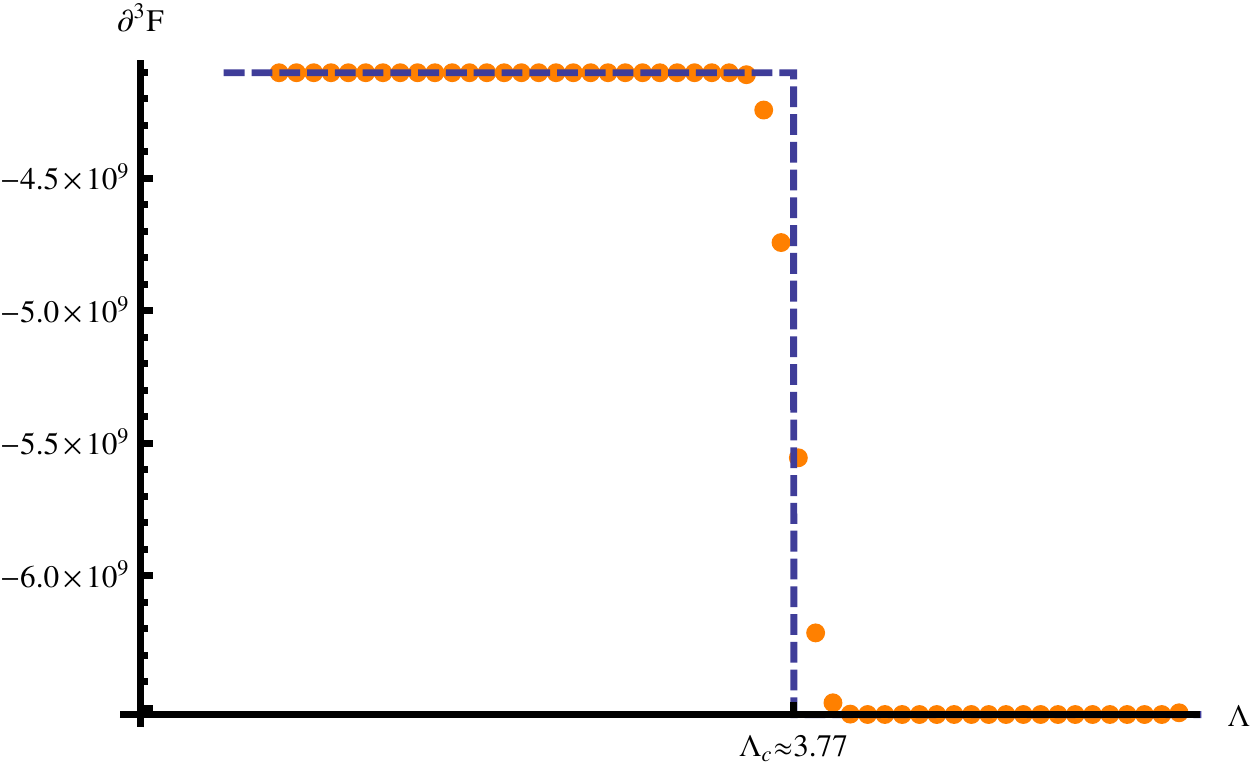}}
\end{center}
\caption{The free energy of $\mathcal{N}=1$ SYM with fundamental hypermultiplets. The orange dots 
show the results for the numerical solution,
while the dashed blue lines represent the analytical solution 
(\ref{free:en:fund})}. The parameters of the theory are taken to be
$r=40,\,m=6,\, \zeta=\frac{3}{4},\,N=100$ corresponding to $\Lambda_c\approx 3.77$.
\label{fund:free:energ:pic}
\end{figure}

We have also checked results for the behavior of the free energy for different parameters numerically. In 
Fig. \ref{fund:free:energ:pic} we show the numerical and analytical results for the certain set of 
the parameters. The orange 
dots show the results of the numerical simulation, while the dashed blue 
lines represent the analytical result (\ref{free:en:fund}).
As we see the analytical results are consistent with the numerical ones. 

However from the Fig.\ref{free:energy:der:fund:pic} we see that the third derivative, obtained numerically, 
is not exactly discontinuous but rather interpolates
smoothly between two constant values, thus leading to the absence of the third-order phase transition.
This happens because numerical calculations were performed for the large, but finite  $r$.
The finite volume of the system leads, as usually, to the absence of  phase transitions, that are 
substituted by the crossover transition. However numerical study of the curve of $\partial_{\Lambda}^3 F$ for different 
values of $r$ have shown that it converges to the step function in the decompactification limit $r\to\infty$, 
signaling about third order phase transition.

\subsection{Wilson Loops}

We also study the behavior of the circular Wilson loops near the critical point.
In the planar limit the exponent term in the matrix model expectation 
value (\ref{wilson:matrix}) does not affect the position of the saddle point, 
which leads to the following integral expression for the Wilson loop vev:
\be
\langle W\rangle = \int d\phi \rho(\phi)e^{2\pi\phi}\,,
\label{wilson:loop:mm}
\ee
Calculating this integral using the eigenvalue density $\rho(\phi)$ from 
(\ref{density:fund:below}) and (\ref{ansatz:fund}) we arrive to
\be
\langle W\rangle^{(I)}=\cosh\left(2\pi r a\right)\,,
\ee
for the phase I and
\be
\langle W\rangle^{(II)}=\left(1-\frac{1}{2}\zeta\right)\cosh\left(2\pi r a\right)+\frac{1}{2}\zeta\cosh(2\pi r m)\,
\ee
for the phase II. 
Using these expressions we find the discontinuity in the second derivative of the Wilson loop 
expectation value with respect to $\Lambda$ at the 
critical point $\Lambda_c$ 
\be\begin{aligned}
\left.\partial_{\Lambda}\left(\langle W\rangle^{(II)}-\langle W\rangle^{(I)}\right)\right|_{\Lambda=\Lambda_c}&=0\,,\\
\left.\partial^{2}_{\Lambda}\left(\langle W\rangle^{(II)}-\langle W\rangle^{(I)}\right)\right|_{\Lambda=\Lambda_c}&=
\frac{\zeta}{2-\zeta}4\pi^2 r^2 \cosh(2\pi r m)\,.
\end{aligned}\ee
which is consistent with the third-order phase transition.

\section{$\NN=1$ SYM with adjoint hypermultiplet}
\label{adjoint:hyper:section}

In this section we consider phase structure of $\NN=1$ SYM with the massive adjoint hypermultiplet 
of the mass $m$.
As we will see this phase structure consists of infinite number of third-order phase transitions on 
the way from weak to strong coupling. 

In the case of the adjoint hypermultiplet using
(\ref{coup-renorm}) we obtain 
\be
g_{eff}^2=g_{YM}^2\,,
\ee
i.e. in this case  the coupling does not get shifts from the UV cut-off $\Lambda$ and 
from now on we use the bare coupling constant $g_{YM}$.

In the planar limit  matrix integral
(\ref{partition-reg}) is dominated by the saddle point
\be
\frac{d F(\tphi)}{d\tphi_k}=\frac{d}{d\tphi_k}\left(\frac{8\pi^3 r}{g_{YM}^2}\sum_i \tphi_i^2+
\sum_{j\neq i}\sum_i \left[F_V(\tphi_i-\tphi_j)+ F_H(\tphi_i-\tphi_j)\right]\right)=0\,.
\ee
Using expressions (\ref{derivtives-F}) for the derivatives of $F_V$ and $F_H$ we obtain
the following equations of motion
\be\begin{aligned}
\frac{16\pi^2 r}{g_{YM}^2}\tphi_i&=\sum_{j\neq i}\left[\left(2-(\tphi_i-\tphi_j)^2\right)\coth(\pi(\tphi_i-\tphi_j))+
\right.\\&\frac{1}{2}\left(\frac{1}{4}+
(\tphi_i-\tphi_j+\tm)^2\right)\tanh(\pi(\tphi_i-\tphi_j+\tm))+\\
&\left.\frac{1}{2}\left(\frac{1}{4}+(\tphi_i-\tphi_j-\tm)^2\right)\tanh(\pi(\tphi_i-\tphi_j-\tm))\right]\,.
\label{eq-adj-full}
\end{aligned}\ee
Restoring the radius dependence (\ref{r:depend}) and taking the decompactification limit, we obtain
\be\begin{aligned}
\frac{16\pi^2}{g_{YM}^2} \phi_i=\sum_{j\neq i}\left[-(\phi_i-\phi_j)^2 {\rm sign}(\phi_i-\phi_j)
+\frac{1}{2}(\phi_i-\phi_j+m)^2 {\rm sign}(\phi_i-\phi_j+m)
+\right.\\\left.\frac{1}{2}(\phi_i-\phi_j-m)^2 {\rm sign}(\phi_i-\phi_j-m)\right]
\end{aligned}\label{eom-adj-disc}\ee
These equations can be directly observed from (\ref{asymptotic_free}). 
Their continuous limit has the following form
\be\begin{aligned}
\frac{16\pi^2}{t}\phi=\int d\psi \rho(\psi)\left[-(\phi-\psi)^2 {\rm sign}(\phi-\psi)
+\frac{1}{2}(\phi-\psi +m)^2 {\rm sign}(\phi-\psi +m)
+\right.\\\left.\frac{1}{2}(\phi-\psi-m)^2 {\rm sign}(\phi-\psi-m)\right]\,,
\end{aligned}\label{eom-adj-int}\ee
where we have introduced 't Hooft coupling constant $t=g_{YM}^2 N$ \footnote{Note that this 't Hooft 
coupling is different from $\lambda=g_{YM}^2 N/r$ used in \cite{Kallen:2012zn,Minahan:2013jwa,Minahan:2014hwa}.
$\lambda$ used before was the dimensionless constant, while $t$ used in this paper has the dimension of length.}.

To solve (\ref{eom-adj-int}) we proceed in the same way as 
while solving (\ref{eom-fund-int}) for the $\NN=1$ SYM 
with the fundamental hypermultiplets. We take three derivatives of (\ref{eom-adj-int}) ,
thus reducing integral equation to the algebraic one. Then we find general solutions 
to this algebraic equation for different phases of the theory, corresponding to 
different values of the parameters. Finally, we fix all integration constants
and free parameters by substituting general solution back into original equation.
Three derivatives of (\ref{eom-adj-int}) are given by
\be
\frac{16\pi^2}{t} &=&\int d\psi \rho(\psi)\left[-2|\phi-\psi|+|\phi-\psi +m|+|\phi-\psi -m|\right]\,,\label{1st:der:adj}\\
0&=&\int d\psi \rho(\psi)\left[-2\,{\rm sign}(\phi-\psi)+{\rm sign}(\phi-\psi +m)+{\rm sign}(\phi-\psi -m)\right]\,,\label{2nd:der:adj}\\
0&=&- 4\rho(\phi)+2\rho(\phi +m)+2\rho(\phi -m)\,. \label{3rd:der:adj}
\ee
Before we start solving these equations let's understand what do we expect to 
obtain. In analogy with $3D$ and $4D$ results \cite{Anderson:2014hxa,Russo:2013qaa,Russo:2013kea}
\begin{itemize}
 \item We start with the theory with $2 a<m$, where 
$a$ is the position of the support endpoint for the solution of (\ref{eom-adj-int}). 
 \item Then we increase the 't Hooft coupling $t$ which leads to the widening of the support (increase of $a$).
At some value of $t$, where $2a=m$, we obtain emergence of two resonances, originating from the nearly massless 
hypermultiplets, i.e. from the terms $|\phi-\psi|\approx m$ in the saddle point equations (\ref{eom-adj-int}).
Appearance of these 
resonances leads to the change of the form of the eigenvalue distribution, signaling about the emergence of a new phase. 
\item If we further increase coupling new secondary resonances appear every 
time when $2a = nm$ with $n\in Z$. Each time this happens
 we expect to obtain phase transition in the theory. 
\item In the limit of very large 't Hooft coupling $t\to\infty$ we expect to obtain an infinite number of resonances 
that will be averaged to the strong coupling solution obtained in \cite{Kallen:2012zn,Minahan:2013jwa}.
\end{itemize}

Our final goal here is to obtain solutions of (\ref{eom-adj-int}) for arbitrary number of resonances $n$. 
However in order to understand how to construct this general solution we start with 
finding explicit solutions for $n=0$ and $n=1$ and only then generalize solution to general $n$.

\subsection{Simple examples}

The phase with no resonances corresponds to the case of $2a<m$. To solve 
(\ref{eom-adj-int}) with these values of the parameters we address (\ref{2nd:der:adj}) corresponding
to the second derivative of the original equation. This equation then is equivalent to
(\ref{fund:below:equation:2nd}) and hence can be solved by 
\be
\rho^{(0)}(\phi)=\frac{1}{2}\left(\delta(\phi-a)+\delta(\phi +a)\right)\,.
\label{density:adj:0}
\ee
Here and further below we treat the $\delta$-functions at the endpoints of the distribution in the same manner as we were doing it 
in the case of the fundamental hypermultiplets, i.e. we always assume the small shift in the argument of the
$\delta$-function that brings whole $\delta$-function inside the eigenvalue support.

Substituting  
ansatz (\ref{density:adj:0}) into the original equation (\ref{eom-adj-int}),
 we obtain the following relation for the position of the support endpoint
\be
a^{(0)}=m-\frac{8\pi^2}{t}\,.
\label{endpoint:adj:0}
\ee
By construction this solution is valid when $2a^{(0)}<m$ or, equivalently $t<\frac{16 \pi^2}{m}$. Thus 
the first critical point is 
\be
t_{c}^{(1)}=\frac{16 \pi^2}{m}\,.
\ee
Similarly to the case of the fundamental hypermultiplets to justify this 
solution we consider solutions to the full equations of motion (\ref{eq-adj-full}) 
in the limit of large but finite $r$, and then obtain (\ref{density:adj:0}) together with 
(\ref{endpoint:adj:0}) in the limit of infinite $r$. For the details of this 
calculation see appendix \ref{finite:R:adj}.

\begin{figure}[!h]
\begin{center}
  \subfigure[$t=12. \left(t<t_c^{(1)}\right)$]{\label{pic:adj:n=0}\includegraphics[width=52mm,angle=0,scale=1.5]{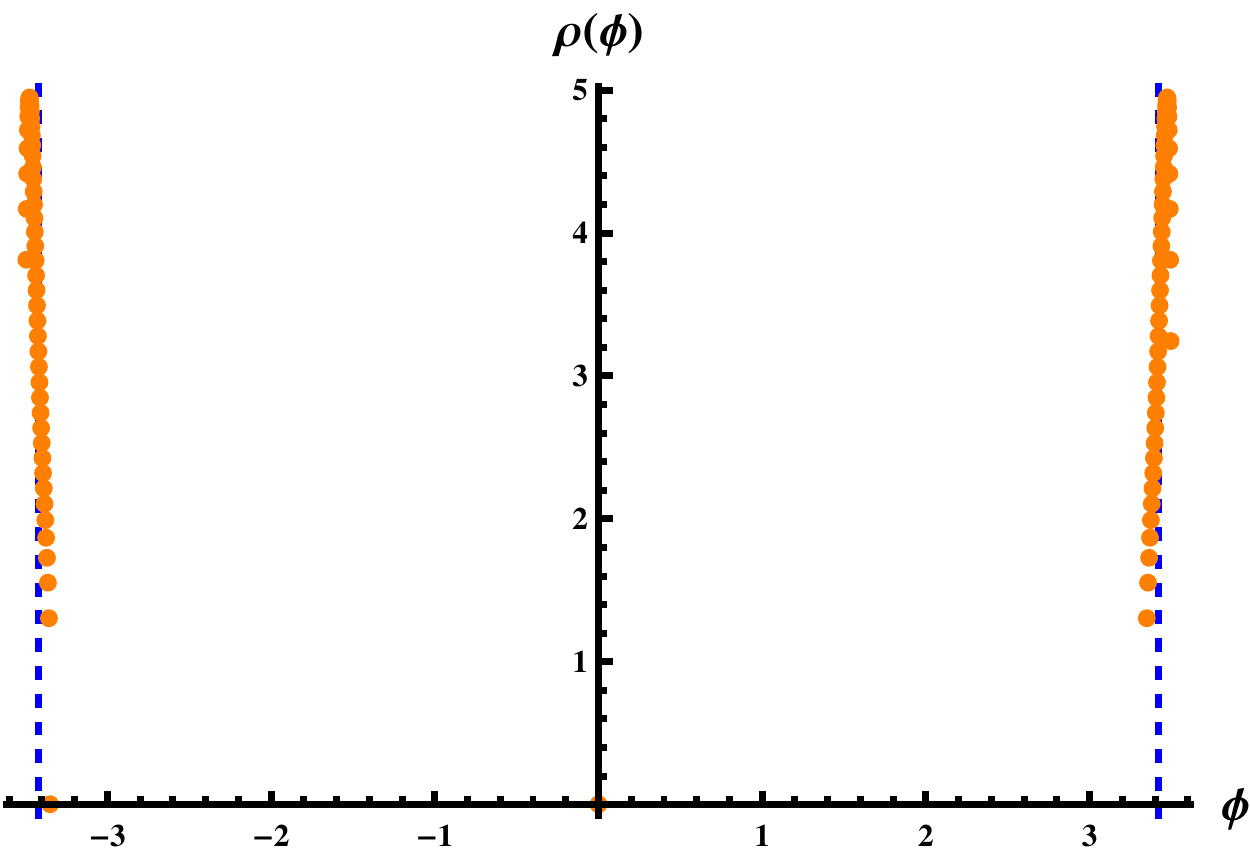}}
  \subfigure[$t=18. \left(t_c^{(1)}<t<t_c^{(2)}\right)$]{\label{pic:adj:n=1}\includegraphics[width=52mm,angle=0,scale=1.5]{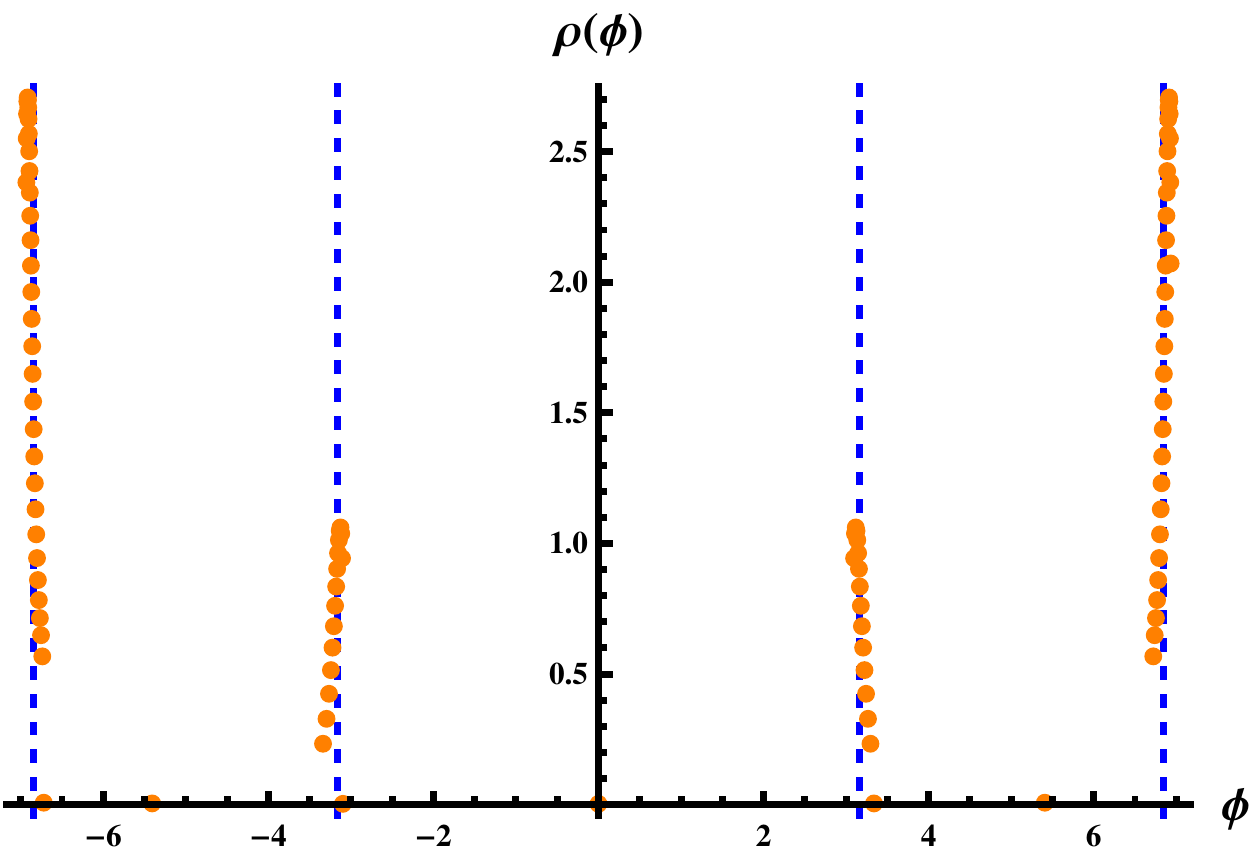}}
\end{center}
\caption{Free energy of the $\mathcal{N}=1$ SYM with adjoint hypermultiplets.
The orange dots show the results for the numerical solution,
while the dashed blue lines on $(a)$ and 
$(b)$ represent positions of the $\delta$-functions in the analytical solutions 
(\ref{density:adj:0}) and (\ref{density:n=1}) respectively. 
The parameters of the theory are taken to be
$r=15,\,m=10,\,N=100$ corresponding to $t_c^{(1)}\approx 15.79$ and $t_c^{(2)}\approx 23.68$.}
\end{figure}

We also found numerical solutions to (\ref{eq-adj-full}).  We show 
these solutions in Fig.\ref{pic:adj:n=0}
with the orange dots. The dashed blue lines on the same plot 
represent positions of the $\delta$-functions in our analytical solution (\ref{density:adj:0}).
As we see from this picture analytical 
solutions reproduce numerical ones very well. All differences between these solutions appear 
due to the effects of finite $r$, which are discussed in appendix \ref{finite:R:adj}.

Now we increase the 't Hooft coupling and pass the critical point $t_c^{(1)}$, so that 
 now $m<2a<2m$. Under this condition we expect two resonances to appear at $\phi=\pm(m-a)$.
To solve equations of motion (\ref{eom-adj-int}) for this case we divide 
the eigenvalue support $[-a,a]$ into three regions
\be\begin{aligned}
                    &~~~ \rho_{A}(\phi),\quad ~~m-a<\phi< a\,,\\
\rho^{(1)}(\phi)=   &~~~ \rho_{B}(\phi),\quad -m+a< \phi< m-a\,,\\
                    &~~~ \rho_{C}(\phi),\quad -a<\phi<-m+a\,,
\label{adj:n=1:regions}
\end{aligned}\ee
and find general solution to (\ref{eom-adj-int}) in every region. We also 
use $\phi\to-\phi$ symmetry of (\ref{eom-adj-int}),
that leads to
\be
\rho_B(\phi)=\rho_B(-\phi)\,,\quad \rho_A(\phi)=\rho_C(-\phi)\,.
\label{symmetry:adj:n=1}
\ee

{\bf Region A} ($m-a<\phi<a$). On this interval equation (\ref{3rd:der:adj}) reads
\be
-4\rho_A(\phi)+2\rho_C(\phi-m)=0\,,
\ee
where we used $\rho(\phi+m)=0$ as $\phi+m>a$ and also $(\phi-m)\in C$.
Using the symmetry property (\ref{symmetry:adj:n=1}) we obtain
\be
2\rho_A(\phi)=\rho_A(m-\phi)\,.
\label{eq1:A:adj:n=1}
\ee
This equation can be satisfied only if $\rho_A(\phi)=0$. In analogy with the previously found 
solutions we assume that this is true up to isolated points on the boundary of the interval at 
$\phi=a,\, m-a$. Then the natural ansatz is
\be
\rho_A(\phi)=c_0 \delta(\phi-a)+c_1 \delta(\phi-m+a)\,.
\label{density:A:n=1:adj}
\ee

{\bf Region B} ($-m+a<\phi<m-a$). In this region  (\ref{2nd:der:adj}) takes the form
\be\begin{aligned}
\int d\psi\rho(\psi)&{\rm sign}(\phi-\psi)\equiv\\
&\int\limits_{-a}^{a-m}d\psi\rho_C(\psi)+
\int\limits_{a-m}^{\phi}d\psi\rho_B(\psi)-\int\limits_{\phi}^{m-a}d\psi\rho_B(\psi)-
\int\limits_{m-a}^{a}d\psi\rho_A(\psi)=0
\end{aligned}\ee
The first and the last terms in this expression cancel each other due to the 
symmetry properties (\ref{symmetry:adj:n=1}) of $\rho(\phi)$.
The remaining part of the equation then reads
\be
\int\limits_{a-m}^{\phi}d\psi\rho_B(\psi)=\int\limits_{\phi}^{m-a}d\psi\rho_B(\psi)\,,
\ee
suggesting, similarly to the case of $n=0$, that solution for the eigenvalue density 
$\rho_B(\phi)$ is sharply peaked around endpoints of the interval $B$ at $\phi=\pm(m-a)$ 
\be
\rho_B(\phi)\sim \delta(\phi-m+a)+\delta(\phi+m-a)
\ee
Note that we have already took one of this $\delta$-functions into account in $\rho_A(\phi)$
in (\ref{density:A:n=1:adj}), while the second $\delta$-function appears in $\rho_C(\phi)$ as we 
will see further.

{\bf Region C} ($-a<\phi<a-m$). Finally, on the last interval $C$ we do not need to solve any 
equations in order to find the eigenvalue density. Instead we utilize our previous result (\ref{density:A:n=1:adj}) 
for the density in the region $A$ together with the symmetry properties (\ref{symmetry:adj:n=1}) to obtain
\be
\rho_C(\phi)=c_0 \delta(\phi+a)+c_1 \delta(\phi +m-a)\,.
\ee
Notice here that, just as we expected, the second $\delta$-function from $\rho_B(\phi)$ showed up here. 

Summing up all our results we find the following ansatz for the eigenvalue density in the 
phase with one pair of the resonances
\be
\rho^{(1)}(\phi)=c_0 \delta(\phi-a)+c_1 \delta(\phi-m+a)+c_1 \delta(\phi +m-a)+c_0 \delta(\phi+a)\,.
\label{density:n=1:ansatz}
\ee
This general solution has three free parameters $c_0,\,c_1$ and $a$. To fix all of them we  
use the normalization condition (\ref{normalization}) together with the equation (\ref{1st:der:adj}).
The consistency of the obtained solution with the original equation (\ref{eom-adj-int}) can then 
be checked by the direct substitution.
From the normalization condition we obtain
\be
c_0+c_1=\frac{1}{2}\,,\label{norm:adj:n=1}
\ee
while from the equation (\ref{1st:der:adj}) it follows 
\be\begin{aligned}
\frac{16\pi^2}{t}=2a \left(-2 c_0+c_1\right)&+2 c_0 (a+m)+2 c_1(2m-a)\\	
&+(-2c_1+c_0)\left(|\phi-a+m|+|\phi+a-m|\right)\,. \label{1st:der:adj:n=1}
\end{aligned}\ee
This equation can be satisfied for any $\phi$ only if 
\be
c_0=2c_1\,.
\ee
 Combined with equation (\ref{norm:adj:n=1}) this gives
\be
c_0=\frac{1}{3},\quad c_1=\frac{1}{6}\,.
\ee
Finally substituting these values back into (\ref{1st:der:adj:n=1}) we obtain the following 
expression for the endpoint position
\be
a^{(1)}=2m-\frac{24 \pi^2}{t}\,.
\label{endp:adj:n=1}
\ee
Corresponding eigenvalue density is given by
\be
\rho^{(1)}(\phi)=\frac{1}{3} \delta(\phi-a)+\frac{1}{6} \delta(\phi-m+a)+\frac{1}{6} \delta(\phi +m-a)+\frac{1}{3} \delta(\phi+a)\,.
\label{density:n=1}
\ee
By construction this solution is valid, when $m<2 a^{(1)}<2m$ or equivalently
$\frac{16\pi^2}{m}<t<\frac{24\pi^2}{m}$. Thus the second critical point is given by
\be
t_c^{(2)}=\frac{24\pi^2}{m}\,.\label{critical:pt:adj:n=1}
\ee
As in the previous cases we justify this solution by finding the solution of (\ref{eq-adj-full})
for large but finite $r$ and taking $r\to\infty$ limit 
in the very end of the calculations. As the result we reproduce solution (\ref{density:n=1}) with the 
right positions and the coefficients of the $\delta$-functions.
For the details of this calculation see appendix \ref{finite:R:adj}.

In Fig.\ref{pic:adj:n=1} we compare the analytical solution (\ref{density:n=1}) with 
the corresponding numerical solution of (\ref{eq-adj-full}) represented by the orange dots.
The dashed blue lines show the positions of the $\delta$-functions in (\ref{density:n=1}). 
As we see the analytical solution coincides with the numerical one up to the 
effects of finite $r$.

\subsection{General solution}

Now, using examples of the solution with $n=0,1$ number of resonance pairs, we 
are ready to find a general solution with an arbitrary number $n$ of the resonance pairs. This solution is valid 
when inequality $mn<2a<m(n+1)$, where $n\in\mathbb{Z}$, is satisfied. In this case the cut can contain $n$ pairs of the resonances. 
The mechanism of the emergence of these resonances is exactly the same as it was in the case of $n=1$.
Primary resonances appear at $\phi=\pm (a-m)$ due to the discontinuities in the kernel of (\ref{eom-adj-int}),
caused by the peaks at the endpoints of the eigenvalue distribution $\psi=\pm a$. 
These resonances in turn create secondary 
resonances at $\phi=\pm(a-2m)$ by the same mechanism and so on up to the last resonance pair at $\phi=\pm(a-n m)$
that fits inside the eigenvalue support. 

Thus generalizing solutions (\ref{density:adj:0}) and (\ref{density:n=1:ansatz}) 
to the arbitrary number of resonances we expect to obtain the eigenvalue density of the following form 
\be
\rho(\phi)=\sum\limits_{k=0}^{n}c_k\left(\delta(\phi-a+mk)+\delta(\phi +a-mk)\right)\,.
\label{density:adj:n:ansatz}
\ee
Now we proceed in the same way as for the 
cases of $n=0,1$. Namely we substitute ansatz (\ref{density:adj:n:ansatz}) into 
equation (\ref{1st:der:adj}). This substitution results in 
\be\begin{aligned}
&\frac{16\pi^2}{t}=\sum\limits_{k=0}^{n}c_{k}\left[-2|\phi -a+mk|+|\phi -a+m(k+1)|+|\phi -a+m(k-1)|+\right.\\
&\left.(a\to -a,\, m\to -m)\right]=\left(-2\sum_{k=0}^{n}c_k+\sum_{k=-1}^{n-1}c_{k+1}+\sum_{k=1}^{n+1}c_{k-1}\right)
\left(|\phi-a+mk|+|\phi +a-mk|\right)\,,
\label{summation:shift}
\end{aligned}\ee
where in the last line we have shifted the summation index for the terms $|\phi\pm a\mp m(k+1)|$ and $|\phi\pm a\mp m(k-1)|$.
Assuming that $c_{-1}=c_{n+1}\equiv 0$ we arrive to the relatively compact equation
\be\begin{aligned}
\frac{16\pi^2}{t}&=\sum_{k=0}^{n}\mathcal{C}_k \left(|\phi-a+mk|+|\phi +a-mk|\right)+\\
&c_n \left(|\phi-a+m(n+1)|+|\phi +a-m(n+1)|\right)+c_0\left(|\phi-a-m|+|\phi +a+m)|\right)\,,
\label{1st:der:adj:gen:n}
\end{aligned}\ee
where for simplicity we introduced
\be
\mathcal{C}_{k}\equiv c_{k+1}-2c_k+c_{k-1}\,.
\ee
Notice now that for the last two terms the following relations are satisfied everywhere on the support $[-a,\,a]$
\be\begin{aligned}
|\phi -a+m(n+1)|+|\phi +a-m(n+1)|&=2(m(n+1)-a)\,,\\
|\phi +a+m|+|\phi -a-m|&=2(m+a)\,.
\label{relation:add:adj:gen:n}
\end{aligned}\ee
However the terms in the summation in (\ref{1st:der:adj:gen:n}) depend on $\phi$ and, in analogy 
with the previously considered cases, we expect the coefficients of these terms to be zero. To prove this statement
precisely we follow the same algorithm as in the case of $n=1$. However, we should consider 
cases of even and odd number of the resonance pairs separately. 
Though they are similar and lead, as we will see
soon, to the same result, there are some differences that should be briefly discussed.

\begin{figure}[!h]
\begin{center}
  \subfigure[Even $n$]{\label{pic:even:n:inter}\includegraphics[width=52mm,angle=0,scale=1.5]{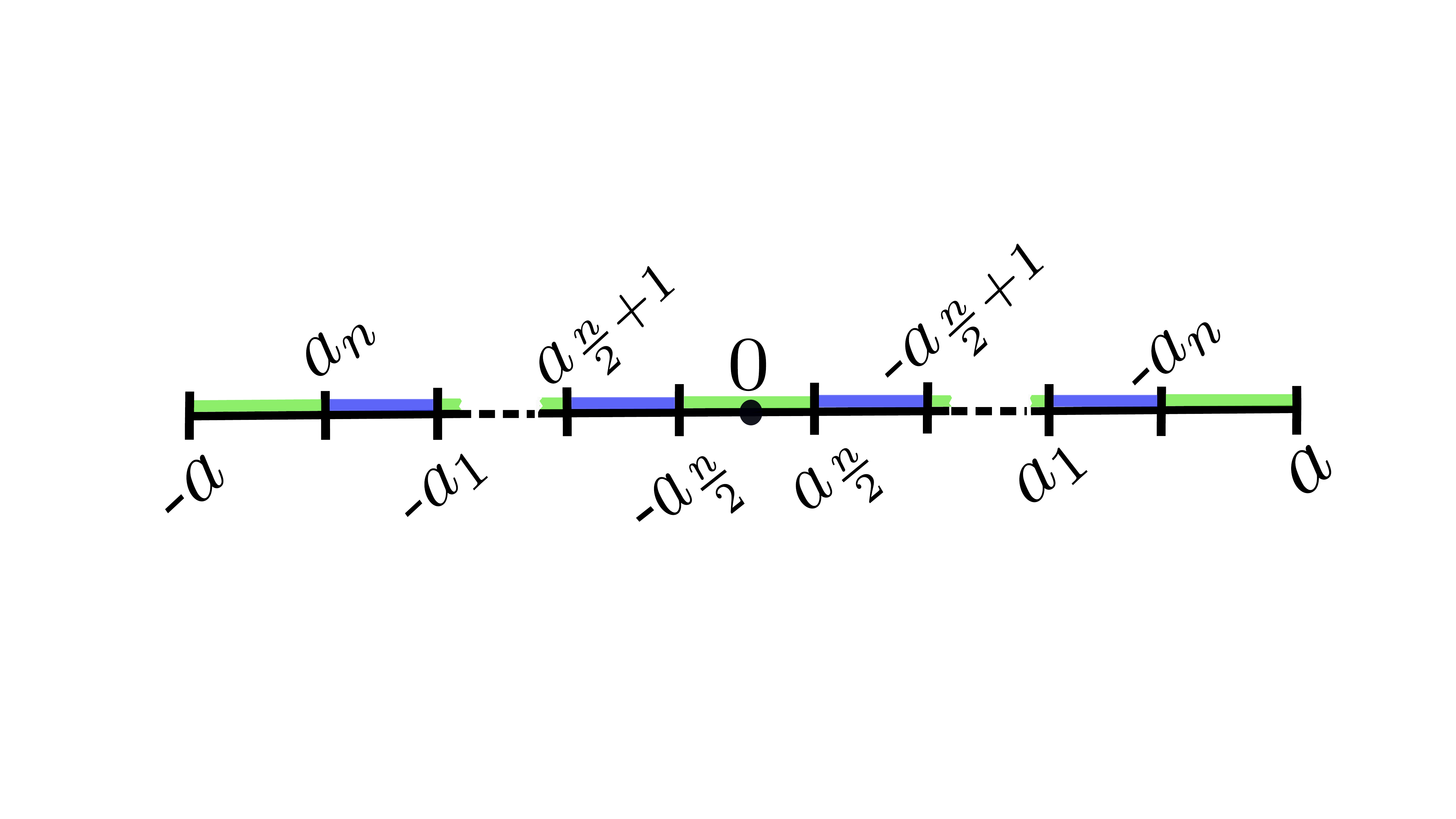}}
  \subfigure[Odd $n$]{\label{pic:odd:n:inter}\includegraphics[width=52mm,angle=0,scale=1.5]{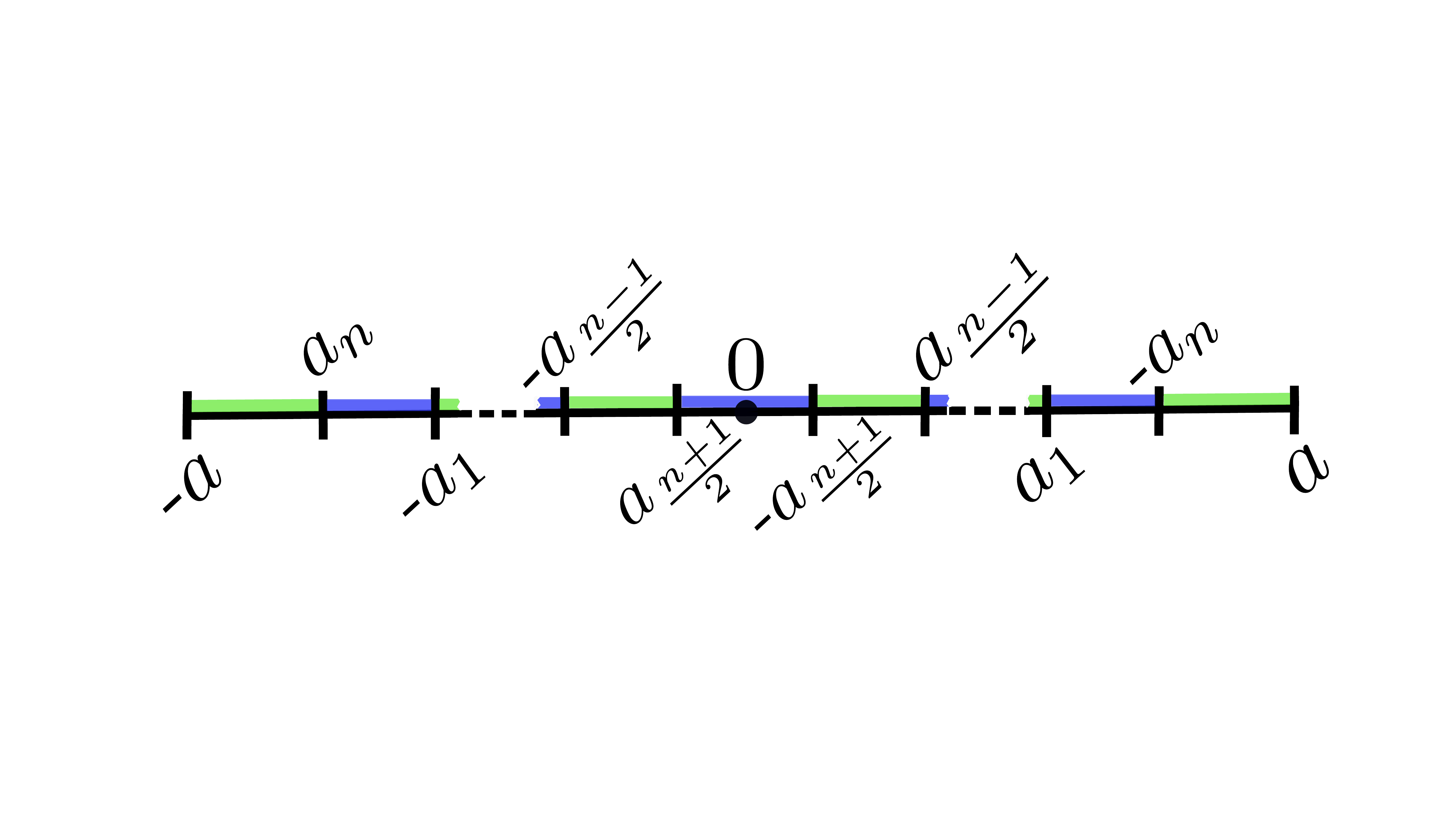}}
\end{center}
\caption{Break of the eigenvalue support into intervals for the cases of even and odd number $n$ of the resonance pairs.}
\label{pic:intervals}
\end{figure}

 In the case of the {\bf even $n$} the entire support $[-a,\,a]$ should be separated into different 
regions with the boundaries corresponding to the resonances as shown in Fig.\ref{pic:even:n:inter}. There 
are two types of intervals we should consider. Intervals of different types are shown with different colors in Fig.
\ref{pic:even:n:inter}. On this picture and everywhere further for shortness we introduce the notation $a_k=a-mk$. 

\underline{\textit{Intervals} $a_{\frac{n}{2}-j}<\phi<-a_{\frac{n}{2}+j+1}$ , $j=0,..,\frac{n}{2}-1$} 
(blue intervals in Fig.\ref{pic:intervals}). If $\phi$ belongs to one of the regions we 
obtain the following relations 
\be\begin{aligned}
                                      &2\phi\,,~~~~~\frac{n}{2}-j\leq k\leq \frac{n}{2}+j\,,\\
|\phi-a+mk|+|\phi +a-mk|\,=~  &2(a-mk)\,,~k\leq \frac{n}{2}-j-1\,,\\
                                   -&2(a-mk)\,,~k\geq \frac{n}{2}+j+1\,,
\label{helpfull:eq:1}
\end{aligned}\ee
 Using (\ref{relation:add:adj:gen:n}) and  (\ref{helpfull:eq:1}) equation (\ref{1st:der:adj:gen:n})
can be rewritten as
\be\begin{aligned}
\frac{16\pi^2}{t}=&2\sum_{k=\frac{n}{2}-j}^{\frac{n}{2}+j}\mathcal{C}_k \phi\\
&+2\left(\sum^{\frac{n}{2}-j-1}_{k=0}-\sum_{k=\frac{n}{2}+j+1}^{n}\right)\mathcal{C}_k(a-mk)+
2c_n(m(n+1)-a) +2c_0(a+m)\,,
\label{1st:der:adj:even:int1}
\end{aligned}\ee

\underline{\textit{Intervals} $-a_{\frac{n}{2}+j}<\phi<a_{\frac{n}{2}-j}$ , $j=1,..,\frac{n}{2}$} 
(green intervals in Fig.\ref{pic:intervals}). In these regions similarly to the previous case we obtain 
\be\begin{aligned}
                                      &2\phi\,,~~~~~\frac{n}{2}-j+1\leq k\leq \frac{n}{2}+j\,,\\
|\phi-a+mk|+|\phi +a-mk|\,=~  &2(a-mk)\,,~k\leq \frac{n}{2}-j\,,\\
                                   -&2(a-mk)\,,~k\geq \frac{n}{2}+j+1\,,
\end{aligned}\ee
and the corresponding equations of motion (\ref{1st:der:adj:gen:n}) turns into 
\be\begin{aligned}
\frac{16\pi^2}{t}=&2\sum_{k=\frac{n}{2}-j+1}^{\frac{n}{2}+j}\mathcal{C}_k \phi\\
&+2\left(\sum^{\frac{n}{2}-j}_{k=0}-\sum_{k=\frac{n}{2}+j+1}^{n}\right)\mathcal{C}_k(a-mk)+
2c_n(m(n+1)-a) +2c_0(a+m)\,.
\label{1st:der:adj:even:int2}
\end{aligned}\ee

Now the algorithm of defining all $(n+1)$ coefficients $c_k$ is the following.
First condition comes from the 
 normalization condition (\ref{normalization}), that reads for the ansatz (\ref{density:adj:n:ansatz}) as 
\be
2\sum_{k=0}^{n}c_k=1\,.
\label{normalization:adj:general:n}
\ee
We then need $n$ more conditions to determine all the coefficients.
These conditions can be obtained by considering the  
intervals shown in Fig.\ref{pic:even:n:inter}. We start from the region closest to 
the origin, i.e. $a_{\frac{n}{2}}<\phi<-a_{\frac{n}{2}+1}$,
corresponding to $j=0$ in (\ref{1st:der:adj:even:int1}).
Then the first sum in this equation is given just by one term
\be
2\sum_{k=\frac{n}{2}-j}^{\frac{n}{2}+j}\mathcal{C}_k \phi\to 2~\mathcal{C}_{\frac{n}{2}}\,\phi\,.
\ee
As before we assume that this $\phi$-dependence should be canceled leading to $\mathcal{C}_{\frac{n}{2}}=0$.

Then we move away from the origin to the next interval $-a_{\frac{n}{2}+1}<\phi<a_{\frac{n}{2}-1}$. This 
interval corresponds to the equation (\ref{1st:der:adj:even:int2}) with $j=1$, which turns the first sum into 
the sum of two terms
\be
2\sum_{k=\frac{n}{2}-j+1}^{\frac{n}{2}+j}\mathcal{C}_k \phi\to 2~\left(\mathcal{C}_{\frac{n}{2}}+\mathcal{C}_{\frac{n}{2}+1}\right)\,\phi\,,
\ee
thus resulting in the condition $\mathcal{C}_{\frac{n}{2}}+\mathcal{C}_{\frac{n}{2}+1}=0$. Repeating this 
procedure for all $n$ intervals on the positive $\phi$ half-axis we obtain $n$ equations
\be
\sum_{k=1}^{n}\mathcal{C}_{k}=0\,.
\label{conditions:C_k}
\ee

For the case of the {\bf odd $n$} all considerations are analogous. As in the 
previous case we consider the division of the whole eigenvalue support into two types of regions, which 
is shown in Fig.\ref{pic:odd:n:inter}. 

On the \underline{\textit{intervals} $a_{\frac{n+1}{2}-j}<\phi<-a_{\frac{n+1}{2}+j}$ , $j=1,..,\frac{n-1}{2}$} (blue intervals
in Fig.\ref{pic:intervals})
equation (\ref{1st:der:adj:gen:n}) can be written in the form 
\be\begin{aligned}
\frac{16\pi^2}{t}=2&\sum_{k=\frac{n+1}{2}-j}^{\frac{n-1}{2}+j}\mathcal{C}_k \phi\\
&+2\left(\sum^{\frac{n-1}{2}-j}_{k=0}-\sum_{k=\frac{n+1}{2}+j}^{n}\right)\mathcal{C}_k(a-mk)+
2c_n(m(n+1)-a) +2c_0(a+m)\,,
\label{1st:der:adj:odd:int1}
\end{aligned}\ee
while on the  \underline{\textit{intervals} 
$-a_{\frac{n+1}{2}+j}<\phi<a_{\frac{n-1}{2}-j}$ , $j=0,..,\frac{n-1}{2}$} (green intervals
in Fig.\ref{pic:intervals}) the same equations can be rewritten as the following
\be\begin{aligned}
\frac{16\pi^2}{t}=2&\sum_{k=\frac{n+1}{2}-j}^{\frac{n+1}{2}+j}\mathcal{C}_k \phi\\
&+2\left(\sum^{\frac{n-1}{2}-j}_{k=0}-\sum_{k=\frac{n+3}{2}+j}^{n}\right)\mathcal{C}_k(a-mk)+
2c_n(m(n+1)-a) +2c_0(a+m)\,.
\label{1st:der:adj:odd:int2}
\end{aligned}\ee
Now as in the case of even number of the resonance pairs we consider intervals and corresponding 
equations (\ref{1st:der:adj:odd:int1}) and (\ref{1st:der:adj:odd:int2}) one by one. 

To obtain $n$ conditions for the coefficients 
$\mathcal{C}_k$ we start with the 
region $-a_{\frac{n+1}{2}}<\phi<a_{\frac{n-1}{2}}$.
This interval corresponds to the equation (\ref{1st:der:adj:even:int1}) with $j=0$
resulting in $\mathcal{C}_{\frac{n+1}{2}}=0$. Repeating procedure for every single interval on the positive
half-axis we obtain the same equations (\ref{conditions:C_k}) for the coefficients $c_k$. 

Now employing equations (\ref{normalization:adj:general:n}) and (\ref{conditions:C_k}) let's determine
the coefficients $c_k$ in our ansatz (\ref{density:adj:n:ansatz}).
To do this notice that the general solution for the recurrence relation 
\be
c_{k+1}-2c_k+c_{k-1}=0\,,\quad k=1,..,n\,.
\ee
 is given by 
\be
c_{k}=\alpha + \beta k\,.
\label{ck:ansatz}
\ee
The easiest way to see this, is to write the recurrence relation for large $k$ in form of the  
differential equation $\frac{d^2}{d k^2}c(k)=0$ with the  general solution given by $c(k)=\alpha+\beta k$.

To determine the coefficients $\alpha$ and $\beta$ we use boundary condition $c_{n+1}=0$. Substituting 
$k=n+1$ into (\ref{ck:ansatz}) we obtain 
\be
\alpha=-\beta (n+1)\,.
\label{relation:alpha:beta}
\ee
Now substituting (\ref{ck:ansatz}) into normalization condition (\ref{normalization:adj:general:n})
we find
\be
\sum_{k=0}^{n}c_k=\alpha(n+1)+\beta \frac{n(n+1)}{2}=\frac{1}{2}\,.
\ee
Finally combining this relation with (\ref{relation:alpha:beta}) we obtain
\be
\alpha=\frac{1}{n+2}\,,\quad \beta=-\frac{1}{(n+1)(n+2)}\,.
\label{alpha:beta:coeff}
\ee

The last ingredient we need for the completeness of the solution is the position of the cut 
endpoint $a$. To determine it we should come back to equation (\ref{1st:der:adj:gen:n}).
As we have shown $\mathcal{C}_k=0$ for $k=1,..,n$, so that corresponding terms do not contribute 
into equations. Using $\alpha$ and $\beta$ determined in (\ref{alpha:beta:coeff}) we obtain 
expression for $\mathcal{C}_0$
\be
\mathcal{C}_0\equiv c_1-2c_0=\beta-\alpha=-\frac{1}{n+1}\,.
\ee
Substituting it into (\ref{1st:der:adj:gen:n}) we find the position of the support endpoint
\be
a^{(n)}=(n+1)m-\frac{4\pi^2}{t}(n+1)(n+2)\,.
\label{endpoint:adj:general:n}
\ee
Obtained solution is valid, when $mn<2a^{(n)}<m(n+1)$ or equivalently when
\be
\frac{8\pi^2}{m}(n+1)<t<\frac{8\pi^2}{m}(n+2)\,,
\label{coupling:interval:gen:n}
\ee
so that the $n^{{\rm th}}$ critical point is located at
\be
t_c^{(n)}=\frac{8\pi^2}{m}(n+1)\,.
\label{critical:pt:adj}
\ee

\begin{figure}[!h]
\begin{center}
  \subfigure[$t=70. \left(t_c^{(3)}<t<t_c^{(4)}\right)$]{\label{pic:adj:n=3}\includegraphics[width=52mm,angle=0,scale=1.5]{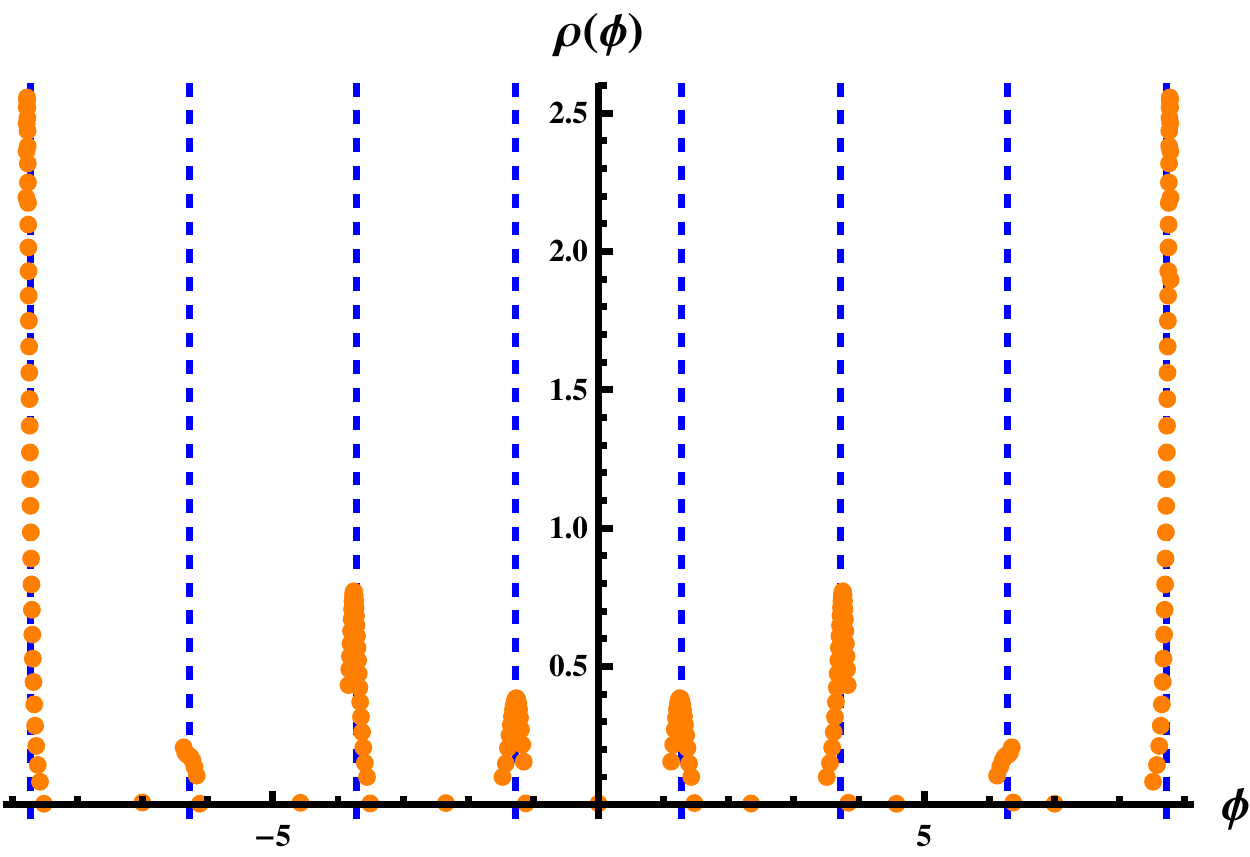}}
  \subfigure[$t=102. \left(t_c^{(5)}<t<t_c^{(6)}\right)$]{\label{pic:adj:n=5}\includegraphics[width=52mm,angle=0,scale=1.5]{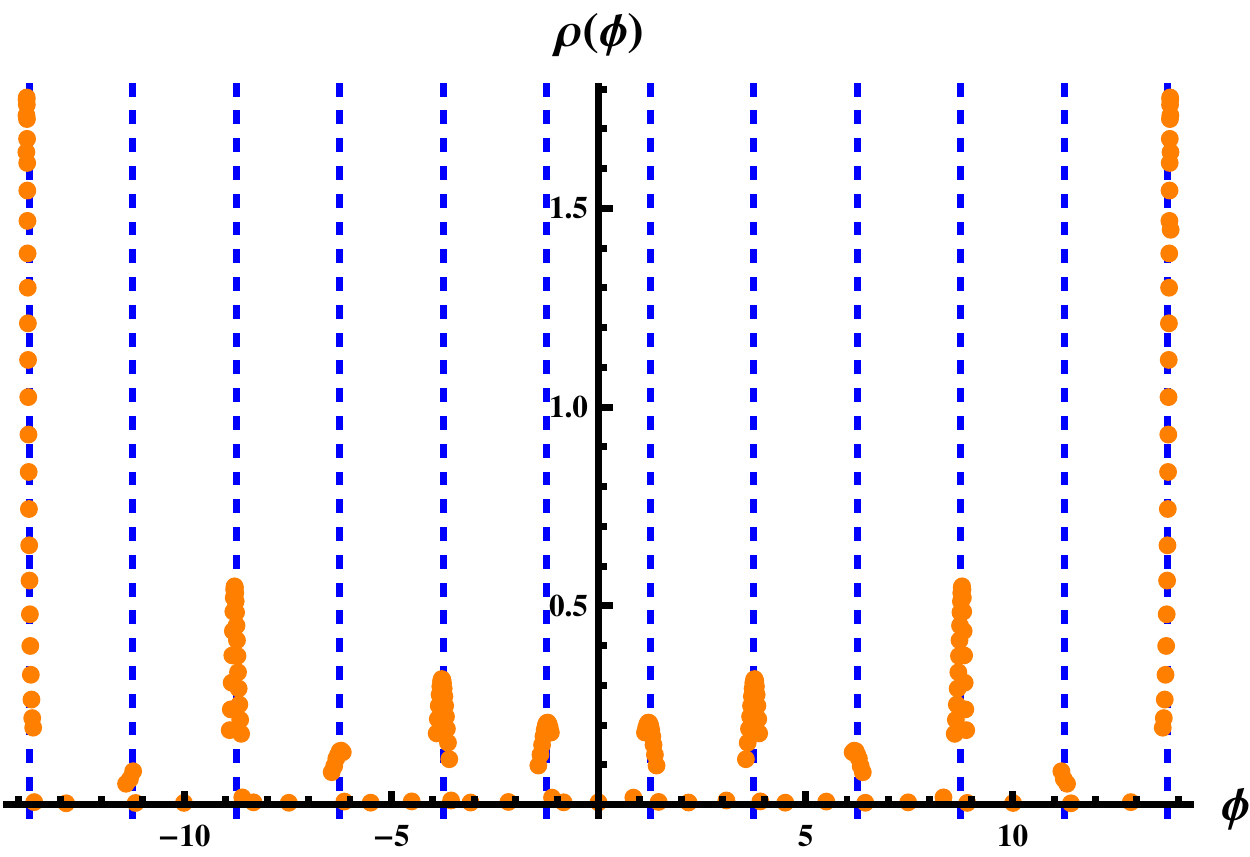}}
\end{center}
\caption{The free energy of $\mathcal{N}=1$ SYM with adjoint hypermultiplets. The orange
dots show the results for the numerical solution,
while the dashed blue lines on $(a)$ and 
$(b)$ represent positions of the $\delta$-functions in the analytical solutions (\ref{density:adj:gen:n}). 
The parameters of the theory are taken to be
$r=25,\,m=5,\,N=200$.}
\label{pic:adj:gen:n}
\end{figure}

Let us summarize the findings of this section. We have obtained the general solution to the matrix 
model saddle-point equation (\ref{eom-adj-int}). If the 't Hooft couping is in
the interval $\frac{8\pi^2}{m}(n+1)<t<\frac{8\pi^2}{m}(n+2)\,$
the solution contains $n$ pairs of resonances. The eigenvalue density for this solution is given by
\be
\rho(\phi)=\sum_{k=0}^n c_k\left(\delta(\phi -a+ mk)+\delta(\phi +a- mk)\right)\,,\quad c_{k}=\frac{n+1-k}{(n+1)(n+2)}\,,
\label{density:adj:gen:n}
\ee
with the position of the eigenvalue support endpoint $a$ given by (\ref{endpoint:adj:general:n}). 

If we take $n=0,1$ in this general solution we will reproduce corresponding formulas (\ref{density:adj:0}),
and (\ref{density:n=1}) obtained previously.
We have also checked general solution (\ref{density:adj:gen:n}) numerically. In Fig.\ref{pic:adj:gen:n} we show
with the orange dots numerical solution to (\ref{eq-adj-full}) for the cases
of $n=3$ and $n=5$.
On the same plots we show the positions of the $\delta$-functions in our analytical solution 
(\ref{density:adj:gen:n})
with the vertical dashed lines. As we see numerical and analytical results are in good agreement. The
positions of the $\delta$-functions reproduces positions of the peaks very well. The height of the resonances also 
decreases when moving away from the endpoint, as predicted in (\ref{density:adj:gen:n}).

\subsection{Free energy and the order of the phase transition}

Now we are ready to address the question of the order of the phase transitions 
taking place at the critical points $t_c^{(n)}$. For this we evaluate the free energy corresponding 
to the eigenvalue density (\ref{density:adj:gen:n}).
We read the free energy in the decompactification 
limit directly from (\ref{asymptotic_free}) choosing appropriate matter content. 
In the continuous limit it can be written as follows
\be\begin{aligned}
\tilde{F}\equiv\frac{1}{2\pi r^3 N^2}F=\frac{4\pi^2}{t}\int d\phi \rho(\phi)\phi^2+\frac{1}{12}\int\int d\phi d\psi
\rho(\phi)\rho(\psi)\left[|\phi-\psi|^3-\right.\\\left.
\frac{1}{2}|\phi-\psi +m|^3-\frac{1}{2}|\phi -\psi-m|^3\right]\,,
\end{aligned}\label{free:adj:int}\ee
where we have also introduced $\tilde{F}$, corresponding to the free energy with excluded 
dependence on the radius $r$ and rank $N$ of the gauge group. 

Integrals in the expression (\ref{free:adj:int}) can be evaluated after substituting our
solution (\ref{density:adj:gen:n}). The first integral is relatively easy to evaluate 
\be\begin{aligned}
I_1^{(n)}&\equiv \frac{4\pi^2}{t}\int d\phi \rho^{(n)}(\phi)\,\phi^2=\frac{4\pi^2}{t}\sum_{k=0}^{n}2 
c_{k}\left(a^{(n)}-mk\right)^2=\\
&\frac{2\pi^2}{3t^3}(1+n)(2+n)\left[96\pi^4 (1+n)(2+n)-16\pi^2 t m (3+2n)+3m^2 t^2\right]\,,
\label{integral:1}
\end{aligned}\ee
where in the last step we have used explicit expressions for the coefficients $c_k$ (\ref{density:adj:gen:n})
and the endpoint position $a^{(n)}$ (\ref{endpoint:adj:general:n}).

The second integral coming from the one-loop determinants is more complicated. Let's consider it in 
 more details here
\be\begin{aligned}
I_2^{(n)}&\equiv -\frac{1}{24}\int \rho(\phi)\rho(\psi)\left[-2|\phi-\psi|^3+\frac{1}{2}|\phi-\psi +m|^3+
\frac{1}{2}|\phi -\psi-m|^3\right]=\\
&-\frac{1}{12}\sum_{k=0}^n\sum_{l=0}^n c_{k}c_{l}\left[-2m^3|k-l|^3-2|2a-m(k+l)|^3+m^3|k-l+1|^3\right.\\
&\left. +m^3|k-l-1|+|2a-m(k+l+1)|^3+|2a-m(k+l-1)|^3\right]\,.
\end{aligned}\ee
In the same manner as it was done when solving the saddle-point equation in (\ref{summation:shift}), 
we shift the summation index $l$ in the last four terms to obtain 
\be\begin{aligned}
I_2^{(n)}&=-\frac{1}{12}\sum_{k=0}^n\sum_{l=0}^n c_k\mathcal{C}_l\left[m^3|k-l|^3+|2a-m(k+l)|^3\right]
-\frac{1}{12}\sum_{k=0}^n c_k\left[c_0 \left(m^3(k+1)^3\right.\right.\\
&\left.\left.+(2a-m(k-1))^3\right)+c_n\left(m^3(n+1-k)^3+(m(k+n+1)-2a)^3\right)\right]\,,
\end{aligned}\ee
where  we have also used inequalities $m n<a<m(n+1)$.
This expression is much easier to work with, because, as we know, $\mathcal{C}_k=0$
for $k=1,..,n$. Then substituting values of the coefficients $c_0=\frac{1}{n+2},\,c_n=\frac{1}{(n+1)(n+2)},\,
\mathcal{C}_{0}=-\frac{1}{n+1}$ and the endpoint position $a^{(n)}$ from (\ref{endpoint:adj:general:n})
we finally obtain
\be
I_2^{(n)}=-\frac{1}{12t^3}\left[512 (1+n)^2(2+n)^2 \pi^6-64mt \pi^4 (1+n)(2+n)(3+2n)+m^3 t^3(3+2n)\right]\,.
\label{integral:2}
\ee
The free energy of the system is then given by the sum of (\ref{integral:1}) and (\ref{integral:2})
\be\begin{aligned}
\tilde{F}^{(n)}=\frac{1}{12 t^3}\left[256\pi^6(1+n)^2(2+n)^2\right.&-\left.64\pi^4 m t(1+n)(2+n)(3+2n)+\right.\\
&\left.24\pi^2t^2m^2(1+n)(2+n)-m^3t^3(3+2n)\right]\,.
\label{free:energy:adj}
\end{aligned}\ee
To find the order of the phase transition we calculate derivatives of the free 
energy (\ref{free:energy:adj}) with respect to the 't Hooft coupling at the critical points 
 (\ref{critical:pt:adj}). After short computation we observe
\be
\left.\partial_{t}\left(F^{(n+1)}-F^{(n)}\right)\right|_{t=t_c^{(n+1)}}&=&
\left.\partial^{2}_{t}\left(F^{(n+1)}-F^{(n)}\right)\right|_{t=t_c^{(n+1)}}=0\,,\\
\left.\partial^{3}_{t}\left(F^{(n+1)}-F^{(n)}\right)\right|_{t=t_c^{(n+1)}}&=&-\frac{m^6}{512 \pi^6 (2+n)^3}\,.
\ee
Hence, when 't Hooft coupling $t$ is increased theory undergoes through an infinite chain of
phase transitions of the third order. Moreover, with the increase of the coupling 
transition becomes weaker transforming into crossover transitions 
or infinite 't Hooft coupling $t$. 
Similar behavior was obtained in the case of the mass-deformed $ABJM$ theory \cite{Anderson:2014hxa,Anderson:2015ioa}.

\begin{figure}[!h]
\begin{center}
  \subfigure[Free Energy $\tilde{F}$.]{\label{pic:free:energ:adj}\includegraphics[width=51mm,angle=0,scale=1.6]{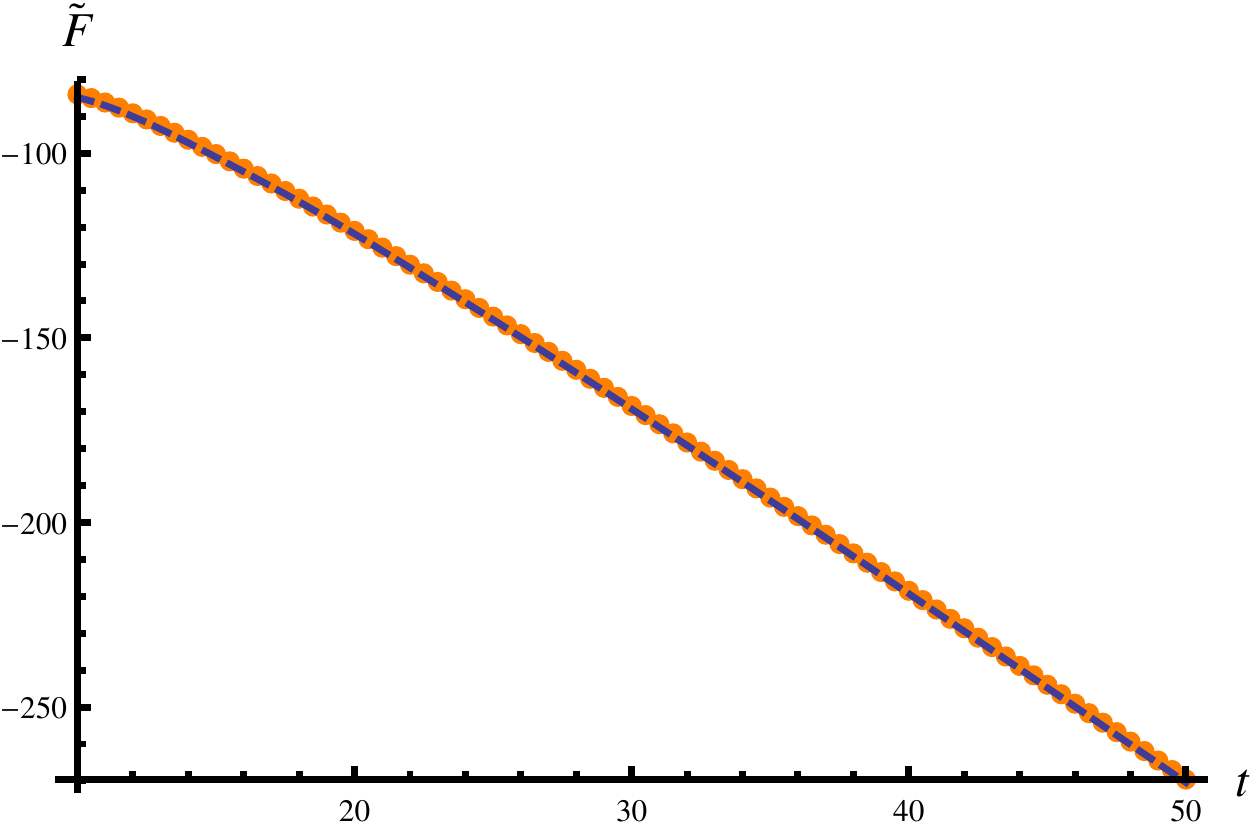}}
  \subfigure[Third derivative of the free energy  $\partial_{t}^3 \tilde{F}$.]
  {\label{pic:free:energy:der:adj}\includegraphics[width=51mm,angle=0,scale=1.6]{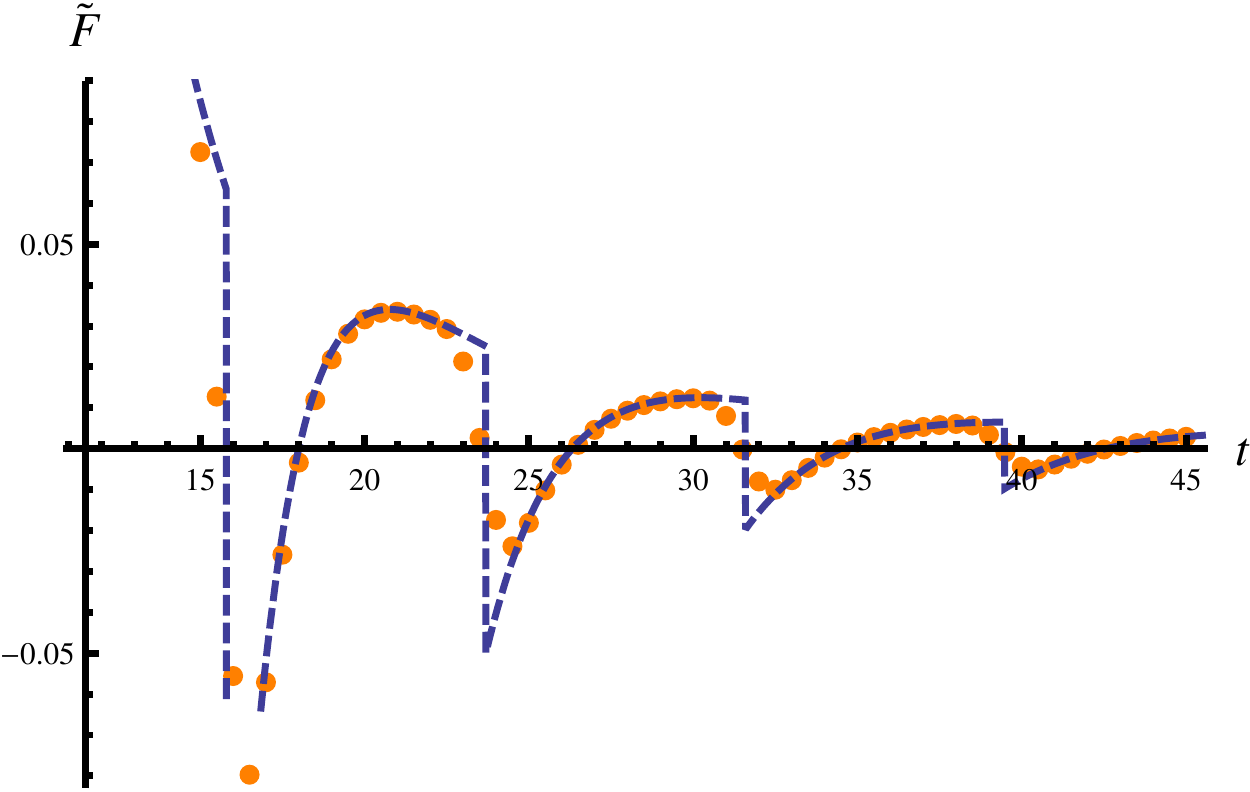}}
\end{center}
\caption{The free energy of $\mathcal{N}=1$ SYM with adjoint hypers. 
On the pictures above $\tilde{F}\equiv \frac{1}{2\pi r^3 N^2}F$
, where $F$ is the free energy defined in the usual way. The orange dots show the results for the numerical solution,
while the dashed blue lines represent the analytical solution (\ref{free:energy:adj})}. 
The parameters of the theory are taken to be $r=25,\,m=10,\,N=100$.
\label{pic:adj:free:energ}
\end{figure}

We have also evaluated the free energy for the full matrix model (\ref{partition-reg}) with 
massive adjoint hypermltiplet numerically. The results of these numerical 
simulations are shown in Fig.\ref{pic:adj:free:energ} with the orange dots. The dashed lines on the same plots
represent our analytical result (\ref{free:energy:adj}) for the free energy. As we see 
numerical and analytical results are in good agreement. The only difference can be obtained in the 
graph of the third derivative of the free energy. As we see the free energy, obtained numerically, does not 
have discontinuities at the critical points and hence leads to the absence of the phase transitions.  
This result is expected because the numerical solution is found for the finite $r$ and, as known,
finite-dimensional systems do not experience phase transitions.

\subsection{Wilson Loops}

We also study the behavior of the circular Wilson loop near the phase transition points. 
In the planar limit contribution of the $e^{2\pi\phi}$ prefactor in (\ref{wilson:matrix})
doesn't effect the position of the saddle point. Thus we
express Wilson loop expectation value in the integral form 
\be
\langle W\rangle = \int d\phi \rho(\phi)e^{2\pi\phi}\,,
\label{wilson:loop:mm2}
\ee
Substituting solution (\ref{density:adj:gen:n}) into this expression we easily obtain
\be
\langle W\rangle^{(n)} = 2\sum_{k=0}^n c_k \cosh\left(2\pi\left[(n+1-k)m-\frac{4\pi^2}{t}(n+1)(n+2)\right]\right)\,,
\ee
where we have used expression (\ref{endpoint:adj:general:n}) for the position of the endpoint.
Calculating derivatives of $\langle W\rangle^{(n)}$ with respect to the coupling constant around 
the critical point $t_c^{(n)}$ we obtain
\be\begin{aligned}
\left. \partial_{t}\left(\langle W\rangle^{(n+1)}-\langle W\rangle^{(n)}\right)\right|_{t=t_c^{(n+1)}}&=0\\
\left. \partial^2_{t}\left(\langle W\rangle^{(n+1)}-\langle W\rangle^{(n)}\right)\right|_{t=t_c^{(n+1)}}&=
\frac{m^4 \sinh\left(\pi m (2+n)\right)}{16\pi^2 (2+n)^2 \sinh(\pi m)}\,.
\end{aligned}\ee
Hence, the first discontinuity appears in the second derivative of the Wilson loop. This result is 
consistent with the third-order phase transitions experienced by the system at the critical 
points $t_c^{(n)}$.

\subsection{The strong coupling limit}

It is interesting to study the behavior of the 
solutions (\ref{density:adj:gen:n}) for a very strong coupling constant, such that $2a\gg m$.
In terms of our solution, this means that the eigenvalue density contains large number of 
resonance pairs $n\gg 1$.

Previously in \cite{Minahan:2013jwa} we have obtained that for the strong 
coupling eigenvalue density at the saddle point is 
\be\begin{aligned}
\rho(\phi)&= \frac{32\pi^2}{\left(9+4 \tilde{m}^2\right)\lambda}\,,~~ |\tilde{\phi}|\leq \tilde{a}\,,\\
          &= 0\,, \qquad\qquad\quad~ |\tilde{\phi}|> \tilde{a}\,
\end{aligned}\ee
where the endpoint position is given by $\tilde{a}=\left(9+4 \tilde{m}^2\right)\lambda/64 \pi^2$ and 
$\lambda$ is the coupling constant defined as $\lambda\equiv g_{YM}^2 N/r= t/r$. Restoring $r$-dependence 
(\ref{r:depend}) and taking the decompactification limit $r\to\infty$ we obtain 
\be\begin{aligned}
\rho(\phi)&= \frac{8\pi^2}{m^2t}\,,~~ |\phi|\leq a\,,\\
          &= 0\,, \quad~~~ |\phi|>a\,,
\end{aligned}\label{density:adj:asymptotic}\ee
where the cut endpoint position is given by 
\be
a=\frac{m^2 t}{16\pi^2}\,.
\label{endpoint:adj:asympt}
\ee

Now let's compare the expressions above with the corresponding limits of our solution (\ref{density:adj:gen:n})
and (\ref{endpoint:adj:general:n}). In order to do this recall that the solution with $n$ pairs of 
resonances is valid when $\frac{8\pi^2}{m}(n+1)<t<\frac{8\pi^2}{m}(n+2)$. Thus for very large $n$ we can approximate 
the number of the resonance pairs by $n\approx \frac{m t}{8\pi^2}$. For the position of the endpoint using 
(\ref{endpoint:adj:general:n}) we can easily obtain 
\be
a\approx mn-\frac{4\pi^2}{t}n^2=\frac{m^2 t}{16\pi^2}\,,
\ee
which exactly reproduces the expected result (\ref{endpoint:adj:asympt}). 

\begin{figure}[!h]
\begin{center}
  \subfigure[$t=300\,~ (n=17)$]{\label{pic:adj:t=300}\includegraphics[width=52mm,angle=0,scale=1.5]{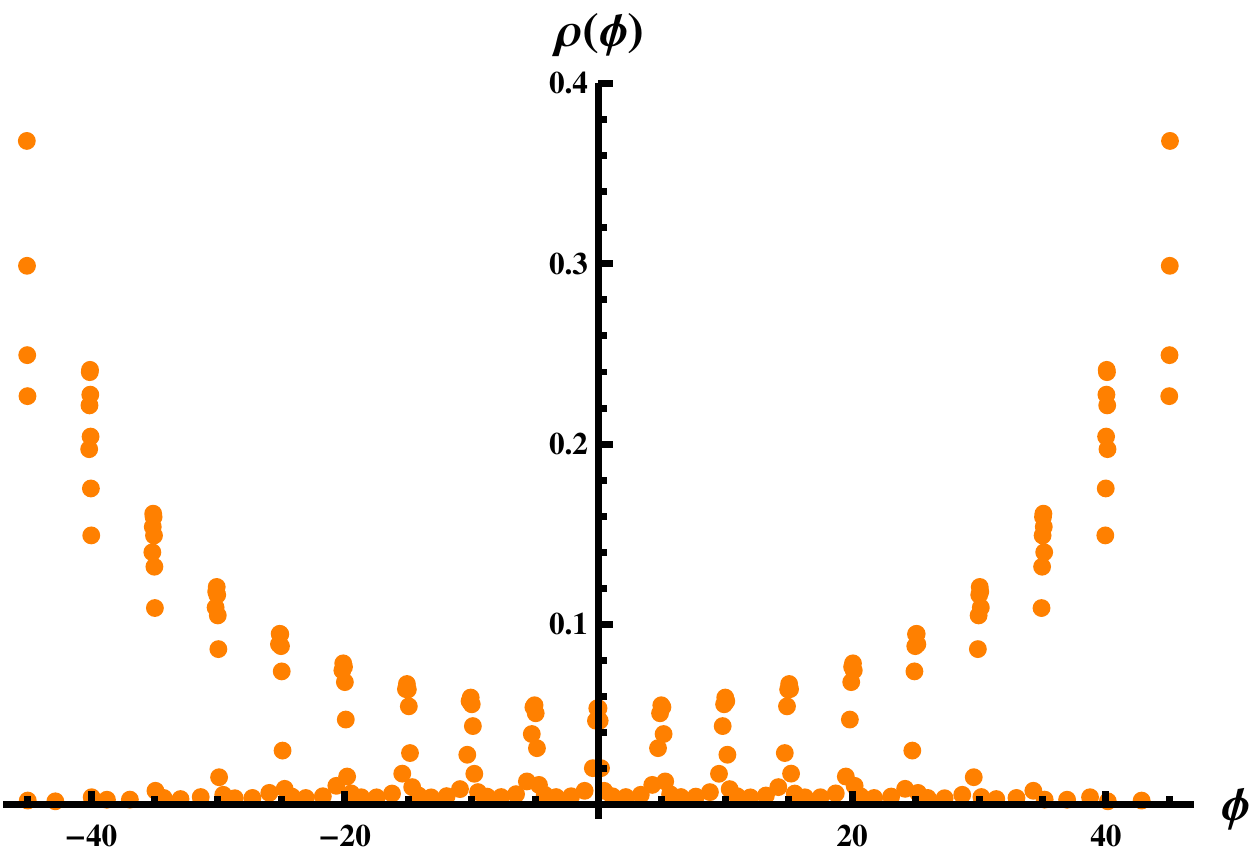}}
  \subfigure[$t=750\,~ (n=46)$]{\label{pic:adj:t=750}\includegraphics[width=52mm,angle=0,scale=1.5]{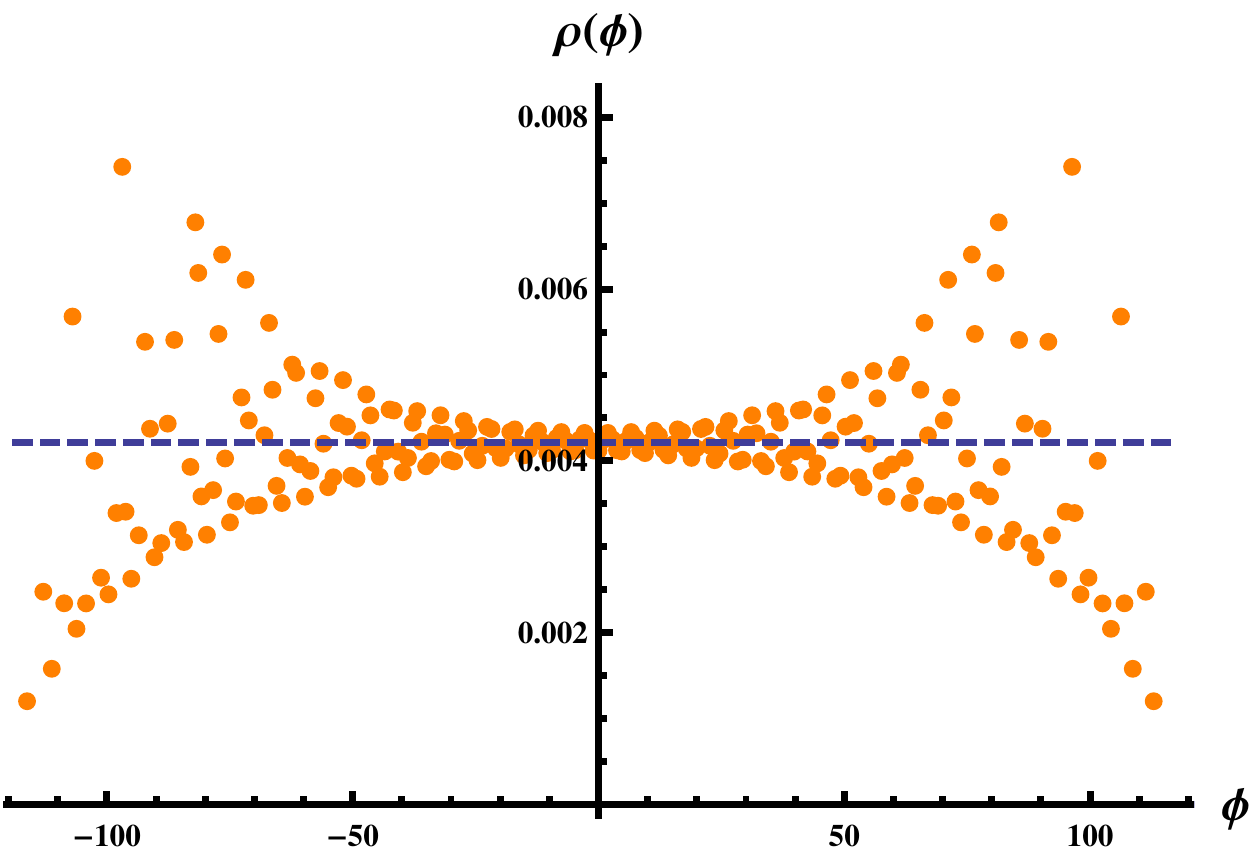}}
  \subfigure[$t=1250\,~ (n=78)$]{\label{pic:adj:t=1250}\includegraphics[width=52mm,angle=0,scale=1.5]{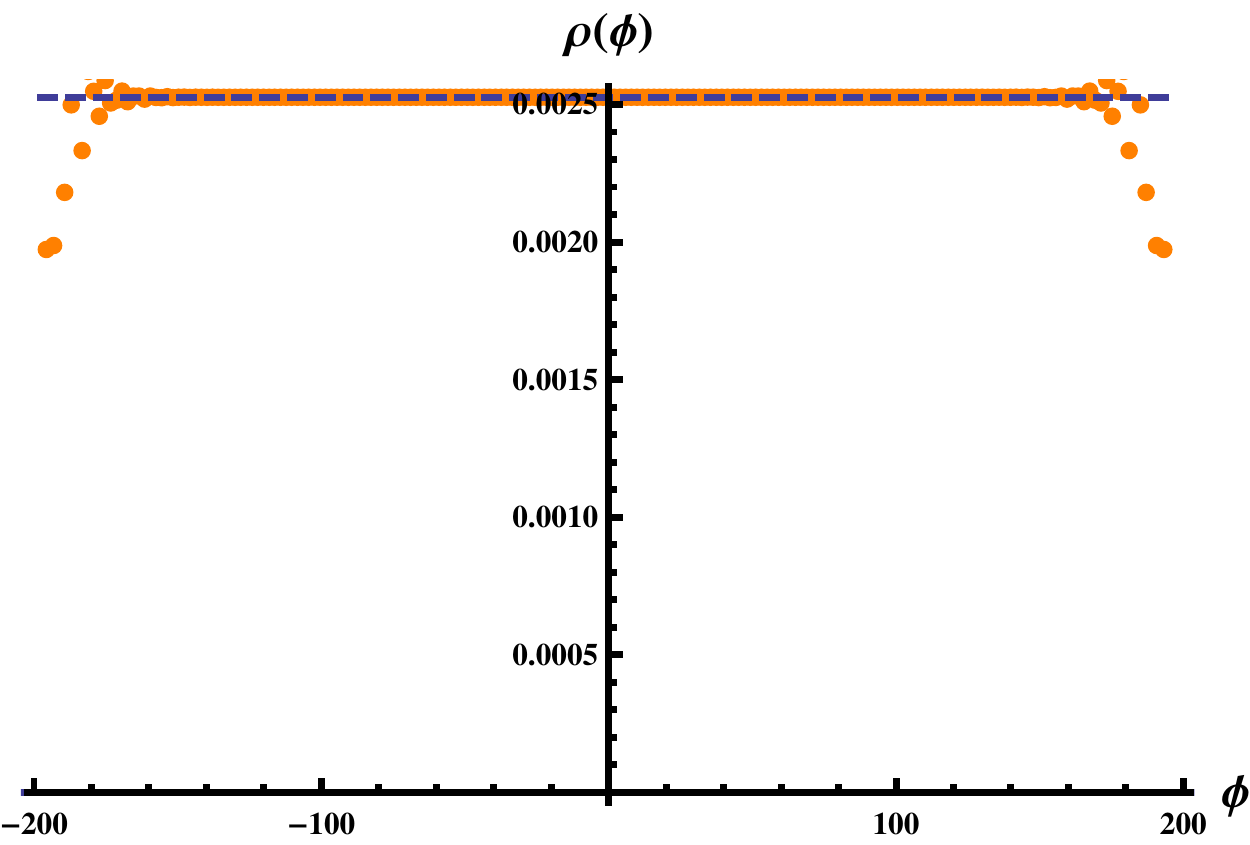}}
\end{center}
\caption{Results of the numerical solution (orange dots) to the equations of motion (\ref{eq-adj-full}) with the parameters 
$m=5$, $r=25$, $N=200$ and very large 't Hooft coupling. As we see for the large $t$ solution
approaches distribution (\ref{density:adj:asymptotic}), which is shown by dashed line}
\label{pic:large:t}
\end{figure}

In order to reproduce the eigenvalue density (\ref{density:adj:asymptotic}) we 
consider the case of large $n$ in (\ref{density:adj:gen:n}). As $n\gg1$ the cut 
 contains a large number of the $\delta$-functions which is then regarded as a continuous 
distribution. In order to understand which distribution is this, let's consider the following integral
\be
\int\limits_{0}^x d\phi \rho^{(n)}(\phi)=\sum_{k=\left[\frac{n}{2}-\frac{x}{m}+1\right]}^{\left[\frac{n}{2}+\frac{x}{m}\right]}c_k\approx
\int\limits_{k=\frac{n}{2}-\frac{x}{m}}^{\frac{n}{2}+\frac{x}{m}}dk\frac{n-k}{n^2}=\frac{x}{mn}\equiv \frac{8\pi^2x}{m^2 t}\,,
\ee
where we have used (\ref{density:adj:gen:n}) for $\rho^{(n)}(\phi)$ and $c_k$, and approximated 
the sum over $k$ by the integral assuming that the sum goes to large enough values of $k$.
As we obtained linear dependence of the integral 
$\int\limits_{0}^x \rho(\phi)$, we conclude that the eigenvalue density is given by 
the constant on the interval $[-a,\,a]$
\be\begin{aligned}
\rho(\phi)&= \frac{8\pi^2}{m^2t}\,,~~ |\phi|\leq a\,,\\
          &= 0\,, \quad~~~ |\phi|>a\,,
\end{aligned}\ee
which exactly reproduces solution (\ref{endpoint:adj:asympt}).

To conclude this section, we have shown that for the large number of resonances we can consider 
our solution as the averaging over these resonances. This averaging results in the 
constant eigenvalue distribution obtained in \cite{Minahan:2013jwa}.
Similar behavior was obtained for the solutions containing resonances in the mass-deformed $ABJM$ theory  
\cite{Anderson:2014hxa,Anderson:2015ioa}  and $4D$ $\mathcal{N}=2^*$ theory \cite{Russo:2013qaa}.  

The same conclusion about behavior of (\ref{density:adj:gen:n}) at very strong 
coupling can be drawn numerically. On the plots in Fig.\ref{pic:large:t} the orange dots represent
numerical solutions to (\ref{eq-adj-full}). On the same plots we show 
the solution (\ref{density:adj:asymptotic}) with the dashed blue line. As we see distribution for $t=300$ still 
looks like a bunch of peaks, while for $t=750$ it already starts approaching  (\ref{density:adj:asymptotic})
and for $t=1250$ completely coincides with it up to small regions of the size $\sim\frac{1}{r}$ around the 
endpoints of the distribution. 
Hence, numerical simulations support the conclusions of this section.

\section{Squashing spheres}

Another issue to discuss is the effect of the geometry on the phase transitions 
described in this paper. Localization methods can be generalized to theories on the space 
with geometry more complicated than the spheres. In particular $5D$ SYM was considered on 
squashed spheres \cite{Imamura:2012xg,Imamura:2012bm}, $Y^{p,q}$ spaces \cite{Qiu:2013pta,Qiu:2013aga}
and even more general toric Sasaki-Einstein manifolds \cite{Qiu:2014oqa}. 
Here we focus on the squashed spheres, which can be embedded into $\mathbb{C}^3$ as follows
\be
\omega^2_1\,|z_1|^2+\omega^2_2\,|z_2|^2+\omega^2_3\,|z_3|^2=r^2\,.
\ee
Then the corresponding matrix model obtained after localizing $5D$ SYM on this space is
\be
 Z=  \int d\tilde\phi e^{-\frac{4\pi^3 r}{g_{YM}^2}\rho  \Tr (\tilde\phi^2)-\frac{\pi k}{3}\rho\text{Tr}(\tilde\phi^3)}
 {\rm det}_{Ad} \Big ( \left.S_3(i\tilde\phi\right|\overline{\omega} ) \Big )\
   {\rm det}^{-1}_{R} \Big ( S_3 \left.\left (i\tilde\phi + \frac{3}{2}  \right|\overline{\omega} \right) \Big )~,
\label{partition:squashed}
\ee
where $\overline{\omega}\equiv (\omega_1\,,\,\omega_2\,,\,\omega_3)$ and $\rho$ is the volume factor of 
the squashed sphere, which is given by the ratio of the volumes of squashed and usual spheres
\be
\rho\equiv\frac{{\rm Vol}\,S^5_{sq.}}{{\rm Vol}\,S^5}=\frac{1}{\omega_1\,\omega_2\,\omega_3}\,.
\label{squashed:volume}
\ee
Finally, $S_3$ denotes triple-sine function that can be defined through the following infinite product
\be
S_3\left(\left.x\right|\overline{\omega}\right)=\prod\limits_{n_{1,2,3}=0}^{\infty}
\left(n_1\,\omega_1+n_2\,\omega_2+n_3\,\omega_3\right)\left((n_1+1)\,\omega_1+(n_2+1)\,\omega_2+(n_3+1)\,\omega_3-x\right)\,.
\ee
The matrix model (\ref{partition:squashed}) was partially studied for the case of $5D$ SCFT, i.e. SYM 
theory with the $USp(N)$ gauge 
group and infinite coupling constant $g_{YM}$ \cite{Alday:2014rxa,Alday:2014bta}. One of the reasons 
to study supersymmetric theories on the squashed spheres is that 
their partition functions are related to the supersymmetric R\'{e}nyi entropy. 
The latter can be evaluated from the partition function of theory on 
the $n$-covering of sphere, which is equivalent to the squashed sphere with the squashing 
parameters $(\frac{1}{n},1,1)$ \cite{Alday:2014fsa,Hama:2014iea}. 

In general the matrix model (\ref{partition:squashed}) is very complicated and can not be solved directly. 
However, we are only interested in the behavior of the matrix model solutions in the decompactification limit 
$r\to\infty$ which implies that the arguments of the triple-sine function are large, $\tilde\phi\gg 1$. 
Thus, we can use asymptotic expression for these functions \cite{Alday:2014bta,Imamura:2012bm}
\be\begin{aligned}
\left.\log S_3\left(\left.ix\right|\overline{\omega}\right)\right|_{x\to\infty}&\sim {\rm sign}(x)\,
\left(\frac{\pi}{6 \omega_1\,\omega_2\,\omega_3}x^3-\frac{i\,\pi\,\omega_{tot.}}{4 \omega_1\,\omega_2\,\omega_3}x^2\right.
\\&\left.+\frac{\pi\,\left(\omega_{tot.}^2+\omega_1\,\omega_2+\omega_1\,\omega_3+\omega_2\,\omega_3\right)}
{12 \omega_1\,\omega_2\,\omega_3}x+ \frac{i\pi \omega_{tot.}(\omega_1^2+\omega_2^2+\omega_3^2)}{24 \omega_1\,\omega_2\,\omega_3}
+O\left(\frac{1}{x}\right)\right)\,,
\label{triple:sin:asymp}
\end{aligned}\ee
where $\omega_{tot.}\equiv \omega_1+\omega_2+\omega_3$. In the case of the matrix model
 (\ref{partition:squashed}), $x=\tilde\phi=r\phi$ and we  
keep only cubic terms so that the partition function can be rewritten in the following form
\be\begin{aligned}
Z&=\int\limits_{Cartan} [d\phi]\,e^{-F(\phi)}\,,\\
\frac{\omega_1\,\omega_2\,\omega_3}{2\pi r^3}F(\phi)&=\frac{4\pi^2}{g_{eff}^2}{\rm tr}\left(\phi^2\right)+\frac{1}{12}{\rm tr}_{Ad}|\phi|^3-\frac{1}{24}\sum\limits_I {\rm tr}_{R_I}\left(|\phi +m|^3+|\phi -m|^3\right)\,.
\label{asymptotic:free:squash}\end{aligned}
\ee
From this expression we can see that the squashing does not affect our conclusions about phase transitions 
neither for fundamental nor for adjoint hypermultiplets. The only contribution of the squashing comes from the  
volume prefactor of the free energy. Similar results were obtained in the strong coupling limit of 
$5D$ ${\cal N}=1$ SYM with adjoint hypermultiplet on $Y^{p,q}$
spaces \cite{Qiu:2013pta} and toric Sasaki-Einstein manifolds \cite{Qiu:2014oqa}.
In particular, it was shown that for both cases the only difference in the prepotential of the theory 
between these spaces and five-sphere is in the overall volume factor.

Note that effects of the squashing on phase transitions were also studied for the case 
of $4D$  $\mathcal{N}=2^*$ SYM in \cite{Marmiroli:2014ssa}. However results obtained in that paper 
differs from ours. The main conclusion is that the squashing of the sphere shifts all critical 
points of the phase transitions chain to the smaller coupling constant. While in our case as we have 
seen above squashing does not affect positions of the critical points. 

\section{Discussion}

In this paper we studied the decompactification limit of the matrix model obtained 
from $5D$ ${\cal N}=1$ SYM theory with different hypermultiplet content on $S^5$.
First we considered $N_f$ massive 
fundamental hypermultiplets coupled to ${\cal N}=1$ SYM theory. Solving the planar limit 
of the corresponding matrix model we found that the eigenvalue density $\rho(\phi)$ at the saddle point 
consists of an isolated peaks. These peaks are located at the boundaries of the distribution at $\phi=\pm a$
 and 
at the points $\phi=\pm m$. Hence, we concluded that there are two phases of the theory. One phase 
corresponds to the case when the support includes $\phi=\pm m$ and another one when it does not. Transition
between these two phases takes place at the critical point
\be
t_c=-\frac{8\pi^2}{m}\left(1-\frac{N_f}{2 N}\right)^{-1}\,,
\ee
where $t=g_{YM}^2 N$ is the 't Hooft coupling used in this paper. Evaluating 
the free energy we find that the transition between these two phases is a third-order
phase transition. 

Second theory we consider is ${\cal N}=1$ SYM  coupled to the massive adjoint 
hypermultiplet. In this case we also find that solution for the eigenvalue density 
consists of isolated peaks. These peaks are located at $\phi=\pm(a-m n)$, where 
$a$ is the endpoint of the distribution support, $m$ is the mass of the hypermultiplet and 
$n$ is an integer number.  As the 
coupling is increased the length of the support grows and it can 
include more and more peaks. As the result, theory goes through an infinite number of phases which differ
by the number of peaks in the solution. Transition between phases with $n$ and $n+1$ pairs of peaks 
take place at 
\be
t_c^{(n)}=\frac{8\pi^2}{m}(n+1)\,.
\ee
Free energy calculation showed that these transitions are third-order phase transitions.
In the limit of very strong coupling the peaks, forming the eigenvalue density, 
smooth out and solution becomes equal to the strong coupling solution derived in \cite{Minahan:2014hwa}.

It has been also shown that the effect of the five-spheres squashing comes 
in the form of an overall prefactor of the free energy in the decompactification limit of the matrix model, 
while saddle point remains the same. Hence, we concluded that the squashing does not affect 
neither the positions of the critical points nor the order of the phase transitions. 

Finally, we  note that the phase structure 
described in this paper seems to be general for 
the supersymmetric theories with massive hypermultiplets, as they arise 
in $D=3,4,5$ dimensions. Hence, in the future it would be interesting 
to understand the nature of these phase transitions better.

\section*{Acknowledgements}

We thank Joseph Minahan and Konstantin Zarembo for many useful discussions.
We also thank Joseph Minahan for commenting on the manuscript of paper.
This research is supported in part by
Vetenskapsr{\aa}det under grant \#2012-3269.

\appendix

\section{Properties of the functions $f(x)$ and $l(x)$}
\label{spec_functions}

In this appendix we consider properties of the functions $f(x)$ and $l(x)$ that appear in the 
expressions for the regularized one-loop contributions (\ref{1-loop-vect-reg}) and (\ref{1-loop-hyp-reg}).
Function $f(x)$,first introduced in \cite{Kallen:2012cs}, can be written in th following form
\be
f(x)=\frac{i\pi x^3}{3}+x^2 \log\left(1-e^{-2\pi i x}\right)+\frac{i x}{\pi}\mathrm{Li}_{2}\left(e^{-2\pi i x}\right)
+\frac{1}{2\pi^2}\mathrm{Li}_{3}\left(e^{-2\pi i x}\right)-\frac{\zeta(3)}{2\pi^2}~.
\label{f:function}
\ee
while $l(x)$, first introduced in \cite{Jafferis:2010un}, is defined by
 \be
l(x)=-x\log\left(1-e^{2\pi i x}\right)+\frac{i}{2}\left(\pi x^2+\frac{1}{\pi}\mathrm{Li}_2(e^{2\pi i x})\right)
-\frac{i \pi}{12}
\label{l:function}
\ee
Using these expressions we can derive the following properties of $l(x)$ and $f(x)$:
\begin{enumerate}
 \item $l(x)$ is an odd function and $f(x)$ is an even function 
 $$l(x)=-l(-x),\qquad f(x)=f(-x)$$
\item 
  The derivatives of the functions are given by
    \be
\frac{d f(x)}{dx}=\pi x^2 \cot(\pi x)\, ;\qquad \frac{d l(x)}{dx}=-\pi x \cot(\pi x) \,;
\label{f:l:derivative}
\ee
\item 
  The asymptotic behavior of the functions is given by
   \begin{eqnarray}
 \lim_{|x|\to\infty} \mathrm{Re}f\left(\frac{1}{2}+i x\right) =  -\frac{\pi}{3}|x|^3+\frac{\pi}{4}|x|\,;  &
 \lim\limits_{ x \to \infty } \mathrm{Im}f\left(\frac{1}{2}\pm i x\right)  =  \pm\frac{\pi}{2}x^2\, ;
\nonumber
\\
 \lim_{|x|\to\infty} \mathrm{Re}~l\left(\frac{1}{2}+i x\right)=  -\frac{\pi}{2}|x|\,;  &
 \lim\limits_{x\to\infty} \mathrm{Im} ~l\left(\frac{1}{2}\pm i x\right) = \mp\frac{\pi}{2}x^2\, ;
 \label{asymptotes}
 \\
 \nonumber
  \lim_{|x|\to\infty} \mathrm{Re}f\left(i x\right) =  -\frac{\pi}{3}|x|^3\,;  &
 \mathrm{Im} f\left(i x\right)  = 0\, .
\end{eqnarray} 
\end{enumerate}
Using these properties we derive similar properties for $F_V(x)$ and $F_H(x)$ functions 
in (\ref{FV-funct}) and (\ref{FH-funct}):
\begin{enumerate}
 \item Both $F_V(x)$ and $F_H(x)$ are symmetric functions
 $$F_V(-x)=F_V(x),\qquad F_H(x)=F_H(x).$$
 \item The derivatives of the functions are given by
 \be\begin{aligned}
 \frac{d F_V(x)}{d x}&=-\frac{\pi}{2}\left(2-x^2\right)\coth(\pi x)\,,\\
  \frac{d F_H(x)}{d x}&=-\frac{\pi}{2}\left(\frac{1}{4}+(m+x)^2\right)\tanh(\pi(x+m))\,.\label{derivtives-F}
 \end{aligned}\ee
\item Finally the asymptotic behavior of these functions is given by
\be\begin{aligned}
\lim\limits_{|x|\to\infty}F_V(x)&=\frac{\pi}{6}|x|^3-\pi |x|\,,\\
\lim\limits_{|x|\to\infty}F_H(x)&=-\frac{\pi}{6}|x+m|^3-\frac{\pi}{8}|x+m|\,.
\label{asymptotic_F}
\end{aligned}\ee
\end{enumerate}

\section{Details of numerical analysis}
\label{numerix:appendix}

To obtain numerical results presented in this paper we have used 
the method introduced in \cite{Herzog:2010hf} in order to solve the $ABJM$ matrix model. 
It was also adopted for solving $5D$ SYM matrix model \cite{Minahan:2013jwa}
and $5D$ Chern-Simons matrix model \cite{Minahan:2014hwa}.

Instead of solving the the saddle point equation 
$\frac{\partial F}{\partial \phi_i}=0$, where $F$ is the prepotential, 
we introduce dependence of the eigenvalues $\phi_i$ on the effective "time" variable $t$. Then we 
substitute our original equation with the effective "heat" equation
\be
\tau \frac{d\phi_i(t)}{d t}=\frac{\partial F}{\partial t}\,.
\label{heat:equation}
\ee
At large time $t$ solution of this differential equation approaches the solution
of the original saddle point equation, provided the choice of $\tau$'s sign is made properly.

However, solving (\ref{eq-fund-full}) which corresponds to  the case of 
the fundamental hypermultiplet, we faced some problems. Our numerical method appeared 
to be unstable and very sensitive towards initial conditions due to the repulsive central potential 
$\sim -\Lambda \phi^2$. 

To avoid this problem we used the fact that the renormalization (\ref{coup-renorm})
of the coupling constant $g_{YM}$ for the case of $\mathcal{N}=1$ SYM  can be reproduced 
by considering the case of one adjoint hypermultiplet with very large mass. The
easiest way to realize this is to consider (\ref{eq-adj-full}). 
If the mass of the hypermultiplet is large we can rewrite this equation 
in the following form
\be
\left(\frac{16\pi^2 r}{g_{YM}^2}-2\tilde{m}N\right)\tphi_i=\sum_{j\neq i}\left(2-(\tphi_i-\tphi_j)^2\right)
\coth(\pi(\tphi_i-\tphi_j))\,,
\label{mass:decoupl}
\ee
where we used  $\tanh\left(\pi(\tphi_i-\tphi_j\pm \tilde{m})\right)={\rm sign}\left(\pm \tilde{m}\pi\right)=\pm 1$
and also $\sum\limits_{j}\phi_j=0$ due to the tracelessness condition. From (\ref{mass:decoupl})
we conclude that the coefficient in front of $\tphi_i$ on the l.h.s. plays the role of effective coupling. To
put (\ref{mass:decoupl}) in the form similar to (\ref{eq-fund-full})  $\Lambda$ should be expressed 
through the bare coupling $g_{YM}$ and the mass of the hypermultiplet $m$ in the following way
\be
\Lambda = m-\frac{8\pi^2 }{t} \,.
\label{effective:coup}
\ee
where $t=g_{YM}^2 N$ is the 't Hooft coupling.
So in order to avoid the problems with the repulsive central potential we introduce 
adjoint hypermultiplet with large enough mass $m$. In this case the central potential itself is attractive as
$g_{YM}^2>0$. However the effective coupling (\ref{effective:coup}) can be made negative as we need it in 
equation (\ref{eq-fund-full}). All numerical simulations of the theory with the fundamental hypermultiplets  
presented in this paper were performed in this way.

\section{Solution for the finite $r$: fundamental matter}
\label{finite:R:fund}

In this appendix we consider the solution of (\ref{eq-fund-full}) for the large enough but finite $r$.
In section \ref{N=1:fund:sec} we have neglected all terms subleading in $1/r$ in this equation
in order to consider properly the decompactification limit $r\to\infty$. As the result we have obtained 
solutions (\ref{density:fund:below}) and (\ref{ansatz:fund}) containing $\delta$-functions. These $\delta$-functions
arise in the solutions because of the kernel in (\ref{eom-fund-int}) 
that looks like $\quad-(\phi-\psi)^{2}{\rm sign}\left(\phi-\psi\right)\quad$ leading to the
attractive potential between the eigenvalues. If not the repulsive central potential,
the eigenvalues would just all condense at the origin $\phi=0$.

Here we will be more accurate with the kernel of integral equation. We now do not  
neglect the subleading term in the kernel of (\ref{eq-fund-full}), which leads to 
\be\begin{aligned}
-2\Lambda r^2 \phi =\frac{1}{4}\zeta r^2 (\phi +m)^2~{\rm sign}(\phi +m)+\frac{1}{4}\zeta r^2 (\phi -m)^2~{\rm sign}(\phi -m)+\\
\int d\psi \rho(\psi)\left(2-r^2(\phi-\psi)^2\right){\rm sign}\left(\phi-\psi\right),
\label{eom:fund:int:finite:r}
\end{aligned}\ee
where $\rho(\phi)$ is the eigenvalue density defined by (\ref{density:def}). Due to the symmetry 
of (\ref{eom:fund:int:finite:r}) we expect the eigenvalue density to be symmetric.
Note that terms subleading in $1/r$ are repulsive, hence they wash out $\delta$-functions 
into peaks of the finite width $\sim 1/r$. At the same time we have not included terms subleading in $1/r$
into the central potential, because they add small central repulsion, thus never playing an important role.
Assuming $\coth\left(\pi r(\phi_i-\phi_j)\right)\approx 
{\rm sign}(\phi_i-\phi_j)$ we also neglect terms exponentially suppressed in large $r$.
However, note that this approximation works pretty well even in the region, where 
$|\phi_i-\phi_j|\approx 1/r$ due to the factor of $\pi$ in the argument of $\coth$. 
Now we consider phases I and II separately.

\subsection{Below the transition point}

We start with the phase below the transition point which corresponds to $a<m$, where 
$a$ is the position of the eigenvalue support endpoint.  In this case equation (\ref{eom:fund:int:finite:r}) 
and its derivatives are given by 
\be
\left(2\Lambda +\zeta m\right)r^2 \phi +
\int d\psi \rho(\psi)\left(2-r^2(\phi-\psi)^2\right){\rm sign}\left(\phi-\psi\right)&=&0\,,\label{eq:orig:finite:r:fund}\\
\left(2\Lambda +\zeta m\right)r^2 +4\rho(\phi)-2 r^2\int d\psi \rho(\psi)|\phi-\psi|&=&0\,,\label{1st:der:fin:r:fund}\\
4\rho'(\phi)-2 r^2\int d\psi \rho(\psi){\rm sign}\left(\phi-\psi\right)&=&0\,,\label{2nd:der:fin:r:fund}\\
4\rho''(\phi)-4r^2 \rho(\phi)&=&0\,,\label{3rd:der:fin:r:fund}
\ee
General symmetric solution to (\ref{3rd:der:fin:r:fund}) is
\be
\rho(\phi)=C~\cosh(r \phi)\,.
\label{sol:gen:below:fund}
\ee
Now we are left to determine constant $C$ and the position of the support endpoint $a$.
To determine these constants we substitute the general solution (\ref{sol:gen:below:fund}) into 
normalization condition (\ref{normalization})
\be
\int\limits_{-a}^{a} d\phi\rho(\phi)=\frac{2}{r}C\sinh(r a)=1
\ee
and into equation (\ref{1st:der:fin:r:fund})
\be
\left(2\Lambda +\zeta m\right)r^2-2 a r^2 +4C\cosh(r a)=0\,.
\ee
This leads to 
\be\begin{aligned}
&C=\frac{r}{2\sinh(r a)}\,,\\
&\left(2\Lambda +\zeta m\right)r^2-2 a r^2 +2 r\coth(r a)=0\,.\label{conditions:Ca:fund}
\end{aligned}\ee
The consistency of these conditions with 
equations (\ref{eq:orig:finite:r:fund}) and (\ref{2nd:der:fin:r:fund}) can be checked by the direct substitution.
In general (\ref{conditions:Ca:fund}) can be solved only numerically for different 
values of the parameters $r, m, \zeta$ and $\Lambda$. However, for the  large radius, when $r a \gg 1$, we can 
simplify the last equation 
\be
a=\Lambda+\frac{1}{2}\zeta m\,,
\ee
which reproduces the support endpoint (\ref{endp:fund:below}) we found in section \ref{N=1:fund:sec}.
Finally, the eigenvalue density below the phase transition point $\Lambda_c$
\footnote{Notice that this critical coupling is in general different
from (\ref{critical:pt:fund}) and can be found numerically using (\ref{conditions:Ca:fund}).
However, in 
the decompactification limit $r\to\infty$ the critical point $\Lambda_c$ approaches (\ref{critical:pt:fund}).} is 
\be\begin{aligned}
\rho^{(I)}(\phi)& = \frac{r}{2\sinh(r a)}\cosh(r \phi)\,,~~|\phi|<a\\
          & = 0,~~~~~~~~~~~~~~~~~~~~~~~~~~|\phi|>a\,.
\label{dens:fund:below:finiter}
\end{aligned}\ee
For the large $r$ this distribution is the function with two 
sharp peaks of width $\sim 1/r$ at $\phi\pm a$. In the decompactification limit 
$r\to\infty$ this density is equivalent to (\ref{density:fund:below}), because the 
width of the peaks goes to zero, while the integral of $\rho(\phi)$ stays
$1$.

\begin{figure}[!h]
\begin{center}
  \subfigure[$\Lambda=2 \left(\Lambda<\Lambda_c\right)$]{\label{fund:below:rfin}\includegraphics[width=52mm,angle=0,scale=1.5]{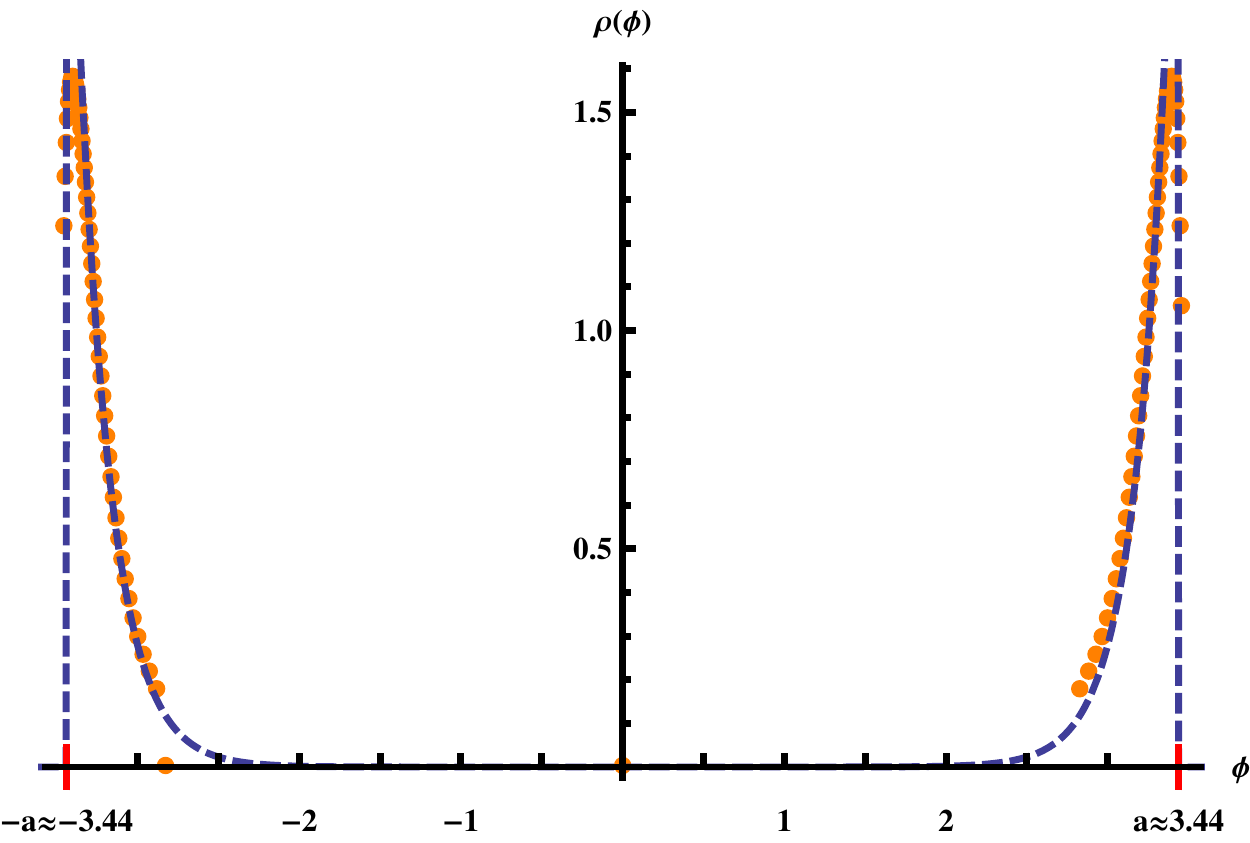}}
  \subfigure[$\Lambda=5.5 \left(\Lambda>\Lambda_c\right)$]{\label{fund:above:rfin}\includegraphics[width=52mm,angle=0,scale=1.5]{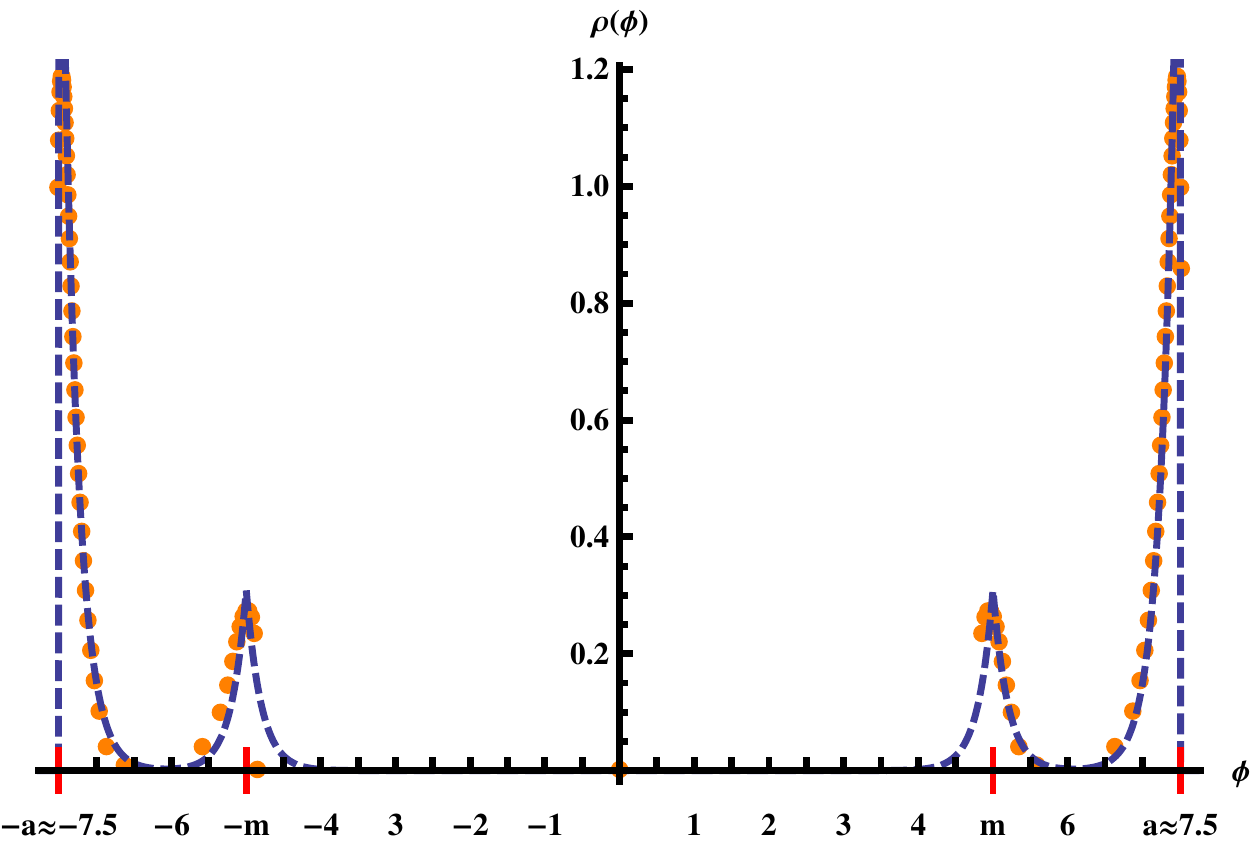}}
\end{center}
\caption{The eigenvalue density $\rho(\phi)$ calculated with the following value of parameters: $r=5,\,m=5,\, \zeta=
\frac{1}{2},\,N=100$ corresponding to $\Lambda_c\approx 3.76$. Orange dots show the results for the numerical solution, dashed blue 
lines show the solutions (\ref{dens:fund:below:finiter}) and (\ref{density:fund:full}) on pictures $(a)$
and $(b)$ correspondingly.}
\label{fund:dens:finiter}
\end{figure}

We have also found the numerical solutions to (\ref{eq-fund-full}), which are shown 
in Fig.\ref{fund:below:rfin} with the orange dots. Dashed blue lines shows the solution 
(\ref{dens:fund:below:finiter}) with the coefficient $C$ and the endpoint $a$ found numerically from the equations 
(\ref{conditions:Ca:fund}). As we see, solutions found above are consistent with the numerical results.

\subsection{Above the transition point}

Now we find the solution to (\ref{eom:fund:int:finite:r}) for the case of 
the system in the phase II. This phase corresponds to $a>m$, and to solve 
equations of motion we should 
divide the whole eigenvalue support $[-a,a]$ into three parts $A,\,B,\,C$ 
\be\begin{aligned}
                    &~~~ \rho_A(\phi),\quad ~~m<\phi<a\,;\\
\rho^{(II)}(\phi)=  &~~~ \rho_B(\phi),\quad -m<\phi<m\,;\\
                    &~~~ \rho_C(\phi),\quad -a<\phi<-m\,;
\end{aligned}\ee
with the symmetry constraints
\be
\rho_B(\phi)=\rho_B(-\phi)\,,\quad \rho_A(\phi)=\rho_C(-\phi)\,.
\label{symm:rho:fund}
\ee
Now we write down equation of motion (\ref{eom:fund:int:finite:r}) in regions $A$ and $B$
and solve them in this regions separately. 

{\bf Region A $(m<\phi<a)$}. In this interval (\ref{eom:fund:int:finite:r}) and 
it's first, second and third derivatives are 
\be
2\Lambda r^2 \phi +\frac{1}{2}\zeta r^2 \left(\phi^2 +m^2\right)+
\int d\psi \rho(\psi)\left(2-r^2(\phi-\psi)^2\right){\rm sign}\left(\phi-\psi\right)&=&0\,,\label{eq:orig:finite:r:fundA}\\
2\Lambda r^2 +\zeta r^2 \phi +4\rho_A(\phi)-2 r^2\int d\psi \rho(\psi)|\phi-\psi|&=&0\,,\label{1st:der:fin:r:fundA}\\
\zeta r^2 + 4\rho_A'(\phi)-2 r^2\int d\psi \rho(\psi){\rm sign}\left(\phi-\psi\right)&=&0\,,\label{2nd:der:fin:r:fundA}\\
4\rho_A''(\phi)-4r^2 \rho_A(\phi)&=&0\,.\label{3rd:der:fin:r:fundA}
\ee
As we do not have any symmetry restrictions on $\rho_A$, the general solution to
(\ref{3rd:der:fin:r:fundA}) is given by 
\be
\rho_A(\phi)=C_2 \cosh\left(r \phi -r \frac{a+m}{2}\right)+C_3 \sinh\left(r \phi -r \frac{a+m}{2}\right)
\label{rhoA:fund}
\ee
Note that we choose solution to be centered around $\phi=(a+m)/2$. This choice 
is made for convenience. The shifts in $\cosh$ and $\sinh$ can be adsorbed 
into constants $C_2$ and $C_3$


{\bf Region B $(-m<\phi<m)$}. On this interval $\phi<m$, so that everything is the same as in the case of the 
phase I. Equation of motion and its three derivatives then can be written in the form of 
 (\ref{eq:orig:finite:r:fund})-(\ref{3rd:der:fin:r:fund}). Similarly to the phase I we 
obtain 
\be
\rho_B(\phi)=C_1 \cosh(r\phi)\,,
\label{rhoB:fund}
\ee

{\bf Region C $(-a<\phi<-m)$}. In this region using 
symmetry properties (\ref{symm:rho:fund}) we find 
\be
\rho_C(\phi)=C_2 \cosh\left(r \phi +r \frac{a+m}{2}\right)-C_3 \sinh\left(r \phi +r \frac{a+m}{2}\right)\,.
\label{rhoC:fund}
\ee

To complete our solution, we need to find all the parameters $C_1,\,C_2,\,C_3,\,a$ in 
(\ref{rhoA:fund}),(\ref{rhoB:fund}) and (\ref{rhoC:fund}). To determine these 
 parameters we substitute solutions (\ref{rhoA:fund})-(\ref{rhoC:fund}) into 
equations (\ref{eq:orig:finite:r:fund})-(\ref{2nd:der:fin:r:fund}), (\ref{eq:orig:finite:r:fundA})-
(\ref{2nd:der:fin:r:fundA}) and normalization condition (\ref{normalization}). This leads to 
\be
C_1\sinh(rm)+2C_2\sinh\left(r\frac{a-m}{2}\right)&=&\frac{r}{2}\,,\label{normal:cond:fund}\\
4C_2 \sinh\left(r\frac{a-m}{2}\right)+4C_3\cosh\left(r\frac{a-m}{2}\right)&=&(2-\zeta)r\,,\label{2nd:der:cond:fundA}\\
4\left(C_2-ar C_3\right)\cosh\left(r\frac{a-m}{2}\right)+
4\left(C_3-ar C_2\right)\sinh\left(r\frac{a-m}{2}\right)&=&-2\Lambda r^2\,,\label{1st:der:cond:fundA}\\
C_1\cosh(m r)-C_2\cosh\left(r\frac{a-m}{2}\right)+C_3\sinh\left(r\frac{a-m}{2}\right)&=&0\,,\label{1st:der:cond:fund}
\ee
where (\ref{normal:cond:fund}) is obtained from the normalization condition, while 
 (\ref{2nd:der:cond:fundA}),(\ref{1st:der:cond:fundA}) and (\ref{1st:der:cond:fund}) are 
obtained after substituting the solution for the eigenvalue density into (\ref{2nd:der:fin:r:fundA}),
(\ref{1st:der:fin:r:fundA}) and (\ref{1st:der:fin:r:fund}) correspondingly. It can be checked by the 
direct calculation that remained equations (\ref{eq:orig:finite:r:fund}),(\ref{eq:orig:finite:r:fundA}),(\ref{2nd:der:fin:r:fund})
will be satisfied automatically if the coefficients $C_1,\,C_2,\,C_3,\,a$ obey equations we have found above.

To summarize we have found that above the phase transition point, when $m<a$, the eigenvalue density 
satisfying equation of motion (\ref{eom:fund:int:finite:r}) can be written in the form
\be\begin{aligned}
                    &~~~ C_2 \cosh\left(r \phi -r \frac{a+m}{2}\right)+C_3 \sinh\left(r \phi -r \frac{a+m}{2}\right),\quad ~~m<\phi<a\,;\\
\rho^{(II)}(\phi)=  &~~~ C_1 \cosh(r\phi)\,\hspace{7.3cm} -m<\phi<m;\\
                    &~~~ C_2 \cosh\left(r \phi +r \frac{a+m}{2}\right)-C_3 \sinh\left(r \phi +r \frac{a+m}{2}\right),\quad -a<\phi<-m\,;
\label{density:fund:full}\end{aligned}\ee
with coefficients $C_1,C_2,C_3$ and the endpoint $a$ of the cut satisfying equations (\ref{normal:cond:fund})-(\ref{1st:der:cond:fund}).

Equations (\ref{normal:cond:fund})-(\ref{1st:der:cond:fund}) 
can be solved exactly in the limit of large radius $r$. In the case, when $rm\gg1$ and $r \frac{a-m}{2}\gg 1$
we approximate these equations with simpler ones
\be
C_1 \exp\left(mr\right)+2C_2 \exp\left(r\frac{a-m}{2}\right)&=&r\,,\nn\\
\left(C_2+C_3\right)\exp\left(r\frac{a-m}{2}\right)&=&\left(1-\frac{1}{2}\zeta\right)r\,,\nn\\
\left(C_2+C_3\right)ar \exp\left(r\frac{a-m}{2}\right)&=&2\Lambda r^2\,,\nn\\
C_1 \exp\left(m r\right)+\left(C_3-C_2\right)\exp\left(r\frac{a-m}{2}\right)&=&0\,.
\ee
This system can be solved leading to 
\be
a&=&\left(1-\frac{1}{2}\zeta\right)^{-1}\Lambda\,,\quad
C_{1}=\frac{1}{4}\zeta r \exp(-mr)\,,\nn\\
C_{2}&=&r\left(\frac{1}{2}-\frac{1}{8}\zeta\right)\exp\left(r\frac{m-a}{2}\right)\,,\quad
C_3=r\left(\frac{1}{2}-\frac{3}{8}\zeta\right)\exp\left(r\frac{m-a}{2}\right)\,.\label{coeff:fund}
\ee
Note that to find these solution we assumed 
$r \frac{a-m}{2}\gg 1$, which means that 
 solutions only makes sense far enough from the critical point where $a\approx m$. However this 
approximation will work very well in the decompactification limit even for the points close to the critical ones. 

As we see from (\ref{coeff:fund}) the position of the cut endpoint is consistent with the one obtained in the 
section \ref{N=1:fund:sec}. If we now come back to the eigenvalue density (\ref{density:fund:below}) and take into account 
coefficients (\ref{coeff:fund}) we can see that in the decompactification limit $r\to\infty$ solutions are nonzero only at 
the points $\phi=\pm a$ and $\phi=\pm m$, where they take infinite value. At the same time from the normalization 
condition (\ref{normalization}) we know that $\rho(\phi)$ integrates to one. Then the natural function to describe this 
limit is (\ref{ansatz:fund}) and the coefficients in front of the $\delta$-functions should be
 \be
 \rho(\phi)=\left(C_2+C_3\right)\delta(\phi-a)+\left(C_1+C_2-C_3\right)\delta(\phi-m)+\left(a\to-a,\, m\to -m\right)\,
 \ee
which leads exactly to (\ref{ansatz:fund}).

We also can find solutions to  (\ref{normal:cond:fund})-(\ref{1st:der:cond:fund}) numerically for the 
particular combinations of the matrix model parameters. In Fig.\ref{fund:above:rfin} the orange dots shows numerical
solution to (\ref{eq-fund-full}), while the dashed blue lines shows the solution 
(\ref{density:fund:full}) with all coefficients found numerically from 
(\ref{normal:cond:fund})-(\ref{1st:der:cond:fund}). As we can see numerical solution perfectly coincides with the 
solution found in this section.

\section{Solution for the finite $r$: adjoint matter}
\label{finite:R:adj}

In this appendix we consider solution to the saddle-point equation of the matrix model 
(\ref{partition-reg}) for the large but finite $r$. We have solved these equations  
in the decompactification limit (see section \ref{adjoint:hyper:section}) neglecting all
terms subleading in $1/r$. As the result we have 
obtained the general solution (\ref{density:adj:gen:n}) containing the $\delta$-functions. In order to show more precise 
how these $\delta$-functions arise we should take into account first subleading terms in the kernels of 
 (\ref{eq-fund-full}), which leads us to 
\be\begin{aligned}
\frac{16\pi^2}{t}r^2\phi&=\int d\psi \rho(\psi)\left[\left(2-r^2 (\phi-\psi)^2\right){\rm sign}(\phi-\psi)\right.\\
&\left. \frac{1}{2}\left(\frac{1}{4}+r^2(\phi-\psi-m)^2\right){\rm sign}(\phi-\psi-m)+(m\to -m)\right]\,.
\label{eq:adj:finite:r}
\end{aligned}\ee
where we restore the radius dependence (\ref{r:depend}) and assume that 
$\coth(\pi r(\phi-\psi))\approx {\rm sign}(\phi-\psi)$
as well as $\tanh(\pi r(\phi-\psi\pm m))\approx {\rm sign}(\phi-\psi\pm m)$ when we consider the limit $r\gg 1$. 
The same equation can be obtained by a minimization of the free energy (\ref{asymptotic_free}).

 Note that subleading terms correspond to the repulsive force between the eigenvalues at $\sim 1/r$ distances.
Hence these terms wash out $\delta$-functions into the peaks of $1/r$ width. 
As we will see further, it is possible to obtain the analytic form of these peaks 
and reproduce (\ref{density:adj:gen:n}) in the decompactification limit $r\to\infty$.

To obtain solutions of (\ref{eq:adj:finite:r}) we first reduce this integral equation 
to the differential one taking three derivatives
\be
\frac{16\pi^2}{t}&=&\frac{4}{r^2}\rho(\phi)+\frac{1}{4r^2}\rho(\phi +m)+\frac{1}{4r^2}\rho(\phi -m)+\int d\psi \rho(\psi)
\left[-2|\phi-\psi|\right.\nonumber\\
&+& \left.|\phi-\psi +m|+|\phi-\psi -m|\right]\,,\label{1st:der:adj:finite:r}\\
0&=&\frac{4}{r^2}\rho'(\phi)+\frac{1}{4r^2}\rho'(\phi +m)+\frac{1}{4r^2}\rho'(\phi -m)
+\int d\psi \rho(\psi)\left[-2\,{\rm sign}(\phi-\psi)\right.\nonumber\\
&+&\left.{\rm sign}(\phi-\psi +m)+{\rm sign}(\phi-\psi -m)\right]\,,\label{2nd:der:adj:finite:r}\\
0&=&\frac{4}{r^2}\rho''(\phi)+\frac{1}{4r^2}\rho''(\phi +m)+\frac{1}{4r^2}\rho''(\phi -m)
- 4\rho(\phi)\nonumber\\
&+&2\rho(\phi +m)+2\rho(\phi -m)\,, \label{3rd:der:adj:finite:r}
\ee
where equations (\ref{1st:der:adj:finite:r}),(\ref{2nd:der:adj:finite:r}) and (\ref{3rd:der:adj:finite:r})
corresponds to the first, second and third derivatives w.r.t. $\phi$. 

Now we are ready to consider solutions for different phases of the theory. In particular, we  
consider only solutions with no resonances and one pair of resonances. 
In the decompactification limit $r\to\infty$ we want to observe 
solutions  (\ref{density:adj:0}) and (\ref{density:n=1}). Unfortunately it is very hard to 
observe solution for the case of arbitrary number of resonance pairs $n$. However 
obtaining particular cases of $n=0$ and $n=1$ will strongly suggest that solutions we 
have found in section \ref{adjoint:hyper:section} are right.

\subsection{Solution with no resonances ($n=0$)}

We start with the case for which we do not expect any resonances. This 
happens when $2 a<m$, where $a$ is the position of the support endpoint as usually.
In this case $\rho(\phi\pm m)=0$ as $\phi\pm m$ appears to be outside the support of the eigenvalue density. 
Hence (\ref{3rd:der:adj:finite:r}) can be written in the form
\be
\frac{4}{r^2}\rho''(\phi)-4\rho(\phi)=0\,.
\ee
General solution to this differential equation is given by
\be
\rho(\phi)=C \cosh(r\phi)\,,
\label{density:adj:n=0:finite:r}
\ee
where $C$ is an integration constant.
Notice that we omit $\sinh(r\phi)$
due to the symmetry $\rho(\phi)=\rho(-\phi)$.

Applying the normalization condition (\ref{normalization}) to 
(\ref{density:adj:n=0:finite:r}) we obtain 
\be
C=\frac{r}{2\sinh(ra)}\,.
\ee
Finally, substituting this solution into (\ref{1st:der:adj:finite:r}) we obtain
the following condition for the position of the support endpoint $a$
\be
\left(m-\frac{8\pi^2}{t}\right)r-ar+\coth(ra)=0\,,
\label{endp:adj:fin:r}
\ee
To summarize, we have found that solutions for the eigenvalue density of $\mathcal{N}=1$
SYM with massive adjoint hypermultiplet and no resonances is given by 
\be\begin{aligned}
\rho(\phi)&=\frac{r}{2\sinh(ra)}\cosh(r\phi)\,, ~~|\phi|<a\,,\\
          &=0,\qquad~~~~~~~~~~~~ \qquad~~|\phi|>a\,,
\label{density:adj:n=0:fin:r}
\end{aligned}\ee
and the eigenvalue support endpoint is defined by (\ref{endp:adj:fin:r}). 

Equation (\ref{endp:adj:fin:r}) can not be solved analytically. 
However in the decompactification limit $r\to\infty$ it simplifies a lot. 
Condition (\ref{endp:adj:fin:r}), defining the position of the support endpoint, reduces to 
\be
a\approx m-\frac{8\pi^2}{t}\,,
\ee
which exactly reproduces (\ref{endpoint:adj:0}) found after neglecting the subleading terms 
from the very beginning.
The eigenvalue density (\ref{density:adj:n=0:fin:r}) becomes highly 
peaked around $\phi=\pm a$ with the width of the peak $\sim \frac{1}{r}$. In the decompactification 
limit $r\to \infty $ (\ref{density:adj:n=0:fin:r}) can be considered as the sum of two $\delta$-functions
$\frac{1}{2}\left(\delta(\phi +a)+\delta(\phi -a)\right)$, because the peaks becomes infinitely high and narrow and 
integrate to one. So we see that (\ref{density:adj:n=0:fin:r}) in the decompactification limit  coincides 
with the solution (\ref{density:adj:0}) found in section \ref{adjoint:hyper:section}.

\begin{figure}[!h]
\begin{center}
  \subfigure[$t=20 ~~\left(t<t_c^{(1)}\right)$]{\label{pic:adj:below:rfin}\includegraphics[width=52mm,angle=0,scale=1.5]{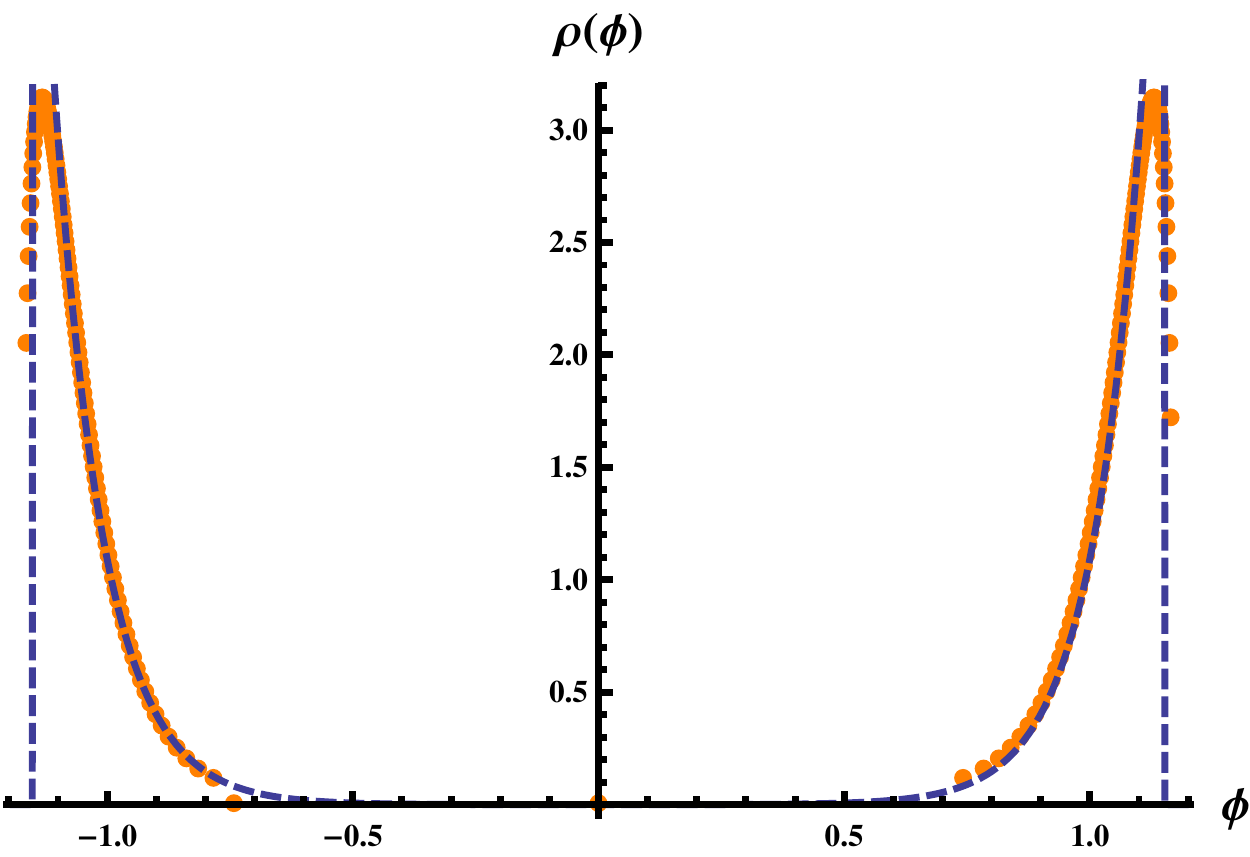}}
  \subfigure[$t=40 ~~\left(t_c^{(1)}<t<t_c^{(2)}\right)$]{\label{pic:adj:above:rfin}\includegraphics[width=52mm,angle=0,scale=1.5]{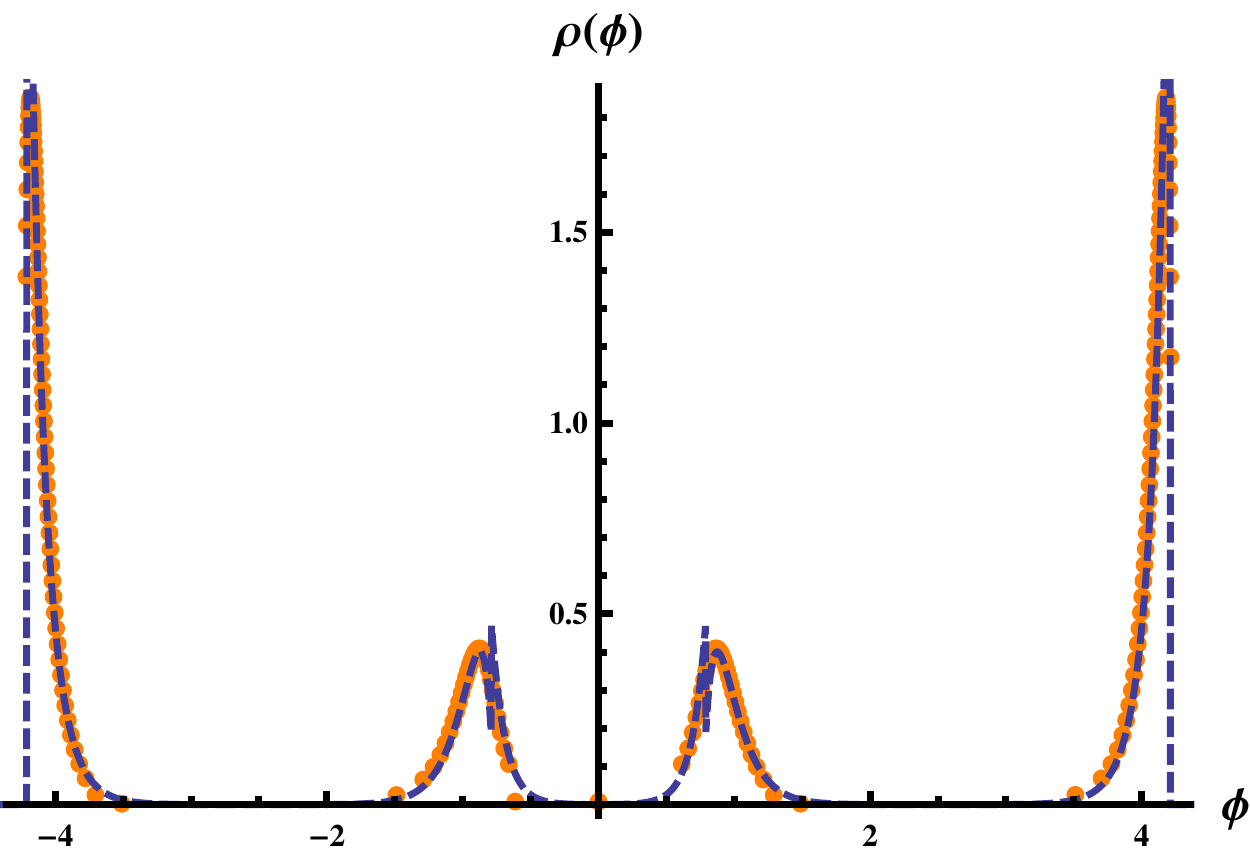}}
\end{center}
\caption{The eigenvalue density $\rho(\phi)$ calculated with the following value of the parameters: $r=10,\,m=5,\,,\,N=200$.
The orange dots show results for the numerical solution, while the dashed blue 
lines show the solutions (\ref{density:adj:n=0:finite:r}) and (\ref{density:adj:fin:r:n=1}) on $(a)$
and $(b)$ respectively.}
\label{adj:dens:finiter}
\end{figure}

We can also compare the solution found here with the results for the numerical simulations of the 
full equation of motion (\ref{eq-adj-full}). In particular in Fig.\ref{pic:adj:below:rfin}
we show the results of this numerical solution with the orange dots. On the same plot we show the analytical solution 
(\ref{density:adj:n=0:fin:r}) with the position of the cut endpoint found from (\ref{endp:adj:fin:r}) numerically. 
 As we see analytical solution is consistent with the numerical results.

\subsection{Solution with one pair of resonances ($n=1$)}

Now we can consider what happens with the increase of the coupling, when the support length 
becomes large enough to contain one pair of resonances, i.e. $m<2a<2m$. To solve equations 
(\ref{eq:adj:finite:r}) we should consider three intervals separated by the positions of the resonances
\be\begin{aligned}
                    &~~~ \rho_A(\phi),\quad ~~m-a<\phi<a\,;\\
\rho(\phi)=         &~~~ \rho_B(\phi),\quad -m+a<\phi<m-a\,;\\
                    &~~~ \rho_C(\phi),\quad -a<\phi<-m+a\,;
\end{aligned}\ee
Symmetry requirement for the eigenvalue density $\rho(\phi)=\rho(-\phi)$ implies following conditions
for the densities on the intervals $A,\,B,\,C$
\be
\rho_A(\phi)=\rho_C(-\phi)\,,\quad \rho_{B}(\phi)=\rho_B(-\phi)\,.
\label{symm:rho:adj}
\ee
Now we consider (\ref{3rd:der:adj:finite:r}) on the intervals $A$ and $B$ in order to find general 
solutions for $\rho_A$ and $\rho_B$ respectively. Then we substitute 
these general solutions into (\ref{1st:der:adj:finite:r}) and (\ref{2nd:der:adj:finite:r}) in order to fix 
integration constants together with the position of the support endpoint $a$.

{\bf Interval A} ($m-a<\phi<a$). Note that on this interval $\phi +m>a$ so that $\rho(\phi+m)=0$ and also 
$-a<\phi-m<a-m$ meaning $\phi - m\in ~C$. Using these remarks and symmetry properties (\ref{symm:rho:adj})
we can rewrite (\ref{3rd:der:adj}) as the following delay differential equation
\be
4\rho''_A(\phi)+\frac{1}{4}\rho''_A(m-\phi)-4r^2 \rho_A(\phi)+2r^2 \rho_A(m-\phi)=0\,.
\label{3rd:der:adj:fin:r:A}
\ee
General solution for this equation is given by
\be
\rho_A(\phi)=C_2 \cosh\left(\alpha r\left(\phi-\frac{m}{2}\right)\right)+
C_3 \sinh\left(\beta r\left(\phi-\frac{m}{2}\right)\right)\,,\quad\alpha^2=\frac{8}{17}\,,\quad \beta^2=\frac{8}{5}\,,
\label{density:A:adj:fin:r}
\ee
where coefficients $\alpha$ and $\beta$
where found by substituting these solutions back into (\ref{3rd:der:adj:fin:r:A}). This procedure is 
similar to writing out characteristic equation for ODE, however still a bit different due to 
delays we have in our equation. In particular single exponents can not satisfy (\ref{density:A:adj:fin:r}) and 
we should instead search for the solution in form of $\cosh\left(\alpha r\left(\phi-\frac{m}{2}\right)\right)$
and $\sinh\left(\beta r\left(\phi-\frac{m}{2}\right)\right)$. Shifts in the arguments can be seen as 
centering of the solution at the middle of the interval $A$ which is made for the convenience and in principle can
be adsorbed in constants $C_2$ and $C_3$.

{\bf Interval B} ($a-m<\phi<m-a$). On this interval situation is simpler then on $A$, 
because $\phi -m<-a$ and $\phi+m>a$ so that both arguments are outside the support and hence 
$\rho(\phi\pm m)=0$ leading to the following form of (\ref{3rd:der:adj:finite:r})
\be
4\rho''_B(\phi)-4r^2\rho_B(\phi)=0\,,
\ee
and its solution
\be
\rho_B(\phi)=C_1 \cosh(r \phi)\,,
\label{density:B:adj:fin:r}
\ee
where symmetry requirements (\ref{symm:rho:adj}) were taken into account.

{\bf Interval C} ($-a<\phi<a-m$). Finally on this interval using
symmetry properties (\ref{symm:rho:adj}) we obtain $\rho_C(\phi)$
directly from the density $\rho_A(\phi)$ given by (\ref{density:A:adj:fin:r}).

To summarize we have found
\be\begin{aligned}
              &= C_2 \cosh\left(\alpha r\left(\phi-\frac{m}{2}\right)\right)+
C_3 \sinh\left(\beta r\left(\phi-\frac{m}{2}\right)\right)\,,~~ m-a<\phi<a\,,\\
\rho(\phi)    &=C_1 \cosh(r \phi)\,,\quad~~~~~~~~~~~~~~~~~~~~~~~~~~~~~~~~~~~~~~~~~~~~~~~~a-m<\phi<m-a\,,\\
              &=C_2 \cosh\left(\alpha r\left(\phi +\frac{m}{2}\right)\right)-
C_3 \sinh\left(\beta r\left(\phi +\frac{m}{2}\right)\right)\,,~ -a<\phi<a-m\,,\\
              &=0\,, ~~\quad~~~~~~~~~~~~~~~~~~~~~~~~~~~~~~~~~~~~~~~~~~~~~~~~~|\phi|>a\,, \quad\alpha^2=\frac{8}{17}\,,\quad \beta^2=\frac{8}{5}\,.
\label{density:adj:fin:r:n=1}
\end{aligned}\ee

Now we should determine integration constants $C_1,\, C_2\,, C_3$ and the position of the support endpoint $a$ using 
normalization condition (\ref{normalization}) together with equations (\ref{eq:adj:finite:r})-(\ref{2nd:der:adj:finite:r}),
written out for different intervals. After some algebra we obtain the following 
system of algebraic equations 
\be
&&C_1\sinh(r(m-a))+2C_2 \alpha^{-1}\sinh\left(\alpha r\left(a-\frac{m}{2}\right)\right)=\frac{r}{2}\,,
\label{normal:adj:fin:r}\\
&&C_3\beta^{-1}\cosh\left(\beta r\left(a-\frac{m}{2}\right)\right)+C_2 \alpha^{-1}
\sinh\left(\alpha r\left(a-\frac{m}{2}\right)\right)=\frac{r}{3}\,,
\label{2nd:der:adj:A}\\
&&\frac{4m}{3}-\frac{2a}{3}+r^{-2}\left[4C_1 \cosh(r(a-m))+8C_3\beta^{-2}
\sinh\left(\beta r\left(a-\frac{m}{2}\right)\right)\right]=\frac{16\pi^2}{t}\,,
\label{1st:der:adj:B}\\
&&\frac{4m}{3}-\frac{2a}{3}+2r^{-2}\left[C_2\alpha^{-2}\cosh\left(\alpha r\left(a-\frac{m}{2}\right)\right)+
C_3\beta^{-2}\sinh\left(\beta r\left(a-\frac{m}{2}\right)\right)\right]=\frac{16\pi^2}{t}\,.\nonumber\\
\label{1st:der:adj:A}
\ee
Here in particular first equation (\ref{normal:adj:fin:r}) is obtained from the normalization condition (\ref{normalization}).
Second equation (\ref{2nd:der:adj:A}) is the result of the substitution of (\ref{density:adj:fin:r:n=1}) 
into (\ref{2nd:der:adj:finite:r})
with the assumption $\phi \in ~A$.
Finally, equations (\ref{1st:der:adj:B}) and (\ref{1st:der:adj:A})
are results of the substitution of (\ref{density:adj:fin:r:n=1}) into (\ref{1st:der:adj:finite:r})
with  $\phi \in~B$ and $\phi\in ~A$ respectively.

Equations (\ref{normal:cond:fund})-(\ref{1st:der:adj:A}) are transcendental and can not be solved analytically.
Instead we can consider these equations and their solutions in the decompactification limit $r\to\infty$.
Assuming that $r(a-\frac{m}{2})\gg1$ and $r(m-a)\gg1$, or equivalently assuming that the radius 
of $S^5$ is large and we are far enough from the critical points, we can rewrite these equations in the following 
form
\be
&&C_1 e^{r(m-a)}+2C_2 \alpha^{-1}e^{\alpha r\left(a-\frac{m}{2}\right)}=r\,,
\label{normal:adj:fin:r:dec}\\
&&C_3\beta^{-1}e^{\beta r\left(a-\frac{m}{2}\right)}+C_2 \alpha^{-1}
e^{\alpha r\left(a-\frac{m}{2}\right)}=\frac{2 r}{3}\,,
\label{2nd:der:adj:A:dec}\\
&&\frac{4m}{3}-\frac{2a}{3}+r^{-2}\left[2C_1 e^{r(m-a)}+4C_3\beta^{-2}
e^{\beta r\left(a-\frac{m}{2}\right)}\right]=\frac{16\pi^2}{t}\,,
\label{1st:der:adj:B:dec}\\
&&\frac{4m}{3}-\frac{2a}{3}+r^{-2}\left[C_2\alpha^{-2}e^{\alpha r\left(a-\frac{m}{2}\right)}+
C_3\beta^{-2}e^{\beta r\left(a-\frac{m}{2}\right)}\right]=\frac{16\pi^2}{t}\,.
\label{1st:der:adj:A:dec}
\ee
Solving these equations we find the following expressions for the integration constants 
\be\begin{aligned}
C_1&=r e^{-r(m-a)}\left[1-2\left(1+\beta^{-1}\right)\left(\alpha^{-1}+3\beta^{-1}+4\right)\right]\,,\\
C_2&=2 r \alpha e^{-\alpha r\left(a-\frac{m}{2}\right)}
\left(1+\beta^{-1}\right)\left(\alpha^{-1}+3\beta^{-1}+4\right)\,,\\
C_3&=2 r \beta e^{-\beta r\left(a-\frac{m}{2}\right)}\left[\frac{1}{3}-
\left(1+\beta^{-1}\right)\left(\alpha^{-1}+3\beta^{-1}+4\right)\right]\,,\label{C123:solution}
\end{aligned}\ee
The last unknown parameter of the solution is the support endpoint position $a$. To find 
it we substitute solutions for $C_2$ and $C_3$ into (\ref{1st:der:adj:A:dec}). Note 
that the last parenthesis containing $C_2$ and $C_3$ is of order $r^{-1}$. Hence 
in the decompactification limit this term is irrelevant and we are left with the relation
\be
a=2m-\frac{24 \pi^2}{t}\,,
\ee
which reproduces the result (\ref{endp:adj:n=1}) obtained in section \ref{adjoint:hyper:section}.
Moreover, we can reproduce solution (\ref{density:n=1}) for the eigenvalue density from (\ref{density:adj:fin:r:n=1}).
To do this lets first notice that $\delta$-function can be defined through the following limit
\be
\delta(\phi-c)=\lim\limits_{r\to\infty} f(r)\,, ~~{\rm where} ~~ f(r)&=& \alpha re^{\alpha r(\phi-c)}\,,~ \phi<c\\
&=& 0\,,~~~~~~~~~~~~\phi>c\,.
\label{delta:fucntion:def}
\ee
Indeed the function $f(r)$ is peaked 
at $\phi=c$ for large $r$ and in the limit $r\to \infty$ this peak becomes infinitely narrow and infinitely 
high, while the integral of $f(r)$ is normalized to $1$. Thus in the decompactification
limit this function can be considered as the $\delta$-function.
We have already used this relation with $\alpha=1$ while considering 
decompactification limit of the solution (\ref{density:adj:n=0:finite:r}) with no resonances. 

The distribution (\ref{density:adj:fin:r:n=1}) has similar exponential peaks 
at $\phi=\pm a$ and $\phi=\pm(m-a)$ and thus we expect to obtain the following eigenvalue 
density in the decompactification limit
\be
\rho(\phi)=c_0 \delta(\phi -a)+c_1\delta(\phi-m+a)+c_1\delta(\phi +m-a)+c_0\delta(\phi + a)\,.
\ee
Coefficients $c_0$ and $c_1$ can be obtained from the solution (\ref{density:adj:fin:r:n=1})
using the normalization prefactor in the definition of the $\delta$-function (\ref{delta:fucntion:def})
\be\begin{aligned}
c_0&=\frac{1}{2}C_2\left(\alpha r\right)^{-1}e^{\alpha r \left(a-\frac{m}{2}\right)}+
\frac{1}{2}C_3\left(\beta r\right)^{-1}e^{\beta r \left(a-\frac{m}{2}\right)}=\frac{1}{3}\,,\\
c_1&=\frac{1}{2}C_2\left(\alpha r\right)^{-1}e^{\alpha r \left(a-\frac{m}{2}\right)}-
\frac{1}{2}C_3\left(\beta r\right)^{-1}e^{\beta r \left(a-\frac{m}{2}\right)}
+\frac{1}{2}C_1 r^{-1}e^{r(m-a)}=\frac{1}{6}\,,
\end{aligned}\ee
where we have also used solutions (\ref{C123:solution}) for the coefficients $C_1\,,\, C_2\,,\, C_3$.
As we see in the decompactification limit we completely reproduce solution obtained
in section \ref{adjoint:hyper:section}.

Finally, we compare analytical solution (\ref{density:adj:fin:r:n=1}) with the numerical results.
We show this comparison in Fig.\ref{pic:adj:above:rfin}. The orange dots represent
the results of the numerical solution, while dashed blue lines on the same picture show analytical solution
(\ref{density:adj:fin:r:n=1}) with the parameters $C_1\,,\,C_2\,,\,C_3$ and $a$
evaluated numerically from the system of equations (\ref{normal:adj:fin:r})-(\ref{1st:der:adj:A}). As we see 
from this picture analytical results coincide with the numerical ones perfectly. Small discontinuities at 
the resonances positions $\phi=\pm (m-a)$ are related to high sensitivity of 
the exponential factors in this solution as well as in the system of equations (\ref{normal:adj:fin:r})-(\ref{1st:der:adj:A}) 
which we solve numerically in the end. Hence these discontinuities are numerical artifacts and can be neglected.

\bibliographystyle{JHEP}
\bibliography{phases}

\end{document}